\documentclass[aps,pra,superscriptaddress,twocolumn,a4paper,reprint,noeprint]{revtex4-1}
\usepackage[utf8]{inputenc}
\usepackage{graphicx}
\usepackage{amsmath}
\usepackage{amsfonts}
\usepackage{amssymb}
\usepackage{bbm}
\usepackage{bm}
\usepackage{bbold}
\usepackage[usenames,dvipsnames,svgnames,table]{xcolor}
\usepackage[USenglish]{babel}
\usepackage{hyperref}
\usepackage{grffile}

\DeclareMathOperator{\Tr}{Tr\,}

\newcommand{\eq}[1]{Eq.~(\ref{#1})} 
\newcommand{\fig}[1]{Fig.~\ref{#1}} 
\newcommand{\Sec}[1]{Sec.~\ref{#1}} 
\newcommand{\ignore}[1]{}
\newcommand{\abs}[1]{\ensuremath{|#1|}}

\allowdisplaybreaks

\begin{document}

\title{Bogoliubov Fermi surfaces: General theory, magnetic
order, and topology}

\author{P. M. R. Brydon}
\email{philip.brydon@otago.ac.nz}
\affiliation{Department of Physics and MacDiarmid Institute for
Advanced Materials and Nanotechnology, University of Otago, P.O. Box
56, Dunedin 9054, New Zealand}
\author{D. F. Agterberg}
\email{agterber@uwm.edu}
\affiliation{Department of Physics, University of Wisconsin,
Milwaukee, WI 53201, USA}
\author{Henri Menke}
\affiliation{Department of Physics  and MacDiarmid Institute for
Advanced Materials and Nanotechnology, University of Otago, P.O. Box
56, Dunedin 9054, New Zealand}
\author{C. Timm}
\email{carsten.timm@tu-dresden.de}
\affiliation{Institute of Theoretical Physics, Technische
 Universit\"at Dresden, 01062 Dresden, Germany}

\begin{abstract}
We present a comprehensive theory for Bogoliubov Fermi
surfaces in inversion-symmetric superconductors which break
time-reversal symmetry. A requirement for such a gap structure is that the
electrons posses internal
degrees of freedom apart from  the spin (e.g., orbital or
sublattice indices), which permits a nontrivial internal structure
of the Cooper pairs. We develop a general theory for such a pairing
state, which we show to be nonunitary. A time-reversal-odd component
of the nonunitary gap product is found to be essential for the
appearance of Bogoliubov Fermi surfaces. These Fermi
surfaces are topologically protected by a
$\mathbb{Z}_2$ invariant. We examine their appearance in a generic
low-energy effective model and then study two
specific microscopic models supporting Bogoliubov Fermi surfaces:
a cubic material with  a $j=3/2$ total-angular-momentum degree of
freedom and a hexagonal material with distinct orbital and spin
degrees of freedom. The appearance of Bogoliubov Fermi surfaces
is accompanied by a magnetization of the low-energy
states, which we connect to the time-reversal-odd component of the
gap product. We additionally calculate the surface spectra associated
with these pairing states and demonstrate that the Bogoliubov Fermi
surfaces are characterized by additional topological indices.
Finally, we discuss the extension of phenomenological theories 
of superconductors to include Bogoliubov Fermi surfaces,
and identify the time-reversal-odd part of the gap product as a
composite order parameter which is intertwined with
superconductivity.
\end{abstract}

\maketitle

\section{Introduction}

A common view of multiband superconductivity is that
the superconducting state is qualitatively like a single-band
superconductor \cite{VoG85,SiU91} but with a
momentum-dependent gap, which in particular can take on different
values on different Fermi surface sheets \cite{suh59}. However,
motivated in part by developments in topological materials
\cite{qi11,sch08,ChS14,CTS16,ScB15}, it has recently  been
realized that the internal electronic degrees of freedom
(i.e., orbital or sublattice) which give rise to the multiband
structure can also appear in the 
Cooper pair wavefunction. Pairing states involving a nontrivial
dependence on these internal degrees of freedom, which we refer to
as ``internally anisotropic'' states, have been proposed for 
many multiband systems, such as the iron-based superconductors
\cite{GSZ10,NGR12,NKT16,nic17,ong16,CVF16,agt17}, nematic
superconductivity in Cu$_x$Bi$_2$Se$_3$ \cite{CuBi2Se3,yon17},
$j=3/2$ pairing in cubic materials motivated by the half-Heusler
compounds 
\cite{BWW16,SRV17,YXW17,RGF17,BoH18,ABT17,kim18,TSA17}, and $j=5/2$
pairing and topological superconductivity in UPt$_3$
\cite{NoI16,NoI16-2,yan16}. These pairing states have also attracted
attention as a way to generate odd-frequency pairing
\cite{bla13,kom15,LiB17} and an intrinsic ac Hall
conductivity that is responsible for the polar magneto-optical
Kerr effect in superconductors with broken time-reversal
symmetry (TRS)~\cite{tay12,BAA18}.    

 Despite this interest, an unambiguous example of an internally
anisotropic pairing state has yet to be
established. A key problem is that in most of the
cases mentioned above, pairing states with trivial and nontrivial
dependence on the internal degrees of freedom can have qualitatively
the same low-energy excitation spectra. The most accessible
experimental probes of unconventional superconductivity, which are
sensitive only to the nodal structure of the excitation gap,
thus cannot distinguish between trivial and
nontrivial pairing. Indeed, the proposed experimental signatures of these
exotic pairing states are quite subtle, e.g., enhanced robustness
against 
disorder~\cite{mif12}, high-energy anomalies in the density of
states~\cite{kom15}, the existence of the polar Kerr
effect~\cite{tay12,BAA18}, and exotic domain structures~\cite{ong16}.
Recently, we have shown that in one important case the consideration
of internal electronic 
degrees of freedom  leads to a unique signature in the
gap structure: in clean, inversion-symmetric (even-parity) superconductors that
spontaneously break 
TRS, the superconducting state is either fully gapped or has
topologically protected Bogoliubov Fermi surfaces \cite{ABT17}. In the
single-band case, the corresponding superconducting state would
not have Bogoliubov Fermi surfaces but rather exhibit point or line
nodes \cite{VoG85,SiU91}.  In the multiband case, these nodes are
replaced by two-dimensional
Fermi surfaces by the inclusion of the internal electronic degrees
of freedom. We note that Bogoliubov Fermi surfaces have
been discussed for other superconductivity and superfluid
systems, to
which our theory does not apply. In particular, they have been
proposed in strong-coupling superconductors \cite{vol89}, in
superconductors and superfluids in which TRS is
broken through an external effective magnetic field
\cite{liu03,gub05}, and in superconductors and superfluids which
break both TRS and inversion symmetry (IS)~\cite{TSA17,yua18,vol18}. 

Candidates for superconductors that break TRS
have been experimentally identified through muon-spin-rotation
and polar-Kerr-effect measurements, and include
UPt$_3$ \cite{luk93,sch14}, Th-doped UBe$_{13}$ \cite{hef90},
PrOs$_4$Sb$_{12}$ \cite{aok03,lev18}, Sr$_2$RuO$_4$ \cite{luk98,xia06},
URu$_2$Si$_2$ \cite{sch15}, SrPtAs \cite{bis13}, and Bi/Ni bilayers
\cite{gon16}. In addition, theory 
has predicted additional possibilities such as graphene
\cite{nan12,abs14}, twisted bilayer graphene \cite{cao18-1,cao18-2},
the half-Heusler compound YPtBi \cite{BWW16}, water-intercalated
sodium cobaltate Na$_x$CoO$_{2}\cdot y$H$_2$O \cite{sar04,kie13},
Cu-doped TiSe$_2$ \cite{gan14}, and monolayer transition-metal
dichalcogenides \cite{hsu16}. In all these cases, multiple
bands either cross
or come close to the Fermi surface, thus meeting the conditions for
the appearance of Bogoliubov Fermi surfaces.
 
In this paper we present a comprehensive theory for the origins
and properties of Bogoliubov Fermi surfaces.
We first develop a general theory for electrons with four-valued
internal degrees of freedom. Our theory is not restricted
to a specific physical origin of these degrees of freedom; they could, for
example, be total-angular-momentum states or a combination of a
two-valued spin and a two-valued orbital degree of freedom.  The
normal state is assumed to be 
invariant under time reversal and inversion so that the spectrum
generically has two doubly degenerate bands. We consider
a generic, inversion-symmetric (even-parity) superconducting state
that preserves IS 
but may break TRS. Using this theory, we establish  the following
results: (\textit{i}) The gap is nonunitary. We define a
time-reversal-odd gap product that describes the contribution to
nonunitary pairing that is needed to understand the
origin of the Bogoliubov Fermi surfaces. (\textit{ii}) The spectrum
of the Bogoliubov-de Gennes (BdG) Hamiltonian
contains Bogoliubov Fermi surfaces when TRS is broken.
(\textit{iii}) These Bogoliubov Fermi surfaces are topologically
protected by a $\mathbb{Z}_2$ invariant, which we give in terms of a
Pfaffian. (\textit{iv}) In an
effective low-energy single-band model, the superconductor generates a
pseudomagnetic field that is closely linked to the time-reversal-odd
gap product. This pseudomagnetic field inflates point and line nodes
into Bogoliubov Fermi surfaces.  

We then apply this generic theory to two specific models:
First, we consider cubic materials with $j=3/2$
electronic degrees of freedom, which can appear in the vicinity of
the $\Gamma$ point in the Brillouin zone. In particular,
we specify the pseudomagnetic fields and the associated magnetization,
the structure and topology of the Bogoliubov Fermi surfaces, and the
surface states that appear in the possible TRS-breaking (TRSB)
superconducting states. Second, we
consider hexagonal superconductors in which the internal electronic
degrees of freedom stem from a two-valued spin and a two-valued
orbital degree of freedom.  We then turn back to a more general discussion,
elucidating the topological invariants
associated with Bogoliubov Fermi surfaces, using the cubic $j=3/2$
system to illustrate the results. We conclude by proposing a
phenomenological Landau theory in which the magnetic and orbital order
appear as an emergent composite order parameter. 
We speculate that the composite order could be present even if
the primary superconducting 
order is absent, providing an example for intertwined order
parameters~\cite{FKT15,FOS18,BAA18}.

\section{General theory}
\label{sec:general_model}

Our starting point is a generic model of a fermionic system with four
internal degrees of freedom that is invariant under
time reversal and inversion. This model includes such
important cases as two-orbital models of the pnictides~\cite{RXS08}and
Sr$_2$RuO$_4$~\cite{tay12}, as well as the $\Gamma_8$ bands of cubic
materials with spin-orbit coupling~\cite{L56}.

The general form of the BdG Hamiltonian reads 
\begin{equation}
H = \frac{1}{2} \sum_{\bf k}\Psi^\dagger_{\bf k}{\cal H}_{\bf
  k}\Psi_{\bf k}\,, \label{eq:genHBdG}
\end{equation}
where $\Psi_{\bf k}=(c^T_{\bf k},c^\dagger_{-{\bf k}})^T$ is a Nambu
spinor, $c_{\bf k}$ is a four component spinor encoding the 
internal degrees of freedom, and  the coefficient matrix is
\begin{equation}
{\cal H}_{\bf k} = \left(\begin{array}{cc}
  H_0({\bf k}) & \Delta({\bf k})\\
  \Delta^\dagger({\bf k}) & -H_0^T(-{\bf k})
  \end{array}\right) .
\label{eq:BdGH}
\end{equation}
The normal-state Hamiltonian $H_0({\bf k})$ can be written as 
\begin{equation}
{H}_0({\bf k}) = (\epsilon_{{\bf k},0}-\mu)\,\mathbb{1}_4 +
\vec{\epsilon}_{\bf k}\cdot\vec{\gamma}\,, \label{eq:genH}
\end{equation}
where $\mathbb{1}_4$ is the $4\times4$ unit matrix and 
$\vec{\gamma}=(\gamma^1,\gamma^2,\gamma^3,\gamma^4,\gamma^5)$ is the
vector of the five anticommuting Euclidean Dirac matrices.
The real functions
$\epsilon_{{\bf k},0}$ and $\vec{\epsilon}_{\bf k}
  =(\epsilon_{{\bf k},1},\epsilon_{{\bf k},2},\epsilon_{{\bf k},3},
  \epsilon_{{\bf k},4},\epsilon_{{\bf k},5})$
are the coefficients of these  
matrices and $\mu$ is the chemical potential. We make the simplifying
assumption that IS $P$ acts trivially on the internal
degrees of freedom  so that the coefficients in \eq{eq:genH} are even
functions of momentum. Time reversal is implemented by
$T = {\cal K}U_T$, where ${\cal K}$ is complex conjugation and the
unitary part can be chosen, without loss of generality, as
$U_T = \gamma^1\gamma^2$. The invariance of the normal-state
Hamiltonian under 
time reversal then implies that $\gamma^1$ and $\gamma^2$ are both
imaginary, and the other three matrices are real.

The normal-state Hamiltonian in \eq{eq:genH} has the doubly degenerate
eigenvalues $E_{{\bf k},\pm}-\mu$, where
\begin{equation}
E_{{\bf k},\pm} = \epsilon_{{\bf k},0} \pm |\vec{\epsilon}_{\bf k}|\,.
\end{equation}
Due to the presence of IS and TRS, we can distinguish the
two states corresponding to each eigenvalue by a pseudospin 
index $s=\pm 1$. The pseudospin-$s$ state $|{\bf k},\pm,s\rangle$ in
the $\pm$ band at momentum ${\bf k}$ then transforms as
\begin{align}
P\,|{\bf k},\pm,s\rangle &= |{-}{\bf k},\pm,s\rangle\,,
\label{eq:ps_inv} \\
T\,|{\bf k},\pm,s\rangle &= -s\,|{-}{\bf k},\pm,-s\rangle\,.
\label{eq:ps_tr}
\end{align}
Although the pseudospin basis only needs to satisfy these two
criteria, it is nevertheless often possible to choose the basis such
that the pseudospin index transforms like a true spin $1/2$ under the
symmetries of the lattice, a so-called
manifestly covariant Bloch basis (MCBB) \cite{F15}. In Appendix
\ref{app.ps}, we present choices of MCBBs for the two   
model systems considered in the rest of the paper. We note, however,
that the analysis in this section requires only that
Eqs.\ (\ref{eq:ps_inv}) and (\ref{eq:ps_tr}) are satisfied.

Topologically stable Bogoliubov Fermi surfaces only appear for 
inversion-symmetric superconducting states. The pairing
potential consistent with this has the general form 
\begin{equation}
{\Delta}({\bf k})
  = \eta_{{\bf k},0}\,U_T + \vec{\eta}_{\bf k}\cdot\vec{\gamma}\,U_T \,,
\label{eq:genpair}
\end{equation}
where the pairing amplitudes $\eta_{{\bf k},0}$ and
$\vec{\eta}_{\bf k}
  =(\eta_{{\bf k},1},\eta_{{\bf k},2},\eta_{{\bf k},3},
  \eta_{{\bf k},4},\eta_{{\bf k},5})$
are even
functions of momentum. The first term in \eq{eq:genpair} describes
standard pairing between time-reversed states. This we call
``internally isotropic'' pairing to describe how the underlying
electronic degrees of freedom are paired. The second term
describes pairing in the five ``internally anisotropic''
channels, where the electronic degrees of freedom
in the Cooper pair do not generally come from Kramers partners. 
  In general, these pairing states transform nontrivially under
lattice symmetries 
due to their dependence on the internal degrees of freedom.
The pairing potential breaks TRS
if the coefficients $\eta_{{\bf k},0}$ and $\vec{\eta}_{\bf k}$ cannot
be chosen as real, up to a common and momentum-independent phase
factor. 

Expressed in the pseudospin basis where the annihilation operator has
the spinor form
$\tilde{c}_{\bf k}^T=(c_{{\bf k},+,\uparrow},c_{{\bf k},+,\downarrow},
  c_{{\bf k},-,\uparrow},c_{{\bf k},-,\downarrow})$,
the pairing Hamiltonian reads
\begin{equation}
\tilde{\Delta}({\bf k}) = \left(\begin{array}{cc}
  \psi_{{\bf k},+}\,i{s}_y &
  (\psi_{{\bf k},I}{s_0}
    +i{\bf d}_{\bf k}\cdot{\bf s})\,i{s}_y \\
  (\psi_{{\bf k},I}{s_0}
    -i{\bf d}_{\bf k}\cdot{\bf s})\,i{s}_y &
  \psi_{{\bf k},-}\,i{s}_y
  \end{array}\right) ,
\label{eq:Deltaps}
\end{equation}
where $\mathbf{s}=(s_x,s_y,s_z)$ is the vector of Pauli matrices
and $s_0$ is the unit matrix in pseudospin space
and all functions in the matrix are even in momentum. The
intraband pseudospin-singlet pairing potentials on the diagonal have
the basis-independent form
\begin{equation}
\psi_{{\bf k},\pm} = \eta_{{\bf k},0}
  \pm \frac{\vec{\epsilon}_{\bf k}\cdot
  \vec{\eta}_{\bf k}}{|\vec{\epsilon}_{\bf k}|}\,.
\label{eq:psiintra}
\end{equation}
The off-diagonal blocks describe unconventional interband pairing,
with both pseudospin singlet and triplet potentials,
$\psi_{{\bf k},I}$ and ${\bf d}_{\bf k}$, respectively.
While the form of the interband pairing potentials depends on
the choice of pseudospin basis in each band, these potentials
must satisfy
\begin{equation}
|\psi_{{\bf k},I}|^2 + |{\bf d}_{\bf k}|^2 = |\vec{\eta}_{\bf k}|^2 -
  \frac{|\vec{\epsilon}_{\bf k}\cdot\vec{\eta}_{\bf k}|^2}
  {|\vec{\epsilon}_{\bf k}|^2}\,.
\end{equation}
The interband terms involve only the internally anisotropic
pairing channels, as the pairing of time-reversed
partners in the conventional state (i.e., the internally
isotropic pairing) implies a purely intraband potential. Note that
the sign change between the pseudospin triplet potentials in the
off-diagonal blocks of \eq{eq:Deltaps} is required by fermionic
antisymmetry; the factor of $i$ ensures that ${\bf d}$ is a real
vector in the case of a time-reversal-symmetric pairing state.

\phantom{x}

\subsection{Nonunitary pairing and time-reversal-odd gap product}

The presence of the five internally anisotropic 
pairing channels in our model generically implies that the pairing
is \emph{nonunitary}. That is, the product $\Delta({\bf k})
\Delta^\dagger({\bf k})$ is not proportional to the unit matrix,
but is instead given by 
\begin{align}
\Delta({\bf k})\Delta^\dagger({\bf k}) &= \left(|\eta_{{\bf k},0}|^2
  + |\vec{\eta}_{\bf k}|^2\right)\mathbb{1}_4
  + 2\, \text{Re}(\eta^\ast_{{\bf k},0}\vec{\eta}_{\bf k})
    \cdot\vec{\gamma} \notag \\
&\quad{}+ \sum_{n>m>0}
  2i\,\text{Im}(\eta_{{\bf k},n}\eta_{{\bf k},m}^\ast)\,
  \gamma^n\gamma^m\,.
\label{eq:nonunitary_gen}
\end{align}
The first term on the right-hand side represents the unitary part of
the gap product, while the next two terms constitute the nonunitary
part. The first of these appears when pairing occurs in both 
the internally isotropic and internally anisotropic
pairing channels and does not
require the breaking of any symmetry. The second nonunitary term is
only present in a TRSB state with a nontrivial phase
difference between at least two internally anisotropic
channels. As we shall
see below, only the latter term is relevant for the appearance of the
Bogoliubov Fermi surfaces.
For later reference, we also give the gap product
in the pseudospin basis,
\begin{widetext}
\begin{align}
&\tilde{\Delta}({\bf k}) \tilde{\Delta}^\dagger({\bf k})
  = \left[\frac{1}{2} \left(|\psi_{{\bf k},+}|^2 + |\psi_{{\bf k},-}|^2\right)
     + |\psi_{{\bf k},I}|^2 + |{\bf d}_{\bf k}|^2\right]\mathbb{1}  \notag \\
&{}+ \left(\begin{array}{cc}
  \tfrac{1}{2}(|\psi_{{\bf k},+}|^2-|\psi_{{\bf k},-}|^2)s_0
  +(i{\bf d}_{\bf k}\times{\bf d}^\ast_{\bf k}
    + 2\,\text{Im}(\psi_{{\bf k},I}{\bf d}^\ast_{\bf k}))
    \cdot{\bf s} &
  (\psi_{{\bf k},+}\psi^\ast_{{\bf k},I}
    + \psi_{{\bf k},-}^\ast\psi_{{\bf k},I})s_0
    + i(\psi_{{\bf k},+}{\bf d}^\ast_{\bf k}
    + \psi_{{\bf k},-}^\ast{\bf d}_{\bf k})\cdot{\bf s} \\
  (\psi_{{\bf k},-}\psi^\ast_{{\bf k},I}
    + \psi_{{\bf k},+}^\ast\psi_{{\bf k},I})s_0
    - i(\psi_{{\bf k},+}^\ast{\bf d}_{\bf k}
    + \psi_{{\bf k},-}{\bf d}^\ast_{\bf k})\cdot{\bf s} &
  \tfrac{1}{2}(|\psi_{{\bf k},-}|^2-|\psi_{{\bf k},+}|^2)s_0
    +(i{\bf d}_{\bf k}\times{\bf d}^\ast_{\bf k}
   - 2\,\text{Im}(\psi_{{\bf k},I}{\bf d}^\ast_{\bf k}))
    \cdot{\bf s}
  \end{array}\right) .
\label{eq:nonunitary_band}
\end{align}
\end{widetext}
The diagonal blocks of the nonunitary part
will play an important role later, in
particular the terms involving the pseudospin vector
${\bf s}$. Since these terms only depend on the interband
pairing potentials in \eq{eq:Deltaps}, they arise from the last term
in \eq{eq:nonunitary_gen} and hence require TRSB pairing in different
internally anisotropic channels. 

To gain insight into the physical meaning of the nonunitary gap
and its relation to broken TRS, we briefly
review the more familiar case of nonunitary pairing in a single
band of spin-$1/2$ electrons \cite{SiU91}. Here it is customary to
write the gap function as
$\Delta_{\bf k}=(\psi_{\bf k}
  + {\bf d}_{\bf k}\cdot\bm{\sigma})\,i\sigma_y$,
where $\psi_{\bf k}$ is the singlet
pairing potential, ${\bf d}_{\bf k}$ describes triplet pairing, and
$\bm{\sigma}$ is the vector of spin Pauli matrices. The gap product is then
\begin{align}
\Delta^{}_{\bf k}\Delta_{\bf k}^\dagger &= (|\psi_{\bf k}|^2+|{\bf d}_{\bf
  k}|^2)\,\sigma_0 + 2\, \text{Re}(\psi_{\bf k}{\bf d}^\ast_{\bf
  k})\cdot\bm{\sigma} \notag \\
&\quad {}+ i\,({\bf d}_{\bf k}\times{\bf d}^\ast_{\bf k})\cdot\bm{\sigma}\,.
\end{align}
The presence of either of the last two terms indicates a nonunitary
gap, which requires the breaking of IS or TRS,
respectively. The presence of the nonunitary part of the gap
product indicates a nonzero value of the spin polarization
$\Tr (\Delta_{\bf k}^\dagger\bm{\sigma}\Delta^{}_{\bf k})$
of the pairing state at ${\bf k}$. This spin polarization has
two contributions, one that breaks TRS and one
that does not. The latter is a consequence of broken
IS and typically the associated spin polarization
is already present in the normal state \cite{smi17}. The
spin polarization due to broken TRS does not
exist in the normal state  but appears spontaneously in the
TRSB superconducting state and is usually taken as the defining
characteristic of nonunitary pairing
\cite{SiU91}. Below we will define a time-reversal-odd gap product
that isolates this contribution, refining the meaning of nonunitary
pairing.

Returning to our four-component system, the
nonunitary part of the gap product in \eq{eq:nonunitary_gen}
can be similarly interpreted as a polarization of
the internal degrees of freedom in
the pairing state. Moreover, the terms proportional to the pseudospin
Pauli matrices in the diagonal blocks of \eq{eq:nonunitary_band} may also
be interpreted as the pseudospin polarization
$\Tr [\Delta^\dagger({\bf k}){\cal P}_{{\bf k},\pm}\check{\bf s}
  {\cal P}_{{\bf k},\pm} \Delta({\bf k})]$
of the pairing state in the $\pm$ band, where ${\cal P}_{{\bf k},\pm}$
are projection operators on the normal-state
Hilbert space which project onto the $\pm$ bands
at momentum ${\bf k}$ and
\begin{equation}
\check{\bf s} \equiv \left(\begin{array}{cc}
  {\bf s} & 0 \\
  0 & {\bf s}
  \end{array}\right) .
\label{eq:checks}
\end{equation}
We thus expect a nonvanishing 
pseudospin polarization of the low-energy states in our model.
As in the single-band case discussed above, there will be a
contribution to this pseudospin polarization that is due solely to
the spontaneous breaking of TRS in the
superconducting state. In the next paragraph we discuss how to
identify its origin.

To link more closely to broken TRS, it is useful to refine
the notion of the nonunitary portion of the gap product and define a
time-reversal-odd gap product.
The time-reversal operator expressed in the Nambu basis is
$T = {\cal K}U_T \tau_0$,
where $\tau_0$ is the unit matrix in particle-hole space.
Time reversal operates as 
\begin{equation} 
{\cal H}_{\bf k}\rightarrow (\tau_0U_T){\cal H}^*_{-\bf k}
  (\tau_0U_T)^{\dagger}\,.
\end{equation}
From this expression, the form of the gap function, and
$U_T=\gamma^1\gamma^2$, we find the key and natural result that the
gap function transforms as 
\begin{equation}
\Delta({\bf k}) \rightarrow
 \Delta_T({\bf k}) \equiv
  U_T\Delta^*({\bf -k})U_T^{\dagger} 
  = (\eta^*_{{\bf k},0} + \vec{\eta}_{\bf k}^*\cdot\vec{\gamma})\,U_T \,.
\end{equation}
Similarly, under time reversal the gap product transforms as
\begin{equation}
\Delta({\bf k})\Delta^{\dagger}({\bf k})\rightarrow U_T
  \Delta^*(-{\bf k})\Delta^T(-{\bf k })U_T^\dagger
  = \Delta_T({\bf k}){\Delta}_T^{\dagger}({\bf k}) \,.
\end{equation}
This justifies the following time-reversal-odd gap product as a measure
of broken TRS:
\begin{align}
\Delta({\bf k})&\Delta^{\dagger}({\bf k})
  -\Delta_T({\bf k})\Delta_T^{\dagger}({\bf k}) \notag \\
&= (\vec{\eta}_{\bf k}\cdot\vec{\gamma})(\vec{\eta}_{\bf
  k}^*\cdot\vec{\gamma})-(\vec{\eta}_{\bf
  k}^*\cdot\vec{\gamma})(\vec{\eta}_{\bf k}\cdot\vec{\gamma})
  \notag \\
&= \sum_{i,j}(\eta_i\eta_j^*-\eta_i^*\eta_j)\, \gamma_i\gamma_j \,,
\label{eq:time_gap_product}
\end{align}
which yields the time-reversal-odd contribution to
\eq{eq:nonunitary_gen}. Applying the same analysis to the gap
function
$\Delta_{\bf k}=(\psi_{\bf k}+{\bf d}_{\bf k}
  \cdot\bm{\sigma})\,i\sigma_y$ of the single-band model,
the time-reversal-odd gap product is
\begin{equation}
\Delta_{\bf k}\Delta_{\bf k}^\dagger
  - \sigma_y\Delta^\ast_{-\bf k}\Delta^T_{-\bf k}\sigma_y
  = 2i\,({\bf d}_{\bf k}\times {\bf d}_{\bf k}^*)\cdot
  \bm{\sigma} \,,
\end{equation}
yielding the term that is usually taken to define a nonunitary
superconductor \cite{SiU91}. Finally, if the time-re\-ver\-sal-odd gap
product is calculated for the gap function expressed in the
pseudospin basis, \eq{eq:Deltaps}, then the terms proportional
to the pseudospin Pauli matrices in the diagonal blocks of
\eq{eq:nonunitary_band} are the only terms that remain
in these blocks. As mentioned earlier, these terms play a central
role in the effective low-energy model.

\subsection{Bogoliubov Fermi surfaces}
\label{sub.BFsurfaces}

The BdG Hamiltonian in \eq{eq:BdGH} possesses both 
par\-ti\-cle-hole symmetry $C$ and IS $P$. Par\-ti\-cle-hole
symmetry dictates that
\begin{equation}
U_C{\cal H}_{-{\bf k}}^\ast U_C^\dagger = -{\cal H}_{{\bf k}}\,,
\end{equation}
where the unitary part is $U_C = \tau_x\otimes \mathbb{1}_4$ and
$\tau_i$ are the Pauli matrices in particle-hole space. Inversion
acts as
\begin{equation}
U_P{\cal H}_{-{\bf k}} U_P^\dagger = {\cal H}_{{\bf k}}\,,
\end{equation}
where $U_P = \tau_0\otimes \mathbb{1}_4$. The product of these
symmetries thus gives
\begin{equation}
U_{CP}{\cal H}_{{\bf k}}^\ast U^\dagger_{CP}=-{\cal H}_{{\bf k}}\,,
\end{equation}
where $U_{CP}=U_{C}U_{P}^\ast = \tau_x\otimes \mathbb{1}_4$. It
hence follows that $(CP)^2=+1$. The existence of this $CP$ symmetry
and the property that it squares to unity guarantees that
the BdG Hamiltonian can be unitarily transformed into an
antisymmetric matrix \cite{ABT17}. For example, defining
\begin{equation}
\tilde{\cal H}_{\bf k} = \Omega {\cal H}_{\bf k}\Omega^\dagger\,,
\end{equation}
where
\begin{equation}
\Omega = \frac{1}{\sqrt{2}}\left(\begin{array}{cc}
  1 & 1 \\
  i & -i
  \end{array}\right)\otimes \mathbb{1}_4\,,
\label{eq:defOmega}
\end{equation}
we find that $\tilde{\cal H}^T_{\bf k} = -\tilde{\cal H}_{\bf k}$.
We can then evaluate the Pfaffian of this matrix, which is given
in compact form by
\begin{align}
P({\bf k}) = \text{Pf}\: \tilde{\cal H}_{\bf k} &= \big(\langle
  \underline{\epsilon}_{\bf k},
  \underline{\epsilon}_{\bf k}\rangle
    - \langle \underline{\eta}_{\bf k},
  \underline{\eta}^\ast_{\bf k} \rangle\big)^2 + 4\, \big|\langle
  \underline{\epsilon}_{\bf k},
    \underline{\eta}_{\bf k}\rangle\big|^2 \quad \notag \\
&\quad {}+ \langle \underline{\eta}_{\bf k},
  \underline{\eta}_{\bf k} \rangle
  \langle \underline{\eta}^\ast_{\bf k},
  \underline{\eta}^\ast_{\bf k} \rangle
  - \langle \underline{\eta}_{\bf k},
  \underline{\eta}^\ast_{\bf k} \rangle^2 \,,
\label{eq:Pfaffian}
\end{align}
where we adopt the ``six-vector'' notation
\begin{equation}
\underline{\epsilon}_{\bf k}
  = (\epsilon_{{\bf k},0}{-}\mu,\vec{\epsilon}_{\bf k})\,,
  \quad\underline{\eta}_{\bf k}
  = (\eta_{{\bf k},0},\vec{\eta}_{\bf k})\,,
\end{equation}
and define $\langle \underline{a}, \underline{b} \rangle
  = a_0b_0 - \vec{a}\cdot\vec{b}$.

Prior to examining the Pfaffian of \eq{eq:Pfaffian}, it is
useful to consider the case of a single-band model to
highlight the new physics that result from \eq{eq:Pfaffian}. In
particular, for a single-band system that is inversion
and time-reversal invariant in the normal state and retains
IS in the superconducting state, the BdG Hamiltonian takes the
usual pseudospin-singlet form
\begin{equation}
{\cal H}_{\bf k} = \left(\begin{array}{cc}
  \xi_0({\bf k})\,\sigma_0 & \psi({\bf k})\,i\sigma_2 \\
  -\psi^*({\bf k})\,i\sigma_2 & -\xi_0({\bf k})\,\sigma_0
  \end{array}\right) .
\label{eq:BdGHsingle}
\end{equation}
Using the same arguments leading to \eq{eq:Pfaffian}, the
Pfaffian for \eq{eq:BdGHsingle} is simply
\begin{equation}
\text{Pf}\:{\cal H}_{\bf k}
  = \xi_0^2({\bf k})+|\psi({\bf k})|^2 \,.
\end{equation}
Notice that this expression is always nonnegative and only
vanishes when (\textit{i}) ${\bf k}$ is on the Fermi surface,
where $\xi_0({\bf k})=0$, and (\textit{ii}) the gap vanishes,
$\psi({\bf k})=0$. Since zeros of the Pfaffian give the zeros of
the excitation spectrum, these two conditions immediately imply
that Bogoliubov Fermi surfaces generically do not appear in
single-band systems, where only point and line nodes are
expected~\cite{SiU91}.

In general, Hamiltonians with $(CP)^2=+1$ can possess Fermi
surfaces with a nontrivial $\mathbb{Z}_2$ topological charge
\cite{KST14,ZSW16,ABT17}. That is, they are stable
against any $CP$-preserving perturbation. The $\mathbb{Z}_2$
invariant is defined in Ref.~\cite{ABT17} in terms of the Pfaffian
in \eq{eq:Pfaffian} as
\begin{equation}
(-1)^l = \text{sgn}\,[P({\bf k}_{-})P({\bf k}_+)] \,,
\end{equation}
where ${\bf k}_{-}$ (${\bf k}_+$) refers to momenta inside
(outside) the Fermi surface, which is characterized by
$P({\bf k})=0$. Fermi surfaces with $l=1$ are topologically
nontrivial, as there must necessarily be
a surface of zeros of the Pfaffian separating regions where it
has opposite sign. In contrast, Fermi surfaces with $l=0$ are
not topologically protected and can be removed by a
$CP$-preserving perturbation.   

One easily sees from \eq{eq:Pfaffian} that the Pfaffian is
always nonnegative in the absence of superconductivity,
i.e., for $\underline{\eta}_{\bf k} = (0,\vec{0})$.
This reflects the fact that the normal-state Fermi surfaces,
given by the zeros of
$\langle \underline{\epsilon}_{\bf k},
  \underline{\epsilon}_{\bf k}\rangle
  = E_{{\bf k},+}E_{{\bf k},-}$,
can be gapped out by the superconductivity, which preserves
IS and particle-hole symmetry. Superconducting states
which preserve TRS, where one can choose a gauge such that
$\underline{\eta}_{\bf k}=\underline{\eta}^\ast_{\bf k}$, also
yield a nonnegative Pfaffian, as the last line of \eq{eq:Pfaffian}
then vanishes. Since the Pfaffian is defined locally in momentum
space, this argument also holds for any TRSB state defined by a
single momentum-dependent phase, i.e., where the pairing satisfies
$\underline{\eta}_{\bf k}
  = \underline{\eta}^r_{\bf k}\, e^{i\phi_{\bf k}}$
with $\underline{\eta}^r_{\bf k}$ entirely real.
Notice also that when the superconductor is time-reversal
invariant the nodes
generally do not lie on the Fermi surface, in contrast to the
single-band case. This follows by observing
that for momenta on the Fermi surface
$\langle \underline{\epsilon}_{\bf k},
  \underline{\epsilon}_{\bf k}\rangle
  = E_{{\bf k},+}E_{{\bf k},-} = 0$
so that
$P({\bf k})=4|\langle \underline{\epsilon}_{\bf k},
  \underline{\eta}_{\bf k}\rangle|^2\ne 0$. 
The position of nodes can therefore change as the
parameters in the Hamiltonian are changed, allowing for the
possibility of annihilating nodes, which has been argued to be
relevant to monolayer FeSe \cite{nic17,CVF16,agt17}.

The last term of the Pfaffian is only nonzero if there is pairing
in multiple superconducting channels with nontrivial phase
difference between them. Writing the pairing potential in each
channel as
$\eta_{{\bf k},n} = |\eta_{{\bf k},n}|\,e^{i\phi_{{\bf k},n}}$,
the last line of \eq{eq:Pfaffian} is then
\begin{align}
\langle \underline{\eta}_{\bf k}&,
  \underline{\eta}_{\bf k} \rangle
  \langle \underline{\eta}^\ast_{\bf k},
  \underline{\eta}^\ast_{\bf k} \rangle
  - \langle \underline{\eta}_{\bf k},
  \underline{\eta}^\ast_{\bf k} \rangle^2 \notag \\
&= 2\sum_{n>0}|\eta_{{\bf k},0}|^2|\eta_{{\bf k},n}|^2
  \big[1 - \cos(2[\phi_{{\bf k},0}-\phi_{{\bf k},n}])
  \big] \notag \\
&\quad{}- 2\! \sum_{n>m>0} \! |\eta_{{\bf k},n}|^2|
  \eta_{{\bf k},m}|^2
  \big[1 - \cos(2[\phi_{{\bf k},n}-\phi_{{\bf k},m}]) \big] \,.
\end{align}
The first term on the right-hand side shows that
coexisting internally isotropic and internally
anisotropic channels always give a nonnegative contribution to
the Pfaffian. In contrast, coexisting internally anisotropic
channels give a nonpositive contribution, which is  strictly
negative if the relative phase differences between the
unconventional channels break TRS. This is the only way to obtain
a negative Pfaffian and thus topologically stable Bogoliubov Fermi
surfaces. It is also equivalent to the presence of a non-vanishing
time-reversal-odd gap product in \eq{eq:time_gap_product},
and thus to a nonvanishing pseudospin polarization.

We now explicitly demonstrate that the Pfaffian can be negative
and hence that Bogoliubov Fermi surfaces exist. 
To simplify the discussion, we consider a TRSB state that
only involves the internally ansiotropic pairing channels.
Nodes are expected to occur on the normal-state Fermi surface
where the intraband pairing potential vanishes, i.e., where
$\vec{\epsilon}_{\bf k}\cdot\vec{\eta}_{\bf k}=0$, see
\eq{eq:psiintra}. If $\vec{\eta}_{\bf k}$ is real up to an
overall phase factor, which corresponds to the
time-reversal-symmetric case, this equation describes a surface
in the Brillouin zone. On the other hand, if $\vec{\eta}_{\bf k}$
has irreducible real and imaginary parts,
corresponding to the TRSB case,
this equation generically decomposes into two independent
real equations and thus describes a line. Restricting ourselves to
momenta where this condition is satisfied, the Pfaffian has the
simpler form  
\begin{align}
P({\bf k})
  &= (E_{{\bf k},+}E_{{\bf k},-}+|\vec{\eta}_{\bf k}|^2)^2 \notag \\
&{}- 2\sum_{n>m>0}|\eta_{{\bf k},n}|^2|\eta_{{\bf k},m}|^2\left[1-
  \cos(2[\phi_{{\bf k},n}-\phi_{{\bf k},m}])\right] .
\label{eq:Pfaffian_restricted}
\end{align}
The product $E_{{\bf k},+}E_{{\bf k},-}$ changes
sign  on the normal-state Fermi surface. For sufficiently
small $|\vec{\eta}_{\bf k}|$, it is therefore
possible to find a point in momentum space  close to the
normal-state Fermi surface, where the first term
in \eq{eq:Pfaffian_restricted} vanishes. The
Pfaffian will then be negative if the second line is nonzero, which
is always true for a nonunitary TRSB state as long as
$\vec{\eta}_{\bf k}\neq0$. That is, the node must arise from the 
projection of the internally anisotropic states onto
the Fermi surface, and not be intrinsic to the internally
ansiotropic pairing potentials $\vec{\eta}_{\bf k}$ themselves.
Since far away from the Fermi surface the product
$E_{{\bf k},+} E_{{\bf k},-}$ should dominate over the terms
involving the gap, there will also be a region in the Brillouin
zone where the Pfaffian is positive. We thus deduce the existence
of a topologically stable Bogoliubov Fermi surface forming
the boundary between the regions of positive and negative Pfaffian.

\subsection{Effective low-energy model}

Further insight into the appearance of Bogoliubov Fermi surfaces can
be obtained from an effective single-band model valid for the states
close to the normal-state Fermi surface. In deriving this model, we
make the weak-coupling assumption that, on the Fermi surface of each
band, the direct energy gap separating the two bands is much 
larger than the pairing potential, i.e.,
$|E_{{\bf k},+} - E_{{\bf k},-}|
  = 2|\vec{\epsilon}_{\bf k}|
  \gg \max(|\eta_{{\bf k},0}|,|\vec{\eta}_{\bf k}|)$.

Without loss of generality, we assume that the $-$ band intersects
the Fermi energy. The Green function $G_-({\bf k},\omega)$ of the
states in the $-$ band satisfies
\begin{equation}
G_{-}^{-1}({\bf k},\omega) = \omega - H_{{\bf k},-}
  - H_{{\bf k},I}^\dagger(\omega - H_{{\bf k},+})^{-1}H_{{\bf k},I}\,,
\label{eq:greensfunction}
\end{equation}
where $H_{{\bf k},\pm}$ and $H_{{\bf k},I}$ are blocks of
the BdG Hamiltonian transformed into the pseudospin basis,
\begin{equation}
\tilde{\cal H} = \left(\begin{array}{cc}
  H_{{\bf k},+} & H_{{\bf k},I} \\[0.5ex]
  H^\dagger_{{\bf k},I} & H_{{\bf k},-}
  \end{array} \right) .
\end{equation}
Note that $H_{{\bf k},I}$ describes the interband coupling due to
superconductivity in the internally anisotropic channels.

To lowest order in the pairing potential, an effective model is
obtained by ignoring the last term in \eq{eq:greensfunction}, which
simply gives the projection of the Hamiltonian onto the low-energy
states. This describes pseudospin-singlet pairing in a
doubly degenerate single band and does not yield the
Bogoliubov Fermi surfaces.
To obtain the leading correction to the projected Hamiltonian, we
analyze the last term in \eq{eq:greensfunction}. Since we are 
interested in the low-energy states in the $-$ band, we
approximate $\omega\approx 0$ in this term.
Furthermore, as the $+$ band is assumed to lie far from the
Fermi surface, we can ignore the effect of the pairing on its
dispersion, i.e., we can write
$H_{{\bf k},+}\approx (E_{{\bf k},+}-\mu)\,{s_0} \tau_z$,
where $\tau_z$ is a Pauli matrix in Nambu space.
Using the fact that $E_{{\bf k},-}\approx \mu$ close to the Fermi
surface of the $-$ band, we hence make the replacement
\begin{equation}
(\omega - H_{{\bf k},+})^{-1}
  \approx (E_{{\bf k},-} - E_{{\bf k},+})^{-1}\, {s_0} \tau_z
  = -\frac{1}{2|\vec{\epsilon}_{\bf k}|}\, {s_0} \tau_z \,.
\end{equation}
Inserting this into \eq{eq:greensfunction}, we obtain the
effective Hamiltonian
\begin{equation}
H^{\text{eff}}_{{\bf k},-}
  = H_{{\bf k},-} + \delta H_{{\bf k},-} \,,
\end{equation}
with the correction term
\begin{align}
\delta& H_{{\bf k},-}
  = - \frac{1}{2|\vec{\epsilon}_{\bf k}|}\,
  H_{{\bf k},I}^\dagger\tau_zH_{{\bf k},I} \notag \\
&= \left(\begin{array}{cc}\delta \epsilon_{{\bf k},-} s_0 +
  \delta{\bf h}_{{\bf k},-}\cdot{\bf s} & 0\\
  0 & -\delta \epsilon_{{\bf k},-} s_0 -
  \delta{\bf h}_{{\bf k},-}\cdot{\bf s}^T
  \end{array}\right) ,
\label{eq:deltaHminus}
\end{align}
where 
\begin{align}
\delta \epsilon_{{\bf k},-} &= \frac{|\psi_{{\bf k},I}|^2
  + |{\bf d}_{\bf k}|^2}{2\,|\vec{\epsilon}_{\bf k}|}\,, \\
\delta {\bf h}_{{\bf k},-} &= \frac{i{\bf d}_{\bf k}
  \times{\bf d}_{\bf k}^\ast
  -2\, \text{Im}(\psi_{{\bf k},I}{\bf d}_{\bf k}^\ast)}
  {2\,|\vec{\epsilon}_{\bf k}|} \notag \\
&= - \frac{\Tr [{\Delta}^\dagger({\bf k})
  {\cal P}_{{\bf k},-} \check{\bf s} {\cal P}_{{\bf k},-}
  {\Delta}({\bf k})]}{4\,|\vec{\epsilon}_{\bf k}|} \,.
\end{align}
The term $\delta\epsilon_{{\bf k},-}$ modifies the 
dispersion and is always present when there is interband
pairing. The term $\delta{\bf h}_{{\bf k},-}\cdot{\bf s}$ is more
interesting:  It
describes an effective ``pseudomagnetic'' field, which is
proportional to the pseudospin polarization of the band states 
in a nonunitary state. It is hence only  present in TRSB states
where the unconventional channels have nontrivial relative phase
differences. It directly leads to the appearance of Bogoliubov
Fermi surfaces, as can be seen in the eigenvalue spectrum of the
low-energy effective model,
\begin{equation}
\pm |\delta{\bf h}_{{\bf k},-}|
  \pm \sqrt{(E_{{\bf k},-}+\delta\epsilon_{{\bf k},-}-\mu)^2
  + |\psi_{{\bf k},-}|^2}\:,
\end{equation}
where the two $\pm$ signs are chosen independently. If the
pseudomagnetic field is nonzero at a node of the  intraband
pairing potential $\psi_{{\bf k},-}$, the dispersion is
split and shifted to finite energies, leading to
Bogoliubov Fermi surfaces. Since $\delta{\bf h}_{{\bf k},-}$
is proportional to the product of the internally
anisotropic pairing potentials, it is therefore necessary that
these potentials are nonzero at the node, consistent with the
condition found above that the node is due to the projection
  into the low-energy states.

We note that the modification of the normal-state dispersion
by $\delta \epsilon_{{\bf k},-}$ can have a nontrivial effect on 
the electronic structure of the superconducting state even when TRS
is not broken. Specifically, as also discussed above for the
general theory, line nodes arising from the  projection of the
pairing potential onto the band basis, i.e., for which
$\vec{\epsilon}_{\bf k}\cdot\vec{\eta}_{\bf k}=0$ but
$\vec{\eta}_{\bf k}=0$, are shifted off the normal-state
Fermi surface by $\delta \epsilon_{{\bf k},-}$, so that they
instead occur on the surface defined by
$E_{{\bf k},-} + \delta \epsilon_{{\bf k},-} = \mu$, allowing for
pairs of nodes to annihilate \cite{nic17,CVF16,agt17}.

\section{Cubic superconductors}

In this section, we consider superconductivity in the $\Gamma_8$
bands of a cubic system as the first concrete example of the generic
model studied above. Here, the four internal degrees of freedom of
the electrons reflect their total angular momentum $j=3/2$, which
arises from the strong atomic spin-orbit coupling.
We present a tight-binding generalization of the
Luttinger-Kohn model of the $\Gamma_8$ bands for the point group
$O_h$~\cite{L56}. Our strategy is to take the simplest nontrivial model that is
consistent with the imposed symmetries. In this spirit, we restrict
ourselves to nearest-neighbor hopping on the fcc lattice and only
consider local Cooper pairing. We determine the Bogoliubov Fermi
surfaces and examine the associated nonunitary gap structure both in
the $j=3/2$ basis and in the pseudospin basis near the Fermi
surface.

The $j=3/2$ effective spin of the
electrons in the cubic superconductor implies that the 
Nambu spinor in~\eq{eq:genHBdG} is $\Psi_\mathbf{k} = (c_{\mathbf{k}}^T,
  c^\dagger_{-\mathbf{k}})^T$ where $c_{\bf k}^T = (c_{{\bf k},3/2},c_{{\bf k},1/2},c_{{\bf k},-1/2},
  c_{{\bf k},-3/2})$. 
The diagonal, normal block of the BdG
Hamiltonian reads 
\begin{align}
H_0(\mathbf{k}) &=  -4t_1 \sum_{\nu} \cos k_\nu \cos k_{\nu+1}
  \notag \\
&{}- 4t_2 \sum_{\nu} \cos k_{\nu} \cos k_{\nu+1}\, J_{\nu+2}^2
  \notag \\
&{}+ 4t_3 \sum_{\nu}  \sin k_{\nu} \sin k_{\nu+1}\,
  (J_\nu J_{\nu+1} + J_{\nu+1}J_\nu) - \mu \,,
\label{h.normal.3}
\end{align}
where $J_\nu$, $\nu=x,y,z$, are the standard $4\times 4$ spin-$3/2$
matrices, $\nu+1$ and $\nu+2$ are understood as cyclic
shift operations on $\{x,y,z\}$,
and we henceforth suppress $4\times4$ unit matrices. The lattice
orientation and unit of length are chosen such that
$\mathbf{R}=(0,0,0)^T$ is a lattice point and its nearest neighbors
are $(1,1,0)^T$, $(-1,1,0)^T$, etc. Note that the
Hamiltonian~\eq{h.normal.3} can be obtained from the one considered in
Ref.\ \cite{TSA17} for half-Heusler compounds by omitting terms that
are odd under inversion and thus forbidden for the point group
$O_h$. The remainder of this subsection closely follows
Ref.\ \cite{TSA17}. For the numerical evaluations in this section we
take $t_1=-0.918\,\mathrm{eV}$, $t_2=0.760\,\mathrm{eV}$,
$t_3=-0.253\,\mathrm{eV}$ (these are the same values as in
Refs.\ \cite{BWW16}, \cite{TSA17}), and $\mu=-0.880\,\mathrm{eV}$.
For these values, one of the two bands meeting at the quadratic
band-touching point $\Gamma_8$ curves upwards and one curves
downwards. Both bands are twofold spin degenerate. The
chemical potential lies $0.5\,\mathrm{eV}$ below the band-touching
point, corresponding to strong hole doping. The resulting relatively
large normal-state Fermi sea is advantageous for real-space
calculations but our results are qualitatively unchanged for smaller
negative $\mu$. The Hamiltonian in \eq{h.normal.3} can be brought
into the general form introduced in
Sec.~\ref{sec:general_model} with the mapping~\cite{A74}
\begin{align}
\gamma^1 &= \tfrac{1}{\sqrt{3}}\, (J_xJ_y+J_yJ_x) \,,
  \label{eq:gammaJ1}\\
\gamma^2 &= \tfrac{1}{\sqrt{3}}\, (J_yJ_z+J_zJ_y) \,, \\
\gamma^3 &= \tfrac{1}{\sqrt{3}}\, (J_xJ_z+J_zJ_x) \,, \\
\gamma^4 &= \tfrac{1}{\sqrt{3}}\, (J_x^2-J_y^2) \,, \\
\gamma^5 &= \tfrac{1}{3}\, (2J_z^2-J_x^2-J_y^2)\label{eq:gammaJ5}\,.
\end{align}
This correctly gives the unitary part of the time-reversal operator
\begin{equation}
U_T = \gamma^1\gamma^2 = \left(\begin{array}{cccc}
    0 & 0 & 0 & 1 \\
    0 & 0 & -1 & 0 \\
    0 & 1 & 0 & 0 \\
    -1 & 0 & 0 & 0
  \end{array}\right) .
\end{equation}
It is straightforward to verify that the normal-state Hamiltonian
$\sum_{\bf k}c^\dagger_{\bf k}H_0({\bf k})c_{\bf k}$ is invariant
under $O_h$ and time reversal.

The off-diagonal, superconducting block in \eq{eq:BdGH} describes
local, on-site, pairing and can be expanded
as~\cite{BWW16,ABT17,TSA17,SRV17,YXW17,RGF17,BoH18}
\begin{equation}
\Delta = \sum_r \Delta_r\, \Gamma_r
\label{Del.ex.2}
\end{equation}
in terms of the six matrices
\begin{align}
A_{1g}\mbox{:} && \Gamma_s &= U_T \,, \\
T_{2g}\mbox{:} && \Gamma_{yz}
  &= \tfrac{1}{\sqrt{3}}\, (J_y J_z + J_z J_y)\, U_T \,, \\
&& \Gamma_{xz}
  &= \tfrac{1}{\sqrt{3}}\, (J_z J_x + J_x J_z)\, U_T \,, \\
&& \Gamma_{xy}
  &= \tfrac{1}{\sqrt{3}}\, (J_x J_y + J_y J_x)\, U_T \,, \\
E_g\mbox{:} && \Gamma_{3z^2-r^2}
  &= \tfrac{1}{3}\, (2J_z^2 - J_x^2 - J_y^2)\, U_T \,, \\
&& \Gamma_{x^2-y^2}
  &= \tfrac{1}{\sqrt{3}}\, (J_x^2 - J_y^2)\, U_T \,,
\end{align}
which belong to the irreps indicated in the first column. In
Eq.\ (\ref{Del.ex.2}), $\Gamma_s$ describes Cooper pairs with total
angular momentum $J=0$ (singlet), whereas the other five matrices
give Cooper pairs with $J=2$
(quintet)~\cite{BWW16,ABT17,TSA17,SRV17,YXW17,RGF17,BoH18}. 
In the context of the general model, the five
quintet channels correspond to the internally anisotropic
states, whereas the singlet channel is internally isotropic.
All six pairing matrices are even under time reversal so that the
pairing state preserves TRS if and only if all amplitudes
$\Delta^0_r$ are real or more generally have the same complex
phase.

\subsection{TRSB pairing states}
\label{sub.bulk}

In a weak-coupling theory, the stable pairing state can be
obtained by minimizing the free energy of the interacting system
within a BCS-like mean-field approximation. We do not pursue this
approach here but instead utilize symmetry arguments to determine
allowed pairing states. Below a transition temperature $T_c$, one
or more of the amplitudes $\Delta_r$ in \eq{Del.ex.2} assume
nonzero values. By symmetry, the critical temperatures for
amplitudes belonging to the same irrep coincide. Hence, below
$T_c$ one generically expects only a single amplitude or several
amplitudes belonging to the same irrep to be nonzero. For
amplitudes belonging to different irreps to be nonzero, multiple
transitions are required. This is possible if the corresponding
pairing interactions are comparable~\cite{RGF17}.

In this subsection, we  present the Bogoliubov Fermi surfaces and
the pseudomagnetic field acting on the low-energy states for
various TRSB combinations of the quintet pairing potentials. The
corresponding surface states are discussed in the following
subsection. For  the purpose of illustration, we take large
numerical values of the 
gap amplitude since this leads to sizable Bogoliubov Fermi
surfaces. Our qualitative results concerning the Fermi surface
topology and the  pseudomagnetic field do not depend on the
specific gap amplitude. 

\begin{table*}[tb]
\caption{\label{tab.sym}Pairing states that break TRS. The first
column specifies the irrep according to which the pairing states
transform. Each list in the second column contains all ordering
vectors $\mathbf{h}$ or $\mathbf{l}$ that are related to one
another by point-group operations or time reversal and thus have
the same free energy. The free energy is also invariant under
multiplication of all components by the same arbitrary phase
factor $e^{i\alpha}$, expressing the broken global
$\mathrm{U}(1)$ symmetry in the superconducting state. The last
column specifies the symmetry broken by the pairing state, i.e.,
the degeneracy of the equilibrium state.}
\begin{ruledtabular}
\begin{tabular}{ccc}
  irrep & ordering vector & degeneracy \\ \hline
$E_g$ & $\mathbf{h} = (1,\pm i)$ \cite{endnote.Eg}
  & $\mathrm{U}(1) \times \mathbb{Z}_2$ \\
$T_{2g}$ & $\mathbf{l} = (1,\pm i,0)$, $(0,1,\pm i)$,
  $(\pm i,0,1)$ & $\mathrm{U}(1) \times \mathbb{Z}_6$ \\
$T_{2g}$ & $\mathbf{l} = (1,\omega^{\pm1},\omega^{\mp 1})$,
  $(1,\omega^{\pm 1},\omega^{\pm 2})$,
  $(1,\omega^{\pm 2},\omega^{\pm 1})$,
  $(1,\omega^{\pm 2},\omega^{\mp 2})$, \quad
  $\omega = e^{i\pi /3}$ & $\mathrm{U}(1) \times \mathbb{Z}_8$
\end{tabular}
\end{ruledtabular}
\end{table*}

\begin{figure*}
\raisebox{1ex}{(a)}\includegraphics[width=0.305\textwidth]
  {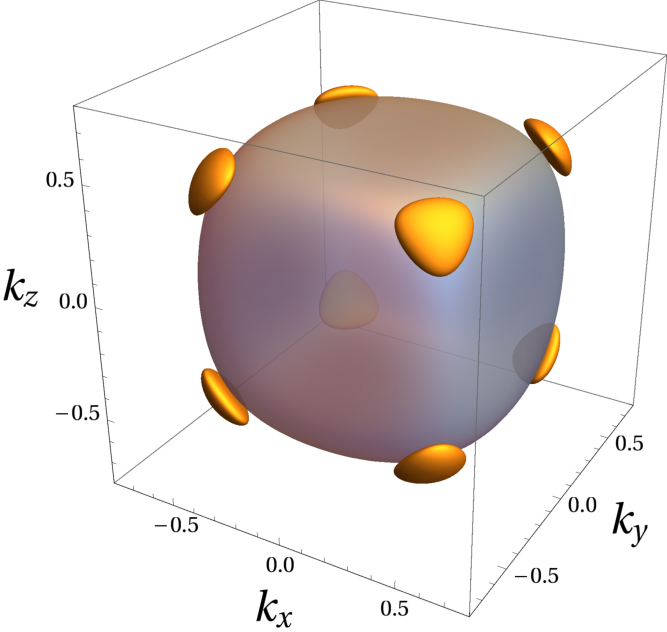}
\raisebox{1ex}{(b)}\includegraphics[width=0.305\textwidth]
  {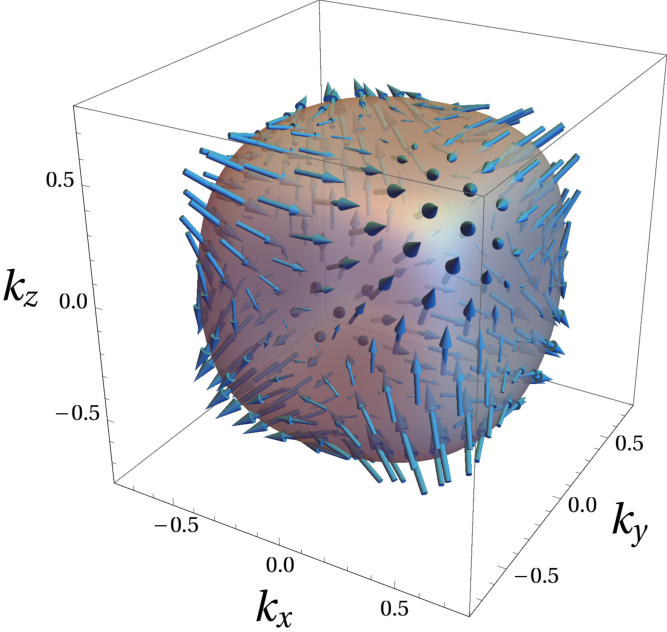}
\raisebox{1ex}{(c)}\includegraphics[width=0.305\textwidth]
  {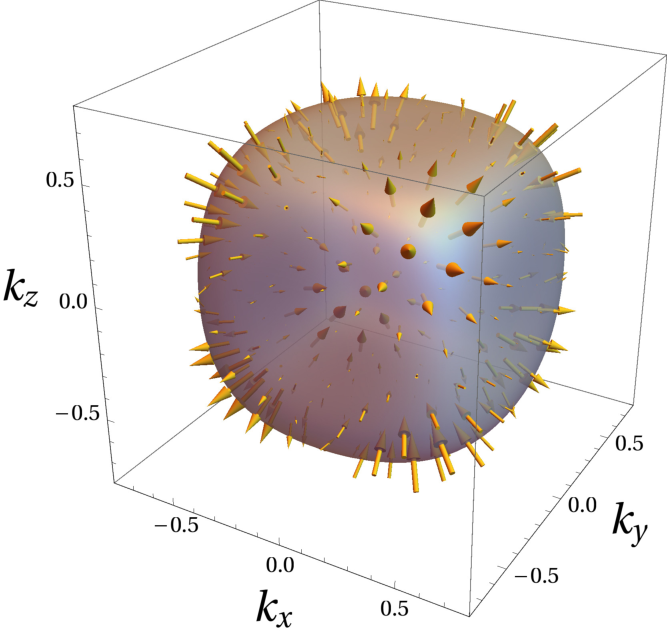}
\caption{\label{fig.bulk.Eg}Low-energy structure of the TRSB
$E_{g}$ pairing state  with ${\bf h}=(1,i)$. (a) Bogoliubov
Fermi surfaces (opaque orange) in comparison to the
normal-state Fermi surface (semi-transparent). A gap amplitude
of $\Delta_0=0.2\,\text{eV}$ has been used. 
(b) Pseudomagnetic field
acting on the states at the Fermi surface due to the nonunitary
pairing state. Note that the orientation is basis dependent and
corresponds to our choice of the MCBB defined in Appendix
\ref{app.ps}. (c) Magnetization of the states at the Fermi
surface arising from the pseudomagnetic field.}
\end{figure*}


\subsubsection{$E_g$ pairing}

For pairing restricted to states from the $E_g$ irrep, the
general potential reads 
\begin{equation}
\Delta = \Delta_0\, (h_{3z^2-r^2} \Gamma_{3z^2-r^2}
  + h_{x^2-y^2} \Gamma_{x^2-y^2})
\end{equation}
so that  the pairing state is characterized by the
ordering vector $\mathbf{h}=(h_{3z^2-r^2},h_{x^2-y^2})$. A
Landau analysis \cite{VoG85,SiU91,BWW16} gives
$\mathbf{h}=(1,0)$, $(0,1)$, and $(1,i)$ as possible equilibrium
solutions as well as symmetry-related vectors $\mathbf{h}$
obtained by applying point-group operations or time reversal.
Since $\Gamma_{3z^2-r^2}$ and $\Gamma_{x^2-y^2}$ cannot be
mapped into each other by any point-group symmetry, the pairing
states $\mathbf{h}=(1,0)$ and $(0,1)$ do not
generically have the same free energy \cite{VoG85,RGF17,SiU91}.
The state with $\mathbf{h}=(1,i)$ breaks TRS.
There are two symmetry partners, which are listed in table
\ref{tab.sym}. As shown in \fig{fig.bulk.Eg}(a), this state
possesses Bogoliubov Fermi surfaces, replacing the eight point
nodes expected along the $(111)$ and equivalent directions
in the single-band case. The pockets and the whole
quasiparticle band structure do not break any lattice
symmetries. The only broken discrete symmetry is TRS.

The time-reversal-odd gap product for the TRSB $E_{g}$
state is given by
\begin{equation}
\Delta\Delta^{\dagger}-\Delta_T\Delta_T^{\dagger} = 
  \frac{8}{\sqrt{3}}\,\Delta_0^2
  \left(J_xJ_yJ_z+J_zJ_yJ_x\right) \,.
\end{equation}
This represents an octopolar magnetic order parameter and
belongs to the irrep $A_{2g}$. It is interesting to note that
this order parameter also appears in the context of
frustrated magnetism on the pyrochlore lattice, where it
describes all-in-all-out (AIAO) magnetic order of spins at the
four corners of the elementary tetrahedra. The term appears in
the single-particle mean-field Hamiltonian of the $j=3/2$
electron bands \cite{SMB14,GRDS17,BoH17}.
The octopolar structure is clearly seen in the pseudomagnetic
field in~\fig{fig.bulk.Eg}(b).

Although the pseudospin is not related in a simple way to the
true spin, the pseudomagnetic field will generally be
accompanied by a polarization of the physical spin since TRS is
broken. We here define the magnetization as the expectation value
of the total angular momentum $\mathbf{J}$. In the current
theory, the magnetization contribution from states close to the
normal-state Fermi surface at momentum $\mathbf{k}$ has the
components
\begin{equation}
m_{\mathbf{k},\mu} = - \frac{1}{|\mathbf{v}_{\mathbf{k},-}|}\,
  \delta\mathbf{h}_{\mathbf{k},-} \cdot
  \Tr (\mathcal{P}_{\mathbf{k},-} \check{\mathbf{s}}
  \mathcal{P}_{\mathbf{k},-} J_\mu) \,.
\label{mphys.2}
\end{equation}
The derivation is relegated to Appendix \ref{app.magnet}. Note
that $m_{\mathbf{k},\mu}$ is independent of the specific choice
of the pseudospin basis since both
$\delta\mathbf{h}_{\mathbf{k},-}$ and $\check{\mathbf{s}}$ are
transformed simultaneously when going from one basis to another.
The magnetization is plotted in \fig{fig.bulk.Eg}(c) and the
octopolar structure is again readily apparent.
The octopolar magnetic order implies that there is no overall
magnetization in this state. Note that a pseudomagnetic field
does not necessarily imply a magnetization at a given momentum:
although $\delta{\bf h}_{{\bf k},-}$ is nonzero in the planes
$k_\mu=0$, $\mu=x,y,z$, the magnetization  vanishes in these
directions.

\subsubsection{$T_{2g}$ pairing}

The general from of a pairing potential involving only states
from the $T_{2g}$ irrep is 
\begin{equation}
\Delta = \Delta_0\, (l_{yz} \Gamma_{yz} + l_{xz} \Gamma_{xz}
  + l_{xy} \Gamma_{xy}) \,.
\end{equation}
The pairing states in the $T_{2g}$ sector are 
characterized by different vectors 
$\mathbf{l}=(l_{yz},l_{xz},l_{xy})$. A Landau analysis
\cite{VoG85,SiU91,BWW16} shows that the possible equilibrium
solutions are $\mathbf{l}=(1,0,0)$, $(1,1,1)$, $(1,i,0)$, and
${\bf l}=(1,e^{2\pi i/3},e^{-2\pi i/3})$, as well as
symmetry-related vectors $\mathbf{l}$.
The two distinct TRSB pairing states correspond to
$\mathbf{l}=(1,i,0)$ (chiral state) and
${\bf l}=(1,e^{2\pi i/3},e^{-2\pi i/3})$ (cyclic
state).  The symmetry partners of these states are listed in
table \ref{tab.sym}. Since $\Gamma_{xz}$, $\Gamma_{yz}$, and
$\Gamma_{xy}$ can be transformed into one another by rotations
contained in $O_h$, these TRSB states are sixfold and
eightfold degenerate, respectively. 

\begin{figure*}
\raisebox{1ex}{(a)}\includegraphics[width=0.3\textwidth]
  {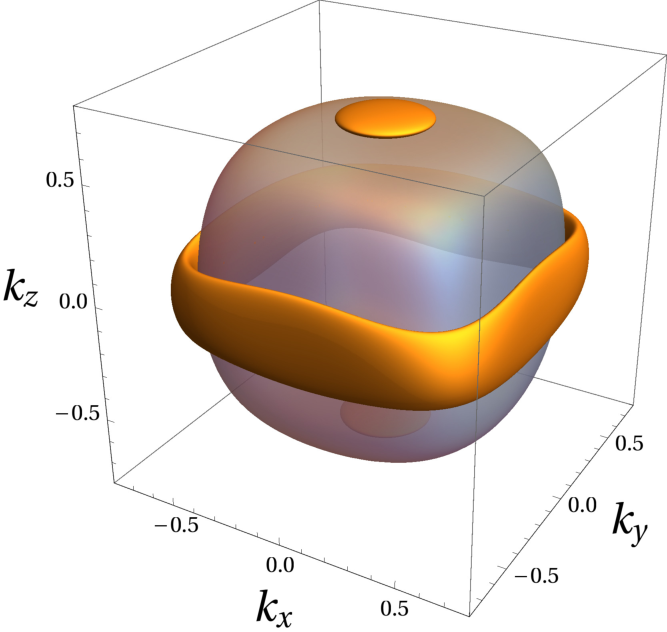}
\raisebox{1ex}{(b)}\includegraphics[width=0.3\textwidth]
  {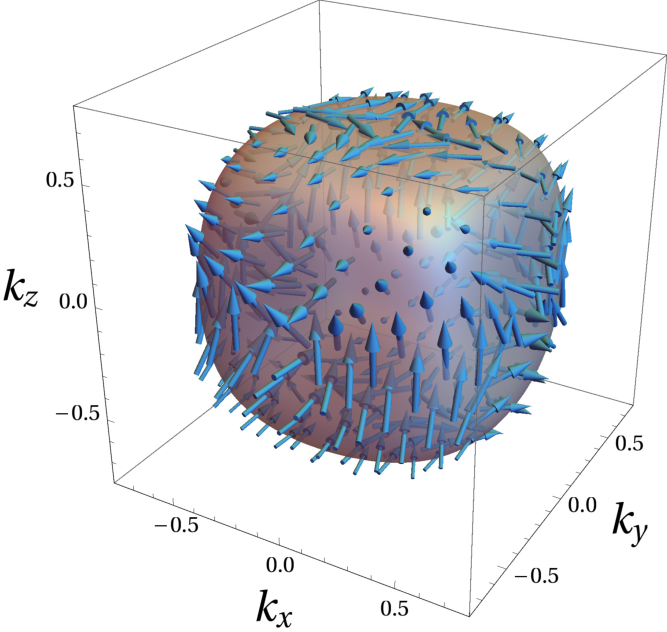}
\raisebox{1ex}{(c)}\includegraphics[width=0.3\textwidth]
  {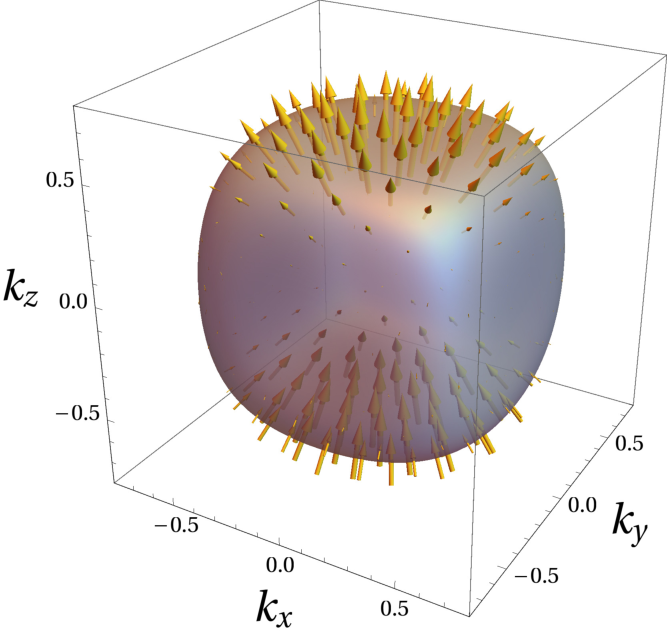}\\[4ex]
\raisebox{1ex}{(d)}\includegraphics[width=0.3\textwidth]
  {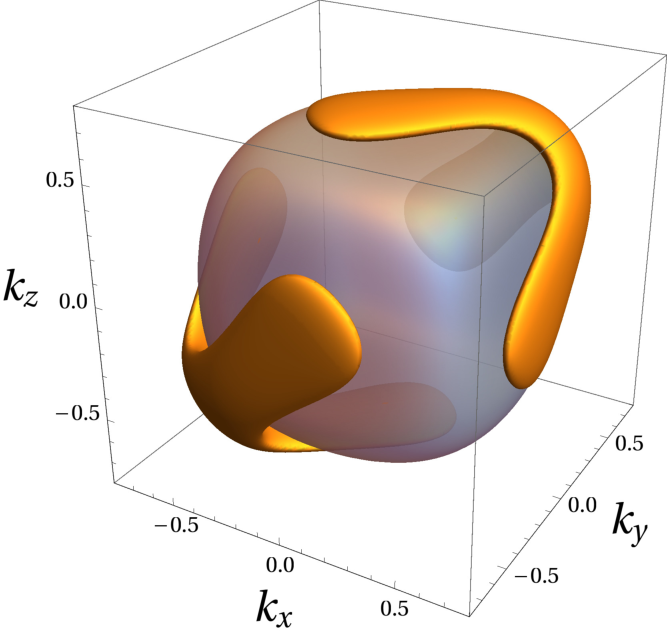}
\raisebox{1ex}{(e)}\includegraphics[width=0.3\textwidth]
  {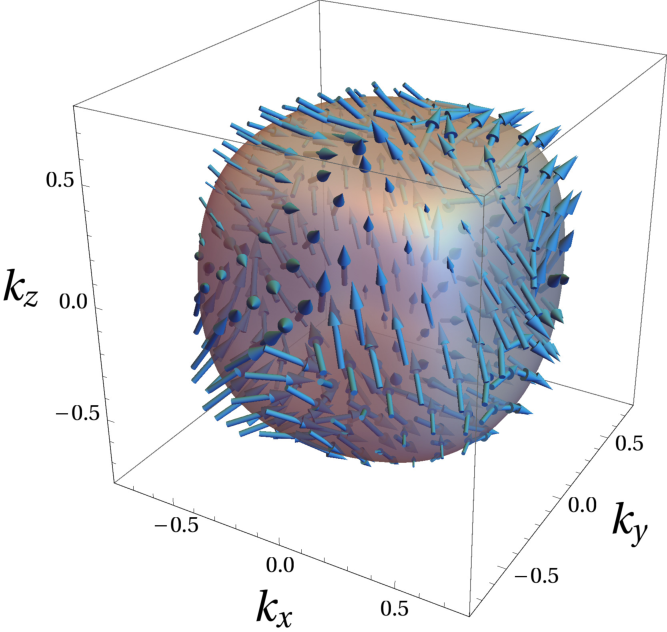}
\raisebox{1ex}{(f)}\includegraphics[width=0.3\textwidth]
  {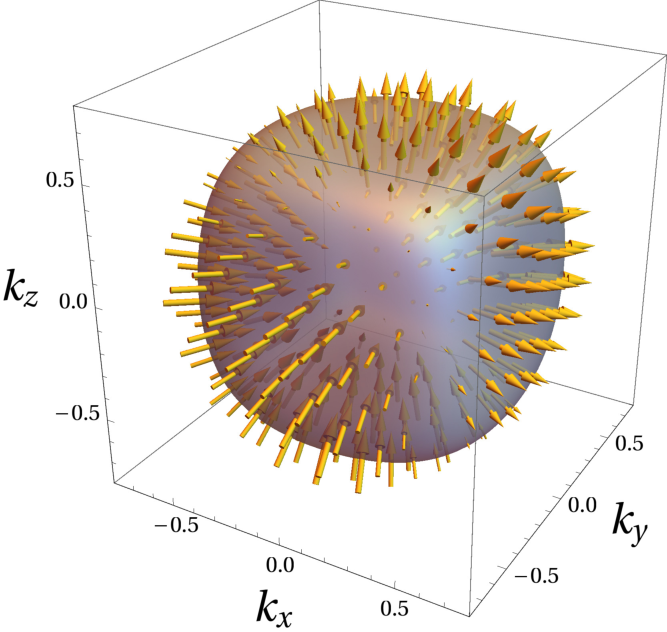}
\caption{\label{fig.bulk.T2g}Low-energy structure of the TRSB
$T_{2g}$ pairing states with ${\bf l}=(1,i,0)$ [(a)--(c)] and
${\bf l}=(1,e^{2\pi i/3},e^{-2\pi i/3})$ [(d)--(f)].
(a), (d) Bogoliubov Fermi surfaces (opaque orange) in comparison
to the normal-state Fermi surface (semi-transparent). (b), (e)
Pseudomagnetic field acting on the states at the Fermi surface.
Note that the orientation is basis dependent and corresponds to
our choice of the MCBB defined in Appendix \ref{app.ps}. (c), (f)
Magnetization of the states at the Fermi surface arising from the
pseudomagnetic field. For panels (a) and (d), a gap amplitude of
$\Delta_0=0.2\,\mathrm{eV}$ has been used.}
\end{figure*} 

We first examine the chiral state, which was previously discussed
in Ref.\ \cite{ABT17}. The line node in the $k_xk_y$ plane and the
point nodes on the $k_z$ axis for the single-band model are
inflated into toroidal and spheroidal pockets, respectively, as
shown in \fig{fig.bulk.T2g}(a). As evidenced by the Bogoliubov
Fermi surfaces, the gap is nonunitary and the
 time-reversal-odd gap product is 
\begin{equation}
\Delta\Delta^{\dagger}-\Delta_T\Delta_T^{\dagger}
  = \frac{4}{3}\, \Delta_0^2\left(7J_z-4J_z^3\right) .
\end{equation}
 The gap product belongs to the irrep $T_{1g}$ (of which the
spin operators $J_\nu$ are irreducible tensor operators), and
involves both dipolar ($\propto J_z$) and octopolar
($\propto J_z^3$) contributions. Intriguingly, these appear in
precisely the same combination as the order parameter of
two-in-two-out order on the elementary tetrahedra of the
pyrochlore lattice, which is associated with a polarization along
the $z$-axis. The two-in-two-out condition constitutes the ice
rule and leads to interesting spin-ice (SI)
physics~\cite{BoH17,GRDS17}.

The pseudomagnetic field of the low-energy states is shown in
\fig{fig.bulk.T2g}(b); although it displays a complicated
vortex-like structure, an overall polarization in the
$z$-direction is apparent,
consistent with the dipole nature of the nonunitary part. The
physical magnetization, presented in \fig{fig.bulk.T2g}(c),
clearly shows a net moment along the $z$-axis. Interestingly,
the pseudomagnetic field
in the vicinity of the toroidal Bogoliubov Fermi surface gives a
negligible magnetization. The symmetry-related ${\bf l}=(0,1,i)$
and ${\bf l}=(i,0,1)$ states have a similar nodal (magnetic)
structure, but with the point nodes (magnetization) oriented
along the $x$- and $y$-axis, respectively. 

\begin{figure}
\includegraphics[width=0.7\columnwidth]{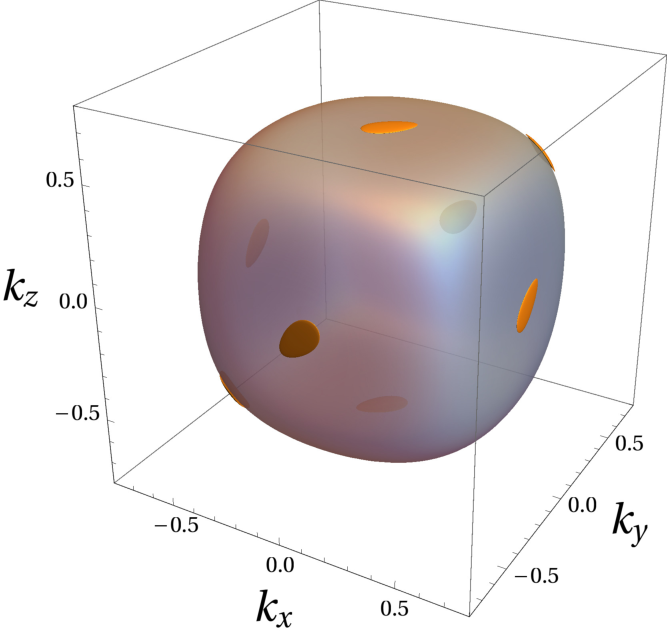}
\caption{\label{fig.bulk.T2gsmall}Small-gap form of the
Bogoliubov Fermi surfaces (opaque orange) of the cyclic
TRSB $T_{2g}$ pairing state with
${\bf l}=(1,e^{2\pi i/3},e^{-2\pi i/3})$ in comparison to the
normal-state Fermi surface (semi-transparent). We have
chosen the gap amplitude $\Delta_0=0.05\,\mathrm{eV}$
for this plot. For larger gap amplitudes, the eight
Bogoliubov Fermi surfaces merge into two, as shown
in \fig{fig.bulk.T2g}(d).}
\end{figure}

In a single-band system, the cyclic state
$\mathbf{l}=(1,\omega^a,\omega^b)$ with $\omega=e^{i\pi/3}$,
see table \ref{tab.sym}, has
point nodes along the three crystal axes, and  two
additional point nodes along the direction
$(-1)^{a+b}\hat{\bf x} + (-1)^b\hat{\bf y} + (-1)^a\hat{\bf z}$,
which remains a threefold rotational axis. However, instead of
the expected eight Bogoliubov Fermi surfaces,
\fig{fig.bulk.T2g}(d) only shows
two. This results from the merging of the Bogoliubov
Fermi surfaces originating from the point nodes along the
$[111]$ direction with the surfaces from the three nearest nodes
along the crystal axes. Despite choosing the same gap magnitude
in all figures, the merging of the Fermi surfaces is not seen in
the $E_g$ or chiral $T_{2g}$ states for two  reasons: First, the
pairing state in the cyclic $T_{2g}$  state has three instead of
two components, and hence the pseudomagnetic field has larger
magnitude near the point nodes. Second, the would-be point nodes
are relatively close  together in momentum space and the
superconducting gap between them remains significantly smaller
than the gap far from the nodes.
Choosing a smaller gap amplitude results in disconnected
Bogoliubov Fermi surfaces: multiplying $\Delta_0$ by a
factor of $1/4$ gives the nodal surfaces shown in
\fig{fig.bulk.T2gsmall}. This is convenient, as we will later
want to exhibit surface states between the projections of
inflated nodes.

The time-reversal-odd gap product for the cyclic state
${\bf l}=(1,\omega^a,\omega^b)$ reads
\begin{align}
\Delta\Delta^{\dagger}-\Delta_T\Delta_T^{\dagger} &=
  - \frac{4}{3}\, \Delta_0^2\, \bigg( \sin \frac{\pi\,(a-b)}{3}\,
  \hat{\bf x} + \sin \frac{\pi b}{3}\, \hat{\bf y} \notag \\
&\quad{}- \sin \frac{\pi a}{3}\, \hat{\bf z}\bigg)
  \cdot\left(7{\bf J}-4{\bf J}^3\right) ,
\end{align}
where ${\bf J}^3$ is the vector $(J_x^3,J_y^3,J_z^3)$.
Again, the nonunitary part has the same form as 
a magnetic order parameter on the pyrochlore lattice, in
this case a spin-ice variant
with magnetization along the threefold rotation axis.
In Ref.\ \cite{GRDS17}, this is referred to as a
three-in-one-out (3I1O) order. For the case of
${\bf l}=(1,\omega^2,\omega^{-2})$, a net
pseudomagnetic field along the threefold rotation axis is
clearly visible in \fig{fig.bulk.T2g}(e), in addition to a
complicated field texture. Plotting the physical magnetization
in \fig{fig.bulk.T2g}(f), the existence of a net
magnetic moment along the $[111]$ direction is evident.

\begin{figure*}
\raisebox{1ex}{(a)}\includegraphics[width=0.3\textwidth]
  {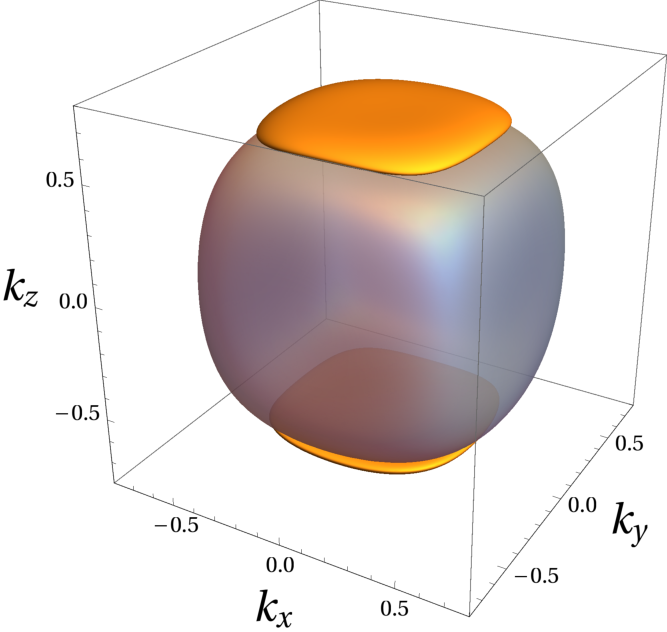}
\raisebox{1ex}{(b)}\includegraphics[width=0.3\textwidth]
  {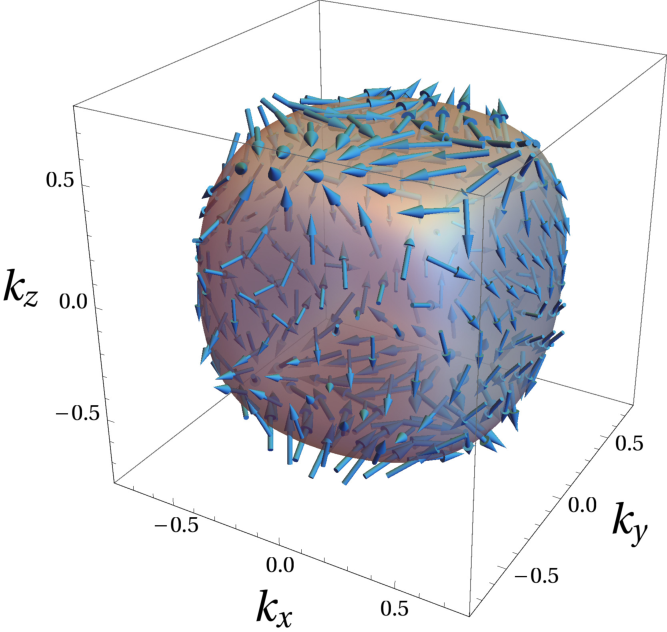}
\raisebox{1ex}{(c)}\includegraphics[width=0.3\textwidth]
  {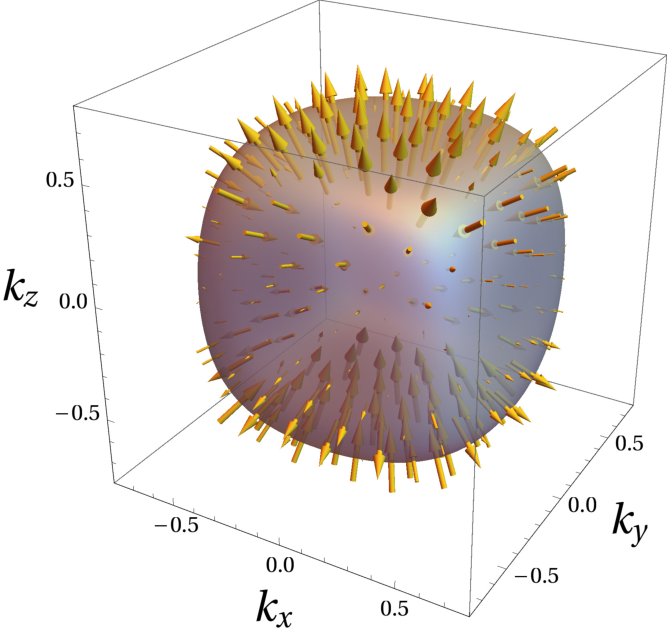}
\caption{\label{fig.bulk.mix}Low-energy structure of the TRSB
mixed $E_{g}$-$T_{2g}$ pairing state of \eq{eq:mix}. (a) Bogoliubov
Fermi surfaces (opaque orange) in comparison to the normal-state
Fermi surface (semi-transparent). A gap amplitude of
$\Delta_0=0.2\,\text{eV}$ has been used. (b) Pseudomagnetic
field acting on the states at the Fermi surface due to the
nonunitary pairing state. Note that the orientation is basis
dependent and corresponds to our choice of the MCBB defined in
Appendix \ref{app.ps}. (c) Magnetization of the states at 
the Fermi surface arising from the  pseudomagnetic field.}
\end{figure*}

\subsubsection{Mixed pairing}
\label{subsub.mixed}

The cases considered so far exhaust the essentially different
TRSB pairing states that belong to a single irrep of $O_h$. As noted
above, pairing amplitudes from different irreps may coexist if the
corresponding pairing interactions are comparable
\cite{RGF17}. An important example of such a state is provided by
\begin{equation}
\Delta = \Delta_0\, (\Gamma_{x^2-y^2} + i \Gamma_{xy})\,,
\label{eq:mix}
\end{equation}
which mixes the $E_g$ and $T_{2g}$ irreps. Projected into the band
states close to the zone center, this realizes a $(k_x+ik_y)^2$-wave
state. In contrast to the pure-irrep states for $O_h$, the
single-band version supports point nodes with
\emph{quadratic} dispersion in the directions along the normal-state
Fermi surface \cite{SiU91,RGF17}, so-called
\emph{double Weyl points}.
Such nodal structures can appear for pure-irrep pairing
in other points groups, e.g., for $D_{6h}$ as discussed below. The
underlying point group is not essential for the topological
properties of the states or the structure of the Bogoliubov Fermi
surfaces, which are shown in~\fig{fig.bulk.mix}(a). For the same 
value for the gap amplitude as in Figs.\ \ref{fig.bulk.Eg}
and \ref{fig.bulk.T2g}, we find much larger inflated 
nodes. The large nodal surfaces reflect the quadratic
dispersion of the double Weyl points of the projected gap along
the normal-state Fermi surface: since the dispersion 
is quadratic, a pseudomagnetic field of the same order as for the
other cases leads to a much larger inflated node. 
The dimensions of the Bogoliubov Fermi surfaces can be estimated as
follows \cite{ABT17}: In the direction perpendicular to the
normal-state Fermi surface, the quasiparticle dispersion is
proportional to $v_F\, \delta k_\perp$,
where $\delta k_\perp$ is the distance from the Fermi surface.
Since the pseudomagnetic field is
proportional to $|\Delta_0|^2$, it follows that the size of the
inflated nodes in the perpendicular direction scales as $\delta
k_\perp \sim |\Delta_0|^2$. In the directions along the
normal-state Fermi surface, the quasiparticle dispersion for a
single-Weyl node (relevant for pure-irrep pairing) is
proportional to $|\Delta_0|\, \delta k_\|$, where $\delta k_\|$
is the distance along the Fermi surface. Comparing this
to the pseudomagnetic field proportional to $|\Delta_0|^2$, we
find that the size of the inflated nodes along the
normal-state Fermi surface scales as
$\delta k_\| \sim |\Delta_0|$. In
the present case of a quadratic point node, however, the energy
in the parallel direction is proportional to
$|\Delta_0|\, (\delta k_\|)^2$ so that the size of the
inflated nodes scales as $\delta k_\| \sim |\Delta_0|^{1/2}$. 

The mixed-irrep state is nonunitary and has the time-reversal-odd
gap product
\begin{equation}
\Delta\Delta^{\dagger}-\Delta_T\Delta_T^{\dagger}
  = \frac{2}{3}\,\Delta_0^2\left(13J_z-4J_z^3\right) .
\label{eq:mixgp}
\end{equation}
This belongs to the $T_{1g}$ irrep and resembles the nonunitary
part for the chiral $T_{2g}$ state. In contrast to the
nonunitary part of the pure-irrep TRSB state, \eq{eq:mixgp} does
not have a straightforward interpretation in terms of
magnetically ordered states of the pyrochlore lattice.
The pseudomagnetic field associated with the  mixed-irrep
pairing state is shown in \fig{fig.bulk.mix}(b). It displays
pronounced vortex-like structures similar to those of the chiral
$T_{2g}$ state. The physical magnetization presented in
\fig{fig.bulk.mix}(c) evidences a net moment
along the $z$-axis.


\begin{figure}[tbh]
\begin{center}
\raisebox{1ex}{(a)}\includegraphics[width=0.9\columnwidth]
  {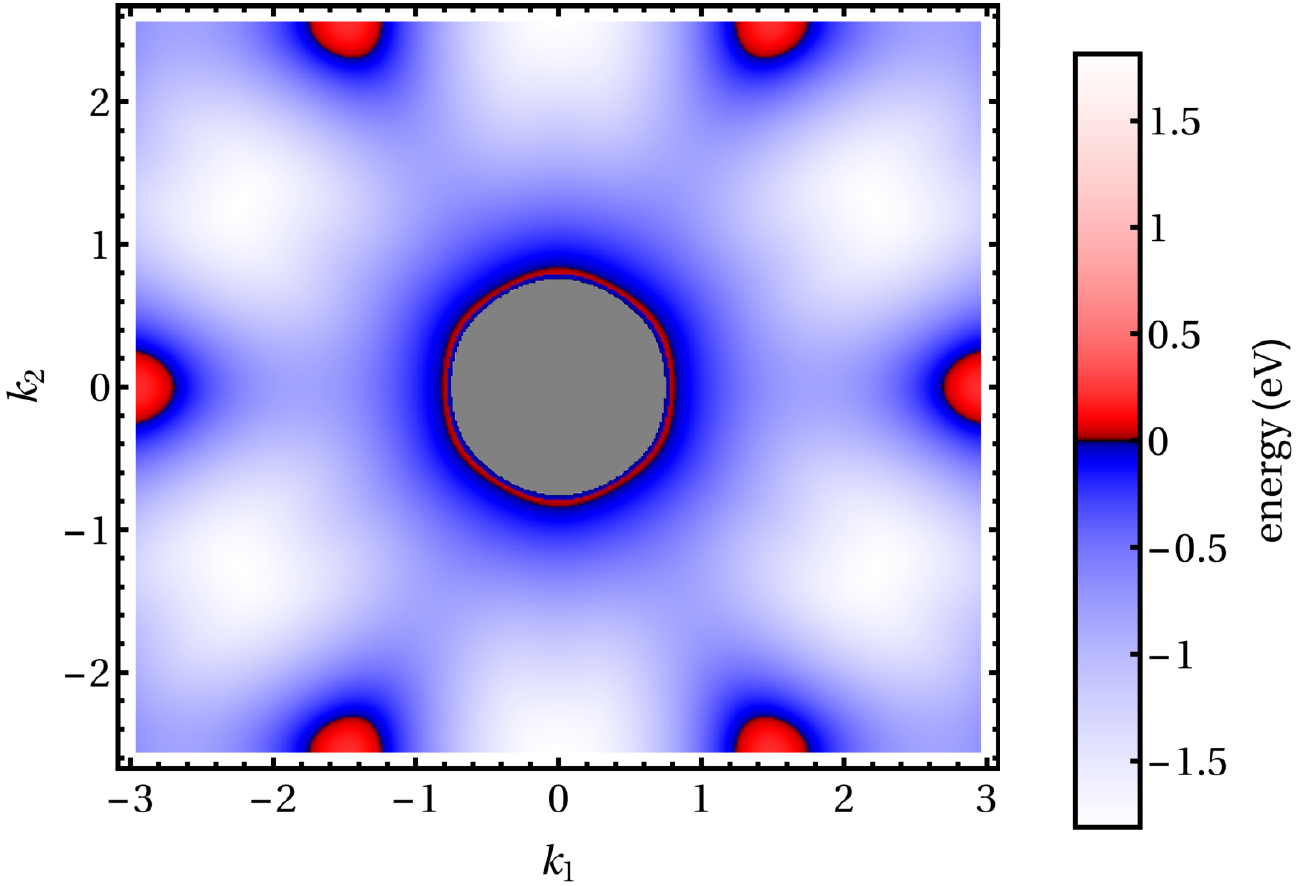}\\[2ex]%
\raisebox{1ex}{(b)}\includegraphics[width=0.9\columnwidth]
  {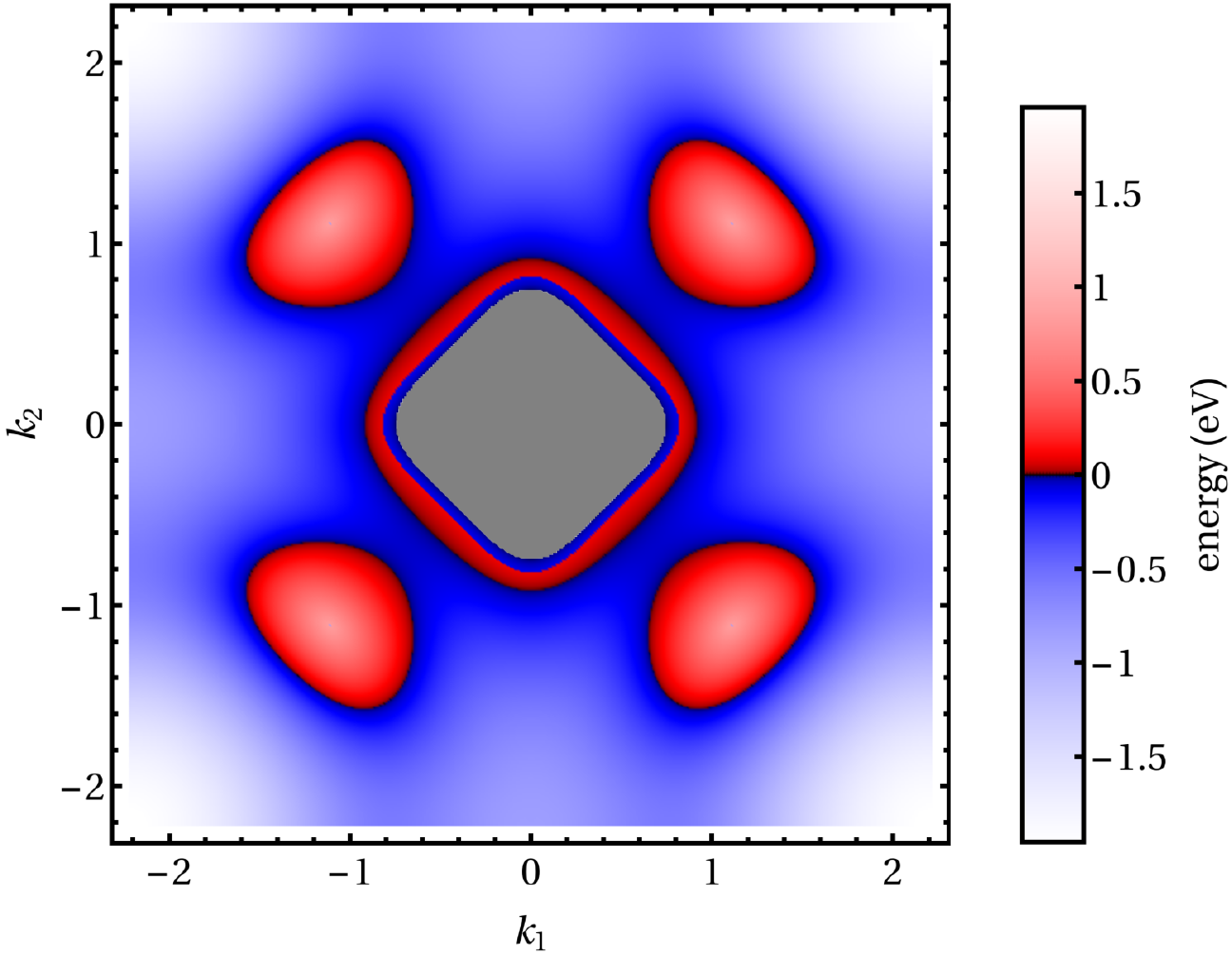}
\end{center}
\caption{Dispersion of the surface bands in the normal state for
(a) the $(111)$ surface and (b) the $(100)$ surface, for a
thickness of $W = 1000$. The projection of the bulk Fermi sea is
shown in gray.}
\label{fig.normal}
\end{figure}

\subsection{Surface states}

Surface states are studied by diagonalizing the BdG
Hamiltonian \eq{eq:BdGH} implemented on a real-space slab
of finite thickness. We will consider $(111)$ and $(100)$ surfaces,
as well as their symmetry partners.
The slabs preserve translation symmetry in two directions so
that the Hamiltonian can be block diagonalized by Fourier
transformation in these directions. Each block has the dimension
$8W\times 8W$, where $W$ is the number of layers in the slab. 
The wave vector parallel to the surfaces is written as
$\mathbf{k}_\|
  = k_1\, (1,-1,0)^T/\sqrt{2} + k_2\, (1,1,-2)^T/\sqrt{6}$
for the $(111)$ case and as
$\mathbf{k}_\|
  = k_1\, (0,1,1)^T/\sqrt{2} + k_2\, (0,-1,1)^T/\sqrt{2}$
for the $(100)$ case. More details can be found in Ref.~\cite{TSA17}.

In the following, we study the surface states of the TRSB
states with Bogoliubov Fermi  surfaces introduced above. The model
system possesses surface states even in the normal phase
\cite{DyaK81}, which form bands that emanate from the
quadratic band-touching point. They are analogous to the surface
bands in noncentrosymmetric half-Heusler materials
\cite{CSL11,LYW16,TSA17}, except that they are twofold spin
degenerate due to the IS in point group $O_h$. For later
comparison with the superconducting states, the dispersion of the
surface band closest to the Fermi energy is shown in
Fig.\ \ref{fig.normal} for the $(111)$ and $(100)$ surfaces. The
plots would of course be identical for directions related by
point-group symmetries; this will not be the case for some of the
superconducting states since these break lattice symmetries. Note
that the $k_1$ and $k_2$ axes for the $(100)$ surface are rotated
by $45^\circ$ relative to the cubic $k_y$ and $k_z$ axes. 

The surface bands in Fig.\ \ref{fig.normal} cross the Fermi
energy, seen as smooth changes from blue through black to red,
i.e., there are one-dimensional Fermi lines of surface states.
These Fermi lines are not protected; the surface bands can be
continuously deformed so that they do not cross the Fermi energy. 
Their presence is nevertheless interesting since it
allows us to study their fate when superconductivity sets in. 

\begin{figure}[tbh]
\begin{center}
\includegraphics[width=0.9\columnwidth]
  {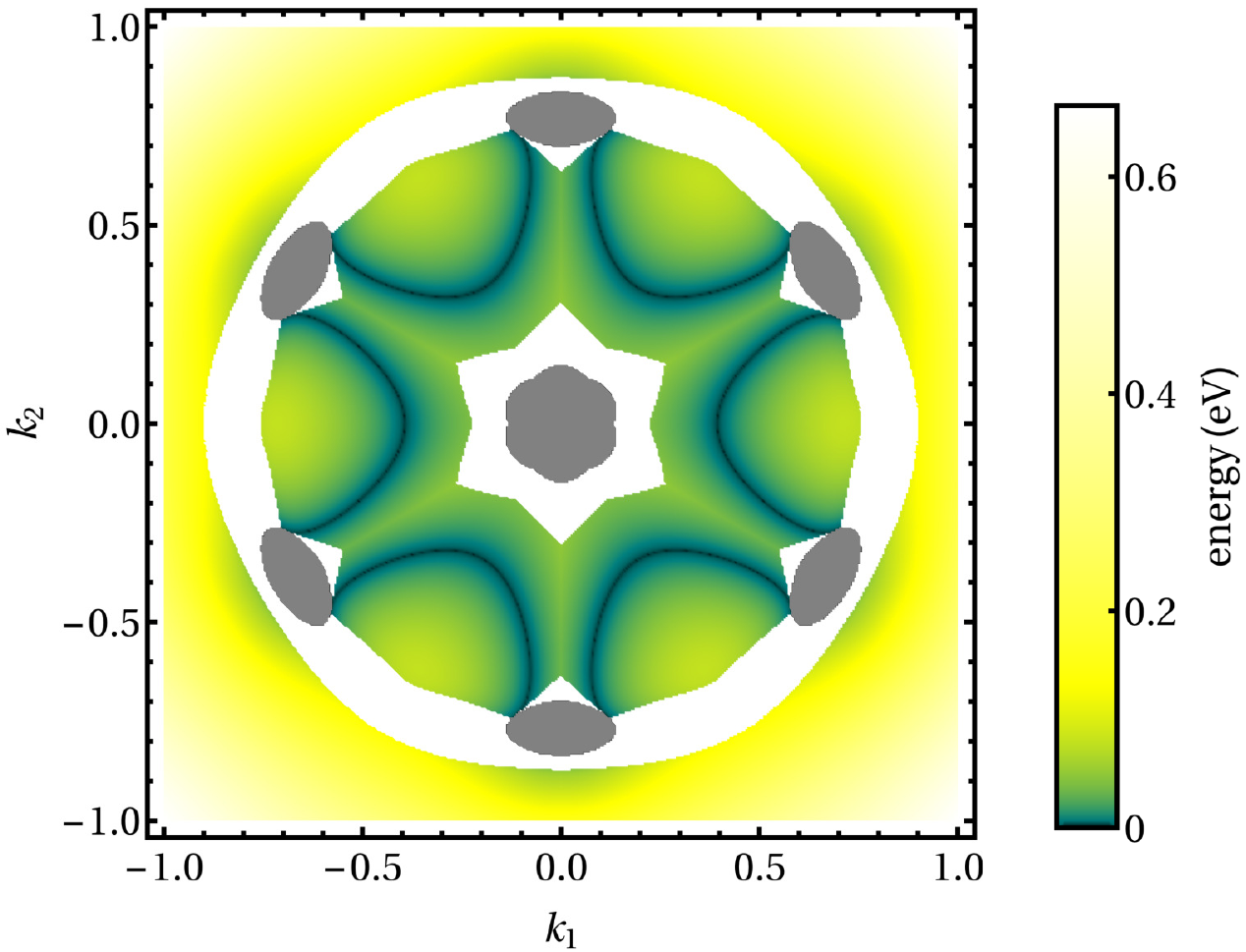}
\end{center}
\caption{Dispersion of the surface bands in the $E_g$ pairing
state with $\mathbf{h}=(1,i)$ for the $(111)$ surface. The
thickness is $W = 1000$. Only a close-up of the region of the
normal-state Fermi sea is shown. The spectrum at each momentum is
symmetric, the color refers to the absolute value of the
corresponding two energies $\pm \epsilon(k_1,k_2)$. The
projections of the Bogoliubov Fermi pockets are shown in gray. In
the white regions, no surface states are found.}
\label{fig.Eg111}
\end{figure}

We first consider the $E_g$ state with $\mathbf{h}=(1,i)$.
The surface dispersion for the $(111)$ surface is shown in
Fig.\ \ref{fig.Eg111}, which should be compared to
Fig.\ \ref{fig.normal}(a) for the normal state. We clearly see
the projections of the eight equivalent spheroidal Fermi pockets,
where the two in the $[111]$ and $[\bar{1}\bar{1}\bar{1}]$
directions are projected on top of each other.
In addition, there are arcs of zero-energy surface states
connecting the other six projected pockets.
Since the spectrum at each $(k_1,k_2)$ consists of pairs
$\pm\epsilon(k_1,k_2)$, two dispersive surface bands with opposite
velocities cross at each arc. An analysis of the corresponding
states shows that the two bands consist of states localized at
opposite surfaces. Hence, there are two arcs  originating
from each of the outer pockets at each surface. The associated
velocities are found to point in the \emph{same} direction for the
two arcs at the same surface, i.e., they have the same chirality.
The  two arcs per surface are
in agreement with the pockets having Chern numbers $\pm 2$, see
Sec.\ \ref{sub.Chern} below. For the central two pockets, no arcs
are present, consistent with their Chern numbers adding up
to zero.

\begin{figure}[tbh]
\begin{center}
\raisebox{1ex}{(a)}\includegraphics[width=0.9\columnwidth]
  {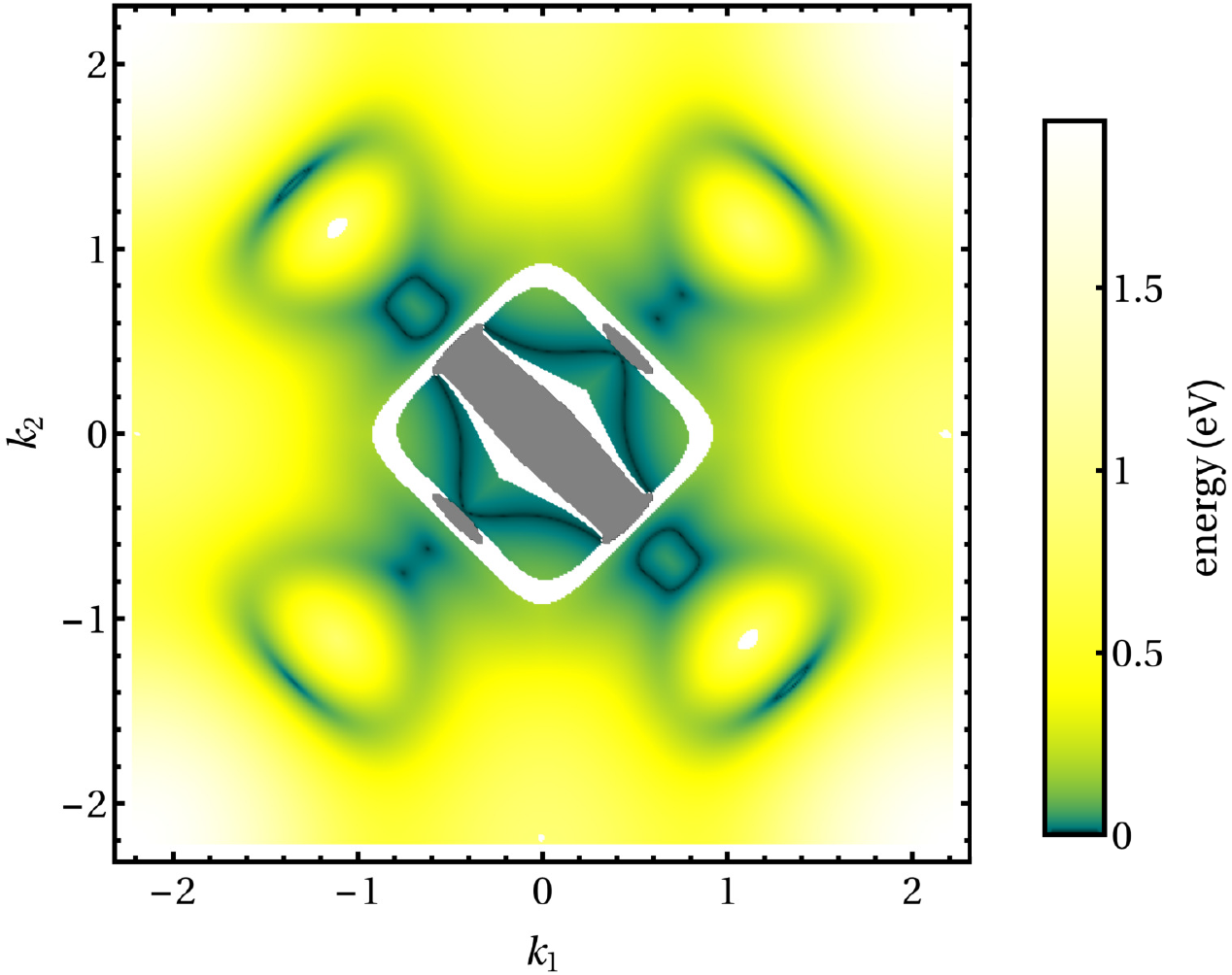}\\[2ex]%
\raisebox{1ex}{(b)}\includegraphics[width=0.9\columnwidth]
  {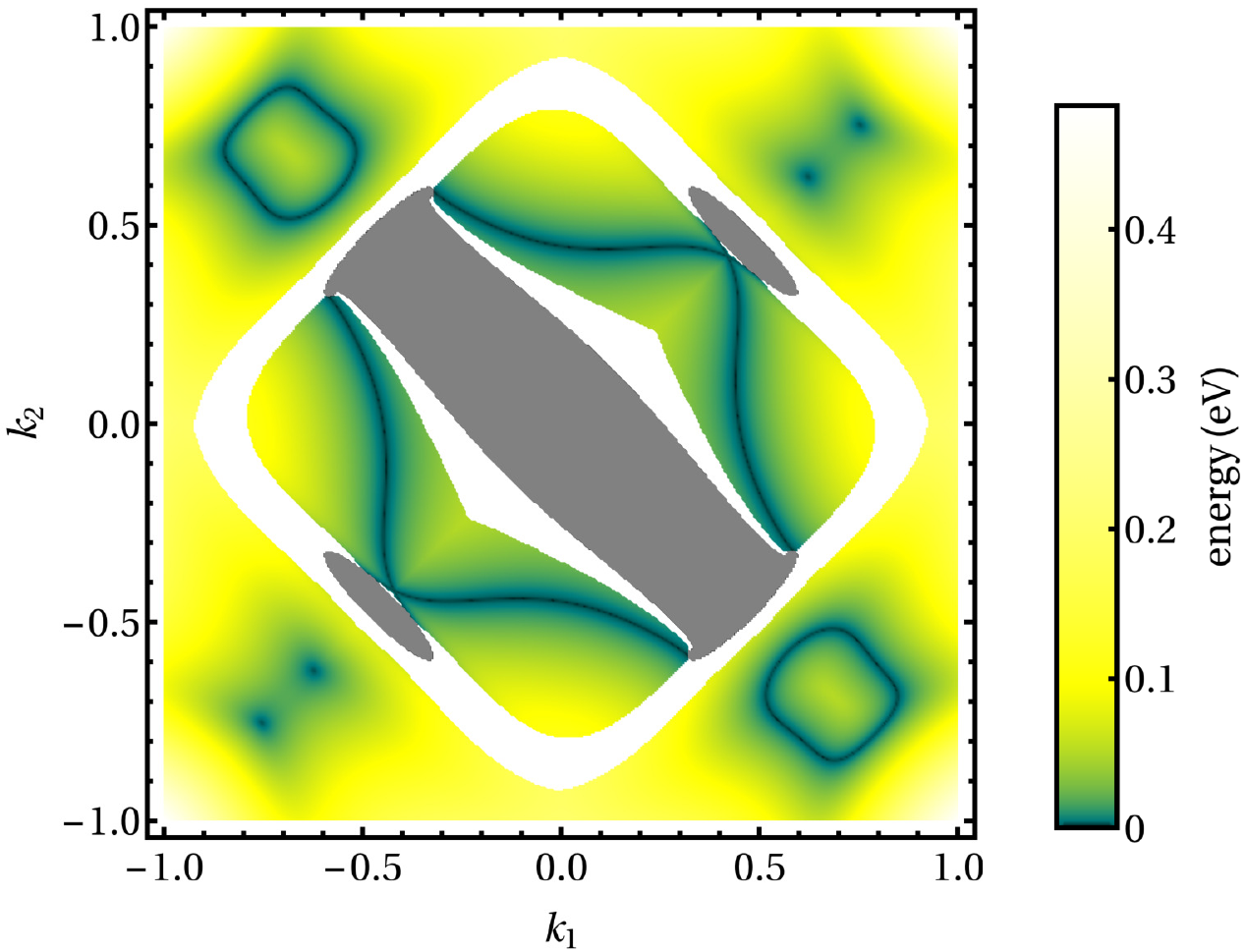}
\end{center}
\caption{(a) Dispersion of the surface bands in the chiral
$T_{2g}$ pairing state with $\mathbf{l}=(1,i,0)$ for the $(100)$
surface. The thickness is $W = 1000$. (b) Close-up of the central
region of panel (a), where in the normal state the Fermi sea
would be found. The spectrum at each momentum is symmetric,
the color refers to the absolute value of the corresponding two
energies $\pm \epsilon(k_1,k_2)$. The projections of the
Bogoliubov Fermi pockets are shown in gray. In the white regions,
no surface states are found.}
\label{fig.super100}
\end{figure}

Next, we turn to the chiral $T_{2g}$ pairing state with
$\mathbf{l}=(1,i,0)$. The chosen gap amplitude is the same as for
the $E_{g}$ state. Figure \ref{fig.super100} shows the surface
dispersion for the $(100)$ surface. The plot should be compared to
Fig.\ \ref{fig.normal} (b) for the normal state. The surface bands
originating from the normal state survive but their Fermi lines are
mostly gapped out and replaced by valleys at nonzero energy.
Instead, the surface bands develop new nodal lines, which are seen
as closed, black loops in Fig.\ \ref{fig.super100}. These lines are
not topologically protected and hence their existence and shape
depends on details of the model. The projections of the two
spheroidal and the toroidal Fermi pockets, see
Fig.\ \ref{fig.bulk.T2g}(a), are also clearly visible in
Fig.\ \ref{fig.super100}. In addition, each projected spheroidal
pocket is connected to the projected toroidal pocket by two Fermi
arcs, in agreement with Chern numbers of $\pm 2$ for the spheroidal
pockets. The two arcs localized at the same surface have the same
chirality, as expected. The arcs can also be understood in terms of
twofold degenerate arcs as found by
Tamura \textit{et al.}\ \cite{TKB17} for a single-band model, which
are split into two by the pseudomagnetic field \cite{ABT17}. As we
will see, the toroidal pocket has Chern number $0$ and thus does
not impose the presence of any arcs. However, its large projection
is in the way of the arcs from the spheroidal pockets. There are
four arcs connected to the projection of the toroidal pocket, with
their chirality summing to zero.

\begin{figure}[tbh]
\begin{center}
\includegraphics[width=0.9\columnwidth]
  {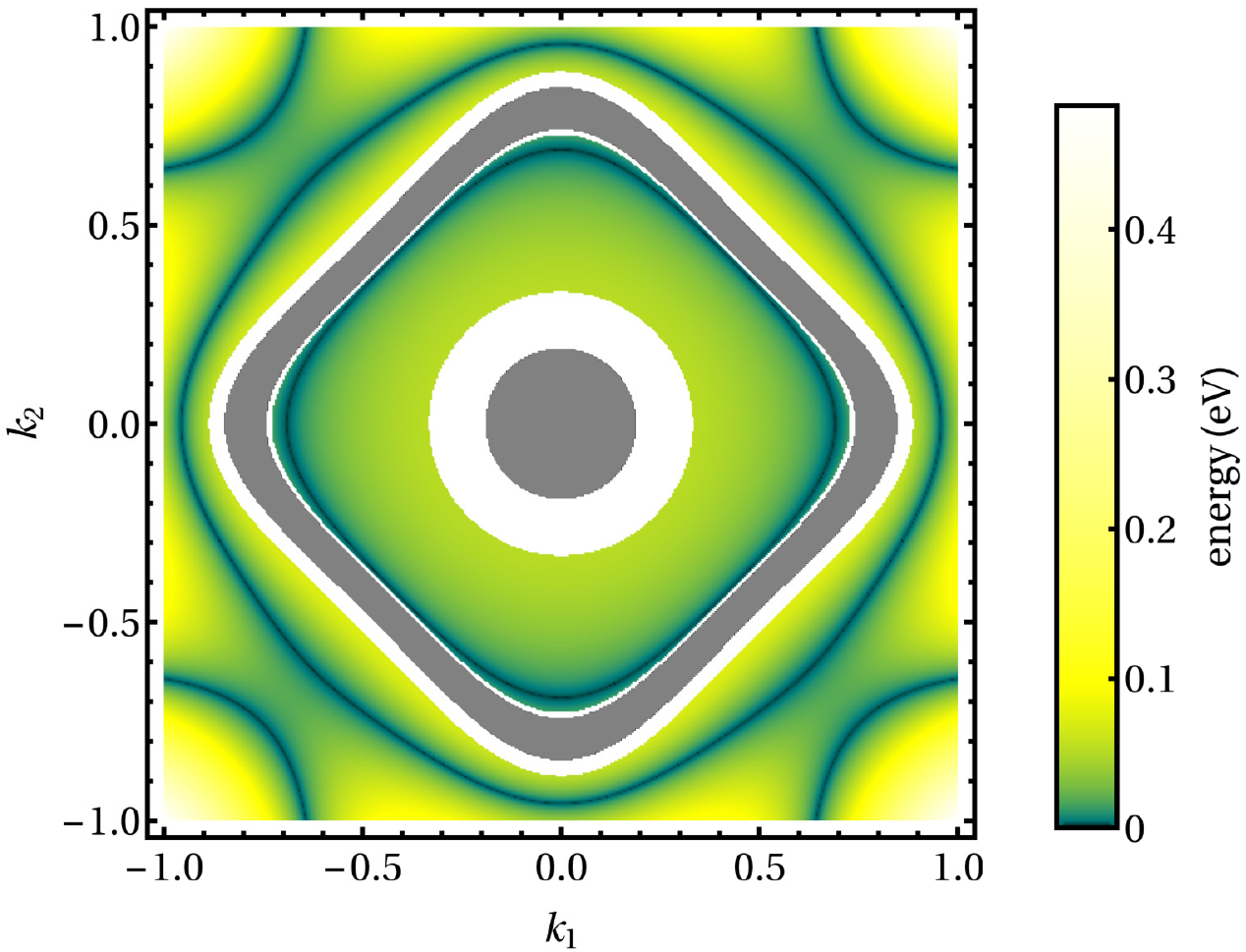}
\end{center}
\caption{Dispersion of the surface bands in the chiral
$T_{2g}$ pairing state with $\mathbf{l}=(1,i,0)$ as in
Fig.\ \ref{fig.super100}(b), but for the $(001)$ surface.}
\label{fig.super001}
\end{figure}

Figure \ref{fig.super100} shows the projection of the inflated line
node on edge. Since line nodes in other topological superconductors,
namely noncentrosymmetric ones that preserve TRS, are accompanied by
flat surface bands \cite{TMY10,STY11,BST11,ScR11,SBT12,ScB15}, one
might ask whether the same is true here. To check this, we plot in
Fig.\ \ref{fig.super001} the surface dispersion for the $(001)$
surface. The projection of the inflated line node is clearly visible
as the rounded gray square. Obviously, there is no flat band
delimited by the projected node. Indeed, a flat band is not expected
since the inflated line node is protected by nontrivial Pfaffians in
each mirror-parity sector, as discussed further in
Sec.\ \ref{sub.Pfaffian}. These Pfaffians are only defined in the
mirror-invariant $k_z=0$ plane. The nature of line nodes in
noncentrosymmetric superconductors with TRS is different: they are
protected by winding numbers calculated along closed loops around
the node. The argument for the existence of flat bands relies on the
deformation of these loops into straight lines perpendicular to the
surface \cite{SBT12,ScB15}. Such a construction is not possible for
the present case of nodes protected by a mirror symmetry.

\begin{figure}[tbh]
\begin{center}
\includegraphics[width=0.9\columnwidth]
  {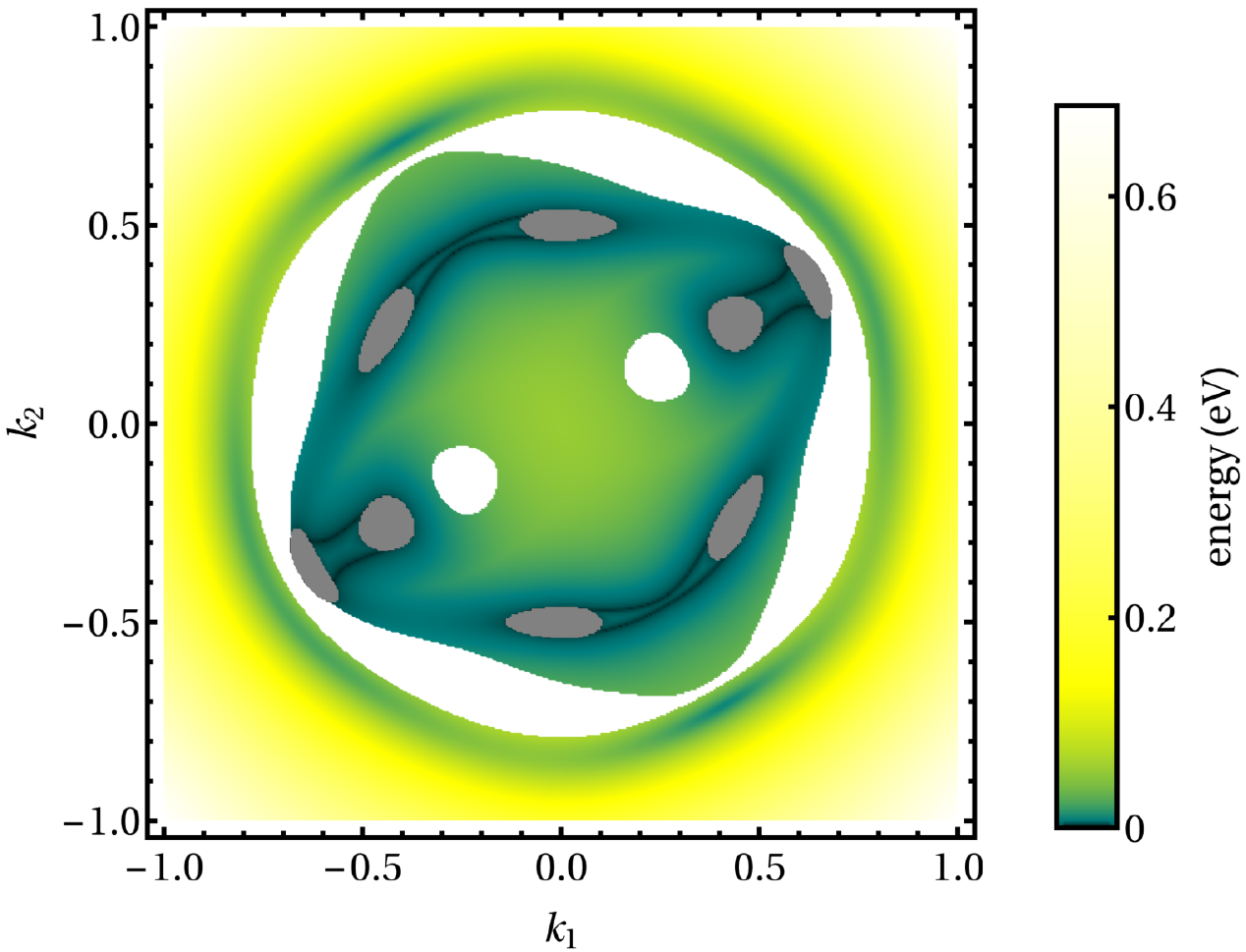}
\end{center}
\caption{Dispersion of the surface bands in the cyclic
$T_{2g}$ pairing state with
$\mathbf{l}=(1,e^{2\pi i/3},e^{-2\pi i/3})$ for the ($1\bar{1}1$)
surface. The thickness is $W = 1000$. Only a close-up of the region
of the normal-state Fermi sea is shown. The spectrum at each momentum
is symmetric, the color refers to the absolute value of the
corresponding two energies $\pm \epsilon(k_1,k_2)$. The projections
of the Bogoliubov Fermi pockets are shown in gray. In the white
regions, no surface states are found.}
\label{fig.T2g4c111}
\end{figure}

\begin{figure}[t]
\begin{center}
\includegraphics[width=0.9\columnwidth]
  {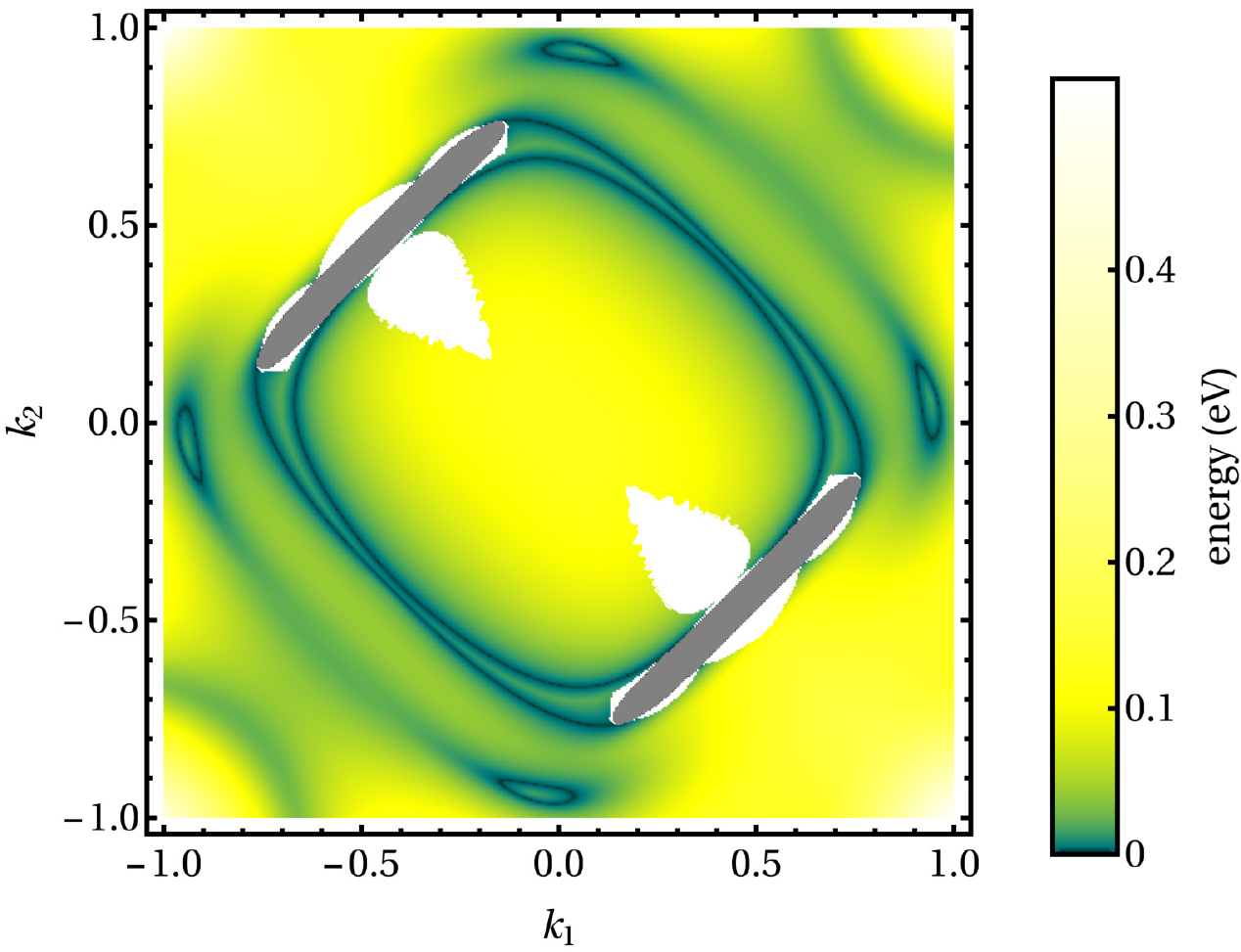}
\end{center}
\caption{Dispersion of the surface bands in the mixed $E_g$-$T_{2g}$
pairing state for the $(100)$ surface. The thickness is $W = 1000$.
Only a close-up of the region of the normal-state Fermi sea is shown.
The spectrum at each momentum is symmetric, the color refers to the
absolute value of the corresponding two energies
$\pm \epsilon(k_1,k_2)$. The projections of the Bogoliubov Fermi
pockets are shown in gray. In the white regions, no surface states
are found.}
\label{fig.mixed100}
\end{figure}

The cyclic $T_{2g}$ pairing state with
$\mathbf{l}=(1,e^{2\pi i/3},e^{-2\pi i/3})$ only has inflated point
nodes. We consider the ($1\bar{1}1$) surface here. This is equivalent
to the $(111)$ surface for $\mathbf{l}=(1,e^{\pi i/3},e^{-\pi i/3})$.
The $(111)$ surface for $\mathbf{l}=(1,e^{2\pi i/3},e^{-2\pi i/3})$
is less instructive since two of the Fermi pockets are projected on
top of each other. The dispersion of surface states is shown in
Fig.\ \ref{fig.T2g4c111} for the smaller gap amplitude
$\Delta_0=0.05\,\mathrm{eV}$, for which the Bogoliubov Fermi pockets
are separated. The Fermi pockets are shown in
Fig.\ \ref{fig.bulk.T2gsmall} above.
The projections of the eight Fermi pockets are clearly visible in
Fig.\ \ref{fig.T2g4c111}. Note that the two at the larger distance
from the center are inequivalent to the other six. All pockets are
connected by Fermi arcs in pairs. There are two arcs associated with
each pocket, as expected for Chern numbers of $\pm 2$.

Finally, we consider the  mixed-irrep pairing state of \eq{eq:mix}.
The surface dispersion for the $(100)$ surface is shown in
Fig.\ \ref{fig.mixed100}. The edge-on projections of the large
Bogoliubov Fermi pockets are clearly  visible.
Four arcs emanate from each of them, unlike for the inflated point
nodes encountered so far.  It is natural to attribute the doubled
number of arcs to  the double-Weyl nature of the original point
nodes. Four arcs would be consistent with Chern numbers
$\mathrm{Ch}_1=\pm 4$. We will see in
Sec.\ \ref{sub.Chern} that this is indeed the correct explanation.

\section{Hexagonal superconductors}
\label{sec.D6h}

In the context of unconventional superconductivity, the $j=3/2$
description of the $\Gamma_8$ bands in a cubic system is rather
unfamiliar. Pairing of four-component fermions is more commonly
encounted when the low-energy electron states have well-defined
orbital and spin degrees of freedom, as in Sr$_2$RuO$_4$ or the
iron-pnictide superconductors. To show how our analysis
works for such a case, we consider the example
of a hypothetical hexagonal superconductor
with point group $D_{6h}$, where the low-energy electron states
arise from orbitals belonging to the two-dimensional irrep
 $E_{1g}$. Selecting orbitals from a two-dimensional
irrep ensures that both orbitals will have equal weight at the
Fermi surface, and therefore represents a more generic origin of
the four-component fermions as compared to the accidental
near-degeneracy of two orbitals from different irreps. Choosing
orbitals which belong to one of the three
other two-dimensional irreps does not introduce qualitatively new
physics.

The normal-state block of the BdG Hamiltonian~\eq{eq:BdGH} reads
\begin{widetext}
\begin{align}
H_0(\mathbf{k}) &= \bigg[ \epsilon_{00} -
  (t_{xy}+t_{xyz}\cos k_z)\bigg(\cos k_x
  + 2\cos\frac{k_x}{2} \cos\frac{\sqrt{3}k_y}{2}\bigg)
  - t_z\cos k_z - \mu\bigg] \chi_{0} \otimes \sigma_0 \notag \\
&\quad{}+ \bigg[ \epsilon_{23} - (t'_{xy}+t'_{xyz}\cos k_z)
  \bigg(\cos k_x
  + 2\cos\frac{k_x}{2} \cos\frac{\sqrt{3}k_y}{2}\bigg)
  - t'_z \cos k_z \bigg] \chi_{2}\otimes\sigma_3 \notag \\
&\quad{}- (t^{\text{inter}}_{xy} + t^{\text{inter}}_{xyz} \cos k_z)
  \bigg[\bigg(\cos k_x
  - \cos\frac{k_x}{2} \cos\frac{\sqrt{3}k_y}{2}\bigg)
  \chi_{3}\otimes\sigma_0
  - \sqrt{3}\sin\frac{k_x}{2} \sin\frac{\sqrt{3}k_y}{2}
  \chi_{1}\otimes\sigma_0\bigg] \notag \\
&\quad{}- 2\,t_{\text{soc}} \sin k_z \bigg[\bigg(2\cos\frac{k_x}{2}
  + \cos\frac{\sqrt{3}k_y}{2}\bigg) \sin\frac{k_x}{2}\,
  \chi_2\otimes\sigma_1
  + \sqrt{3}\cos\frac{k_x}{2} \sin\frac{\sqrt{3}k_y}{2}\,
  \chi_2\otimes\sigma_2\bigg] \,, \label{eq:hexham}
\end{align}
\end{widetext}
where $\sigma_\mu$ and $\chi_\mu$ are Pauli matrices
describing the spin
and the orbital degree of freedom, respectively.
This tight-binding model includes
nearest-neighbor hopping in the \textit{xy} plane 
and normal to the plane and also next-nearest-neighbor
hopping out of plane. We note that the third line describes
orbitally nontrivial hopping, while the second and fourth lines
describe spin-orbit coupling. Their matrix structure is
determined by the transformation properties of the orbitals
under point-group operations. The momentum-dependent prefactors
are constrained by the periodicity in reciprocal space and by
$H_0(\mathbf{k})$ having to transform trivially, i.e.,
according to the irrep $A_{1g}$.

The fundamental difference of the normal-state band structure of
the hexagonal model compared to the cubic case is that there is no
symmetry-protected band-touching point at $\mathbf{k}=0$ since the
double group $D_{6h}$ does not have any four-dimensional irreps
\cite{DDJ08}. The five nontrivial Kronecker products appearing
in \eq{eq:hexham} are a representation of the Euclidean Dirac
matrices. We can set
\begin{align}
\gamma^1 &= \chi_{2}\otimes\sigma_{3}\,,\\
\gamma^2 &= \chi_{2}\otimes\sigma_1\,,\\
\gamma^3 &= \chi_{2}\otimes\sigma_{2}\,,\\
\gamma^4 &= \chi_{1}\otimes\sigma_{0}\,,\\
\gamma^5 &= \chi_{3}\otimes\sigma_{0}
\end{align}
so that the unitary part of the time-reversal operator is
\begin{equation}
U_T = \gamma^1\gamma^2 = \chi_{0}\otimes i\sigma_{2}\,.
\end{equation}
The orbital degree of freedom is therefore invariant under
time reversal. For the numerical calculations we take
$\epsilon_{00}=6.625\,\mathrm{eV}$,
$\epsilon_{23}=5.3\,\mathrm{eV}$,
$t_z=1.5\,\mathrm{eV}$, $t_{xy}=0.667\,\mathrm{eV}$,
$t_{xyz}=-0.333\,\mathrm{eV}$, 
$t_z'=2\,\mathrm{eV}$, $t_{xy}'=0.8\,\mathrm{eV}$,
$t_{xyz}'=0.133\,\mathrm{eV}$,
$t_{xy}^{\text{inter}}=0.462\,\mathrm{eV}$,
$t_{xyz}^{\text{inter}}=-0.231\,\mathrm{eV}$,
$t_{\text{soc}}=-0.231\,\mathrm{eV}$,  
and $\mu=3.300\,\mathrm{eV}$, yielding an ellipsoidal Fermi
surface for the $-$ states at the zone center.

The local pairing is described in terms of the six matrices
\begin{align}
A_{1g}\mbox{:} && \Gamma_s &= U_T \,, \phantom{abcdefghijklm}
  \\
&& \Gamma_{s'} & = (\chi_{2}\otimes\sigma_3)\, U_T \,,\\
E_{1g}\mbox{:} && \Gamma_{xz} &= (\chi_{2}\otimes\sigma_{1})\,
  U_T \,,
\label{Gamma.E1g.1}\\ 
&& \Gamma_{yz} &= (\chi_{2}\otimes\sigma_{2})\, U_T \,, \\
E_{2g}\mbox{:} && \Gamma_{x^2-y^2}
  &= (\chi_{3}\otimes\sigma_{0})\, U_T \,,
\label{Gamma.E2g.1} \\
&& \Gamma_{xy} &= (\chi_{1}\otimes\sigma_{0})\, U_T \,.
\end{align}
The labeling of the matrices reflects the form of the gap when
projected  onto the states near the zone center: the $A_{1g}$
states are $s$-wave-like, whereas the $E_{1g}$ and the $E_{2g}$
states resemble $(k_xk_z,k_yk_z)$-waves and
$(k_x^2-k_y^2,k_xk_y)$-waves,
respectively. The unconventional $A_{1g}$ state and the
$E_{1g}$ states represent orbital-singlet spin-triplet pairing,
whereas the $E_{2g}$ states involve orbital-triplet
spin-singlet pairing.

\begin{figure*}
\raisebox{1ex}{(a)}\includegraphics[width=0.3\textwidth]
  {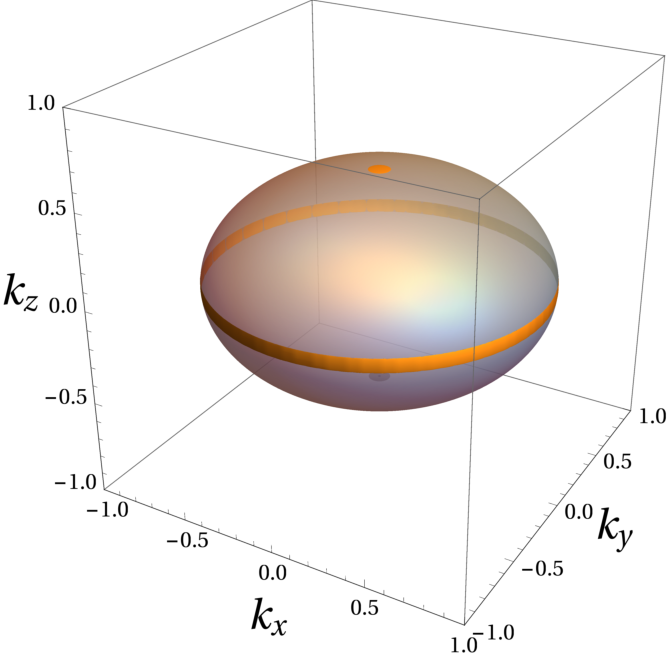}
\raisebox{1ex}{(b)}\includegraphics[width=0.3\textwidth]
  {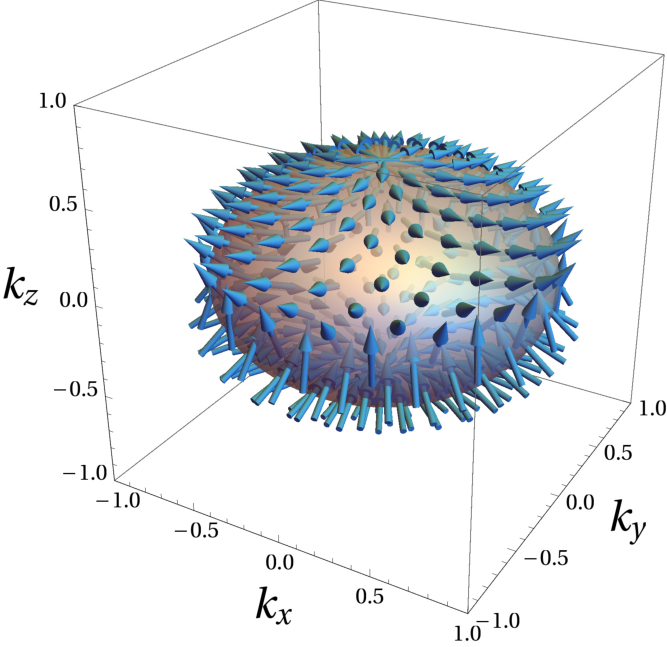}
\raisebox{1ex}{(c)}\includegraphics[width=0.3\textwidth]
  {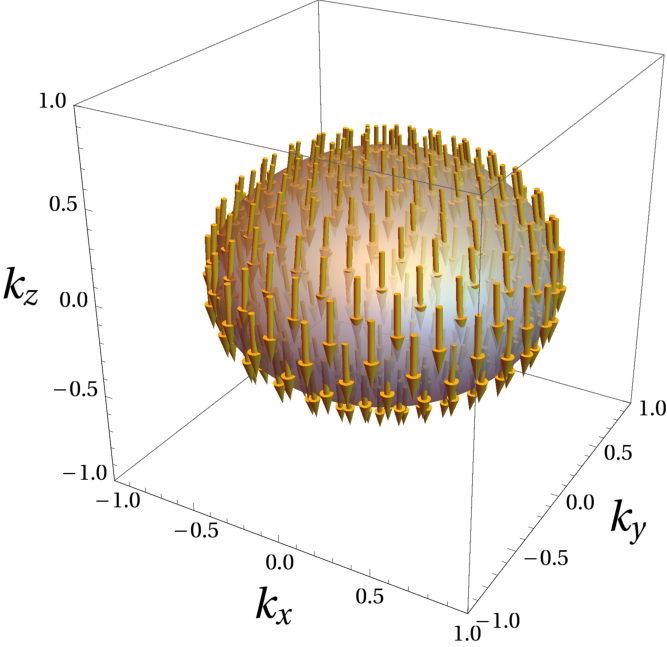}\\
\raisebox{1ex}{(d)}\includegraphics[width=0.3\textwidth]
  {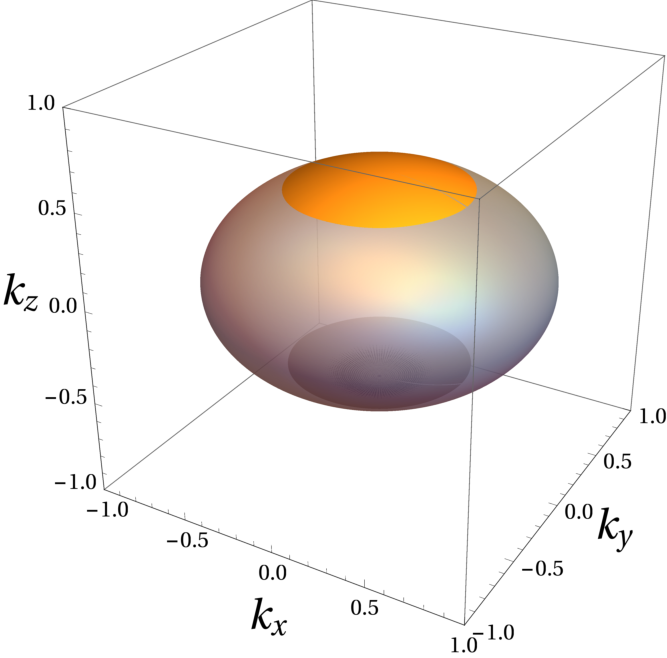}
\raisebox{1ex}{(e)}\includegraphics[width=0.3\textwidth]
  {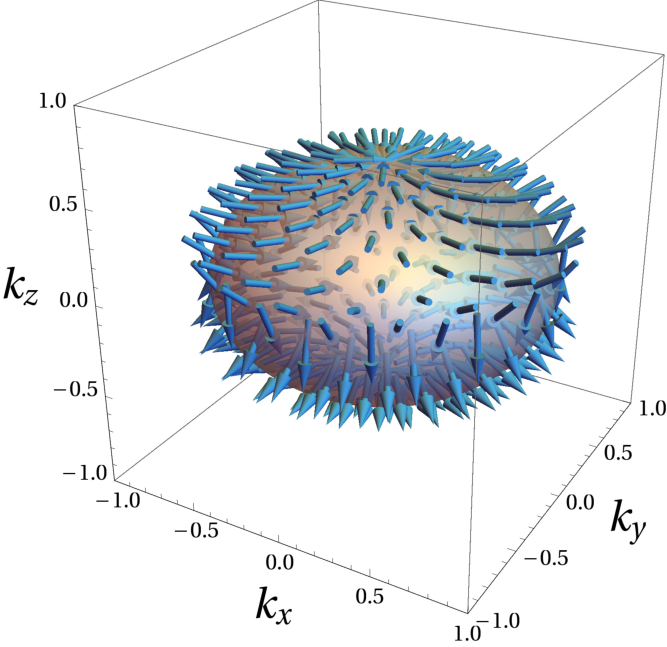}
\raisebox{1ex}{(f)}\includegraphics[width=0.3\textwidth]
  {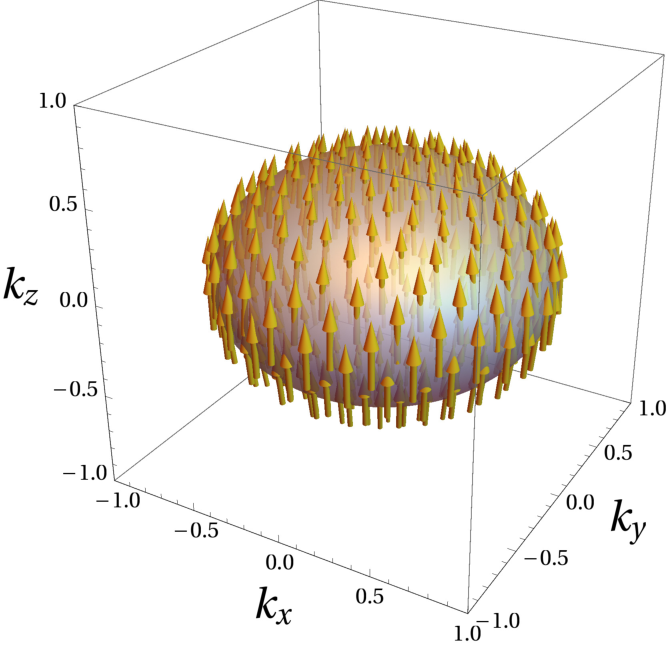}
\caption{\label{fig.bulk.hex}Low-energy structure of the TRSB
$E_{1g}$ [(a)--(c)] and $E_{2g}$ [(d)--(f)] pairing states.
(a), (d) Bogoliubov Fermi surfaces (opaque orange) in
comparison to the normal-state Fermi surface (semi-transparent).
(b), (e) Pseudomagnetic field acting on the states at
the Fermi surface. Note that the orientation is basis dependent
and corresponds to our choice of the MCBB defined in Appendix
\ref{app.ps}. (c), (f)  Magnetization of the states at
the Fermi surface arising from the pseudomagnetic field. For
panels (a) and (d), a gap amplitude of
$\Delta_0=0.02\,\mathrm{eV}$ has been used.}
\end{figure*}

\subsection{TRSB pairing}

Restricting ourselves to pure-irrep pairing, two TRSB pairing
states are allowed in our model:
\begin{align}
\Delta_{E_{1g}} &= \Delta_0\, (\Gamma_{xz} + i\Gamma_{yz}) \,,\\
\Delta_{E_{2g}} &= \Delta_0\, (\Gamma_{x^2-y^2} + i\Gamma_{xy})
  \,.
\end{align}
The single-band analogue of the $E_{1g}$ state is a
$k_z(k_x+ik_y)$-wave state~\cite{SiU91,VoG85}, similar to the
chiral $T_{2g}$ state of the cubic superconductor. It is thus
expected to have line nodes in the $k_z=0$ plane and point nodes
on the $k_z$ axis. In the
single-band limit, the nodal structure of the $E_{2g}$ pairing
resembles the $(k_{x}+ik_y)^2$-wave state~\cite{SiU91,VoG85}, with
double Weyl points on the $k_z$-axis. In the two-band system
considered here, these nodes are inflated into Bogoliubov
Fermi surfaces, which are shown in
Figs.\ \ref{fig.bulk.hex}(a) and (d).
The Fermi surfaces are similar to the corresponding chiral
$T_{2g}$ and mixed-irrep states, respectively, for the $O_h$ point
group. Since for the $E_{2g}$ state the single-band variant has a
double Weyl point, the mechanism explained in \Sec{subsub.mixed}
leads to large inflated nodes in the multiband case. For this
reason, a small gap amplitude has been chosen for \fig{fig.bulk.hex}.

An interesting distinction between the two states is provided by the
time-reversal-odd gap product:
\begin{align}
\Delta_{E_{1g}}^{\phantom{\dagger}}\Delta_{E_{1g}}^\dagger
  -\Delta_{T,E_{1g}}^{\phantom{\dagger}}\Delta_{T,E_{1g}}^\dagger
  &= 4 \Delta_0^2\, \chi_{0}\otimes\sigma_3\,,\\
\Delta_{E_{2g}}^{\phantom{\dagger}}\Delta_{E_{2g}}^\dagger 
  -\Delta_{T,E_{2g}}^{\phantom{\dagger}}\Delta_{T,E_{2g}}^\dagger
  &= 4 \Delta_0^2\, \chi_{2}\otimes\sigma_0\,.
\end{align}
Although in both cases this product belongs to the
irrep $A_{2g}$, for the $E_{1g}$ case it represents a purely magnetic
order, whereas in the $E_{2g}$ case it corresponds to chiral orbital
order. This reflects the spin- and orbital-triplet nature of these
pairing states, respectively.

The pseudomagnetic field is shown in \fig{fig.bulk.hex}(b) and (e).
Remarkably, the two states have almost identical
$\delta{\bf h}_{{\bf k},-}$, albeit with opposite sign,
although their nodal structure and spin-orbital character are quite
different. From this, we obtain the physical magnetization
in analogy to the cubic case,
\begin{equation}
m_{\mathbf{k},\mu} = - \frac{1}{|\mathbf{v}_{\mathbf{k},-}|}\,
  \delta\mathbf{h}_{\mathbf{k},-} \cdot
  \Tr \mathcal{P}_{\mathbf{k},-} \check{\mathbf{s}}
  \mathcal{P}_{\mathbf{k},-}
  \chi_0\otimes\sigma_\mu \,.
\end{equation}
As expected from the pseudomagnetic field, the physical
magnetization is also very similar: in both cases an almost uniform
magnetization across the Fermi surface is
observed with net moment along the $z$-axis, but with opposite
sign. Although such a polarization is not surprising for the
$E_{1g}$ state in view of the explicitly magnetic form of
the time-reversal-odd gap product, it is less obvious for the
$E_{2g}$ state, for which the  time-reversal-odd gap product
corresponds to orbital order. The origin of the
magnetization in the latter case is the strong spin-orbit coupling,
in particular the term  in the second line of \eq{eq:hexham}, which
converts the orbital polarization into a spin polarization. 

\begin{figure}[tbh]
\begin{center}
\raisebox{1ex}{(a)}\includegraphics[width=0.95\columnwidth]
  {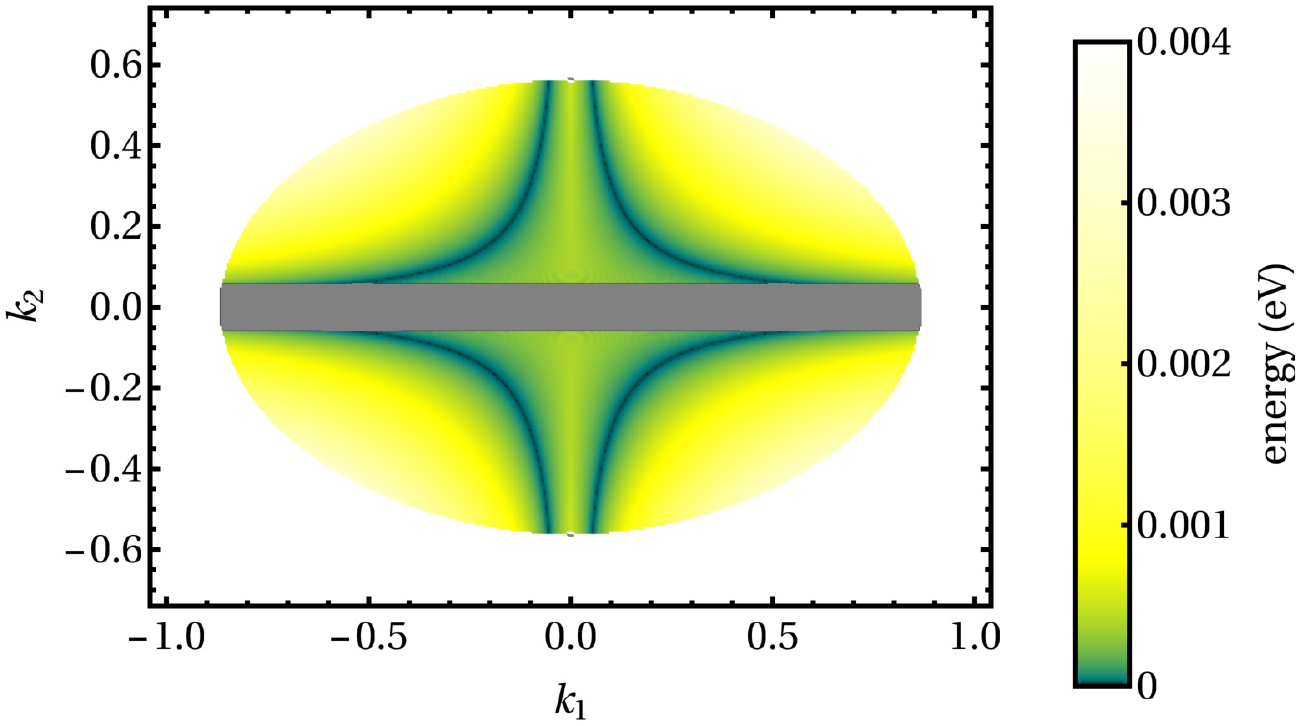}\\[2ex]%
\raisebox{1ex}{(b)}\includegraphics[width=0.95\columnwidth]
  {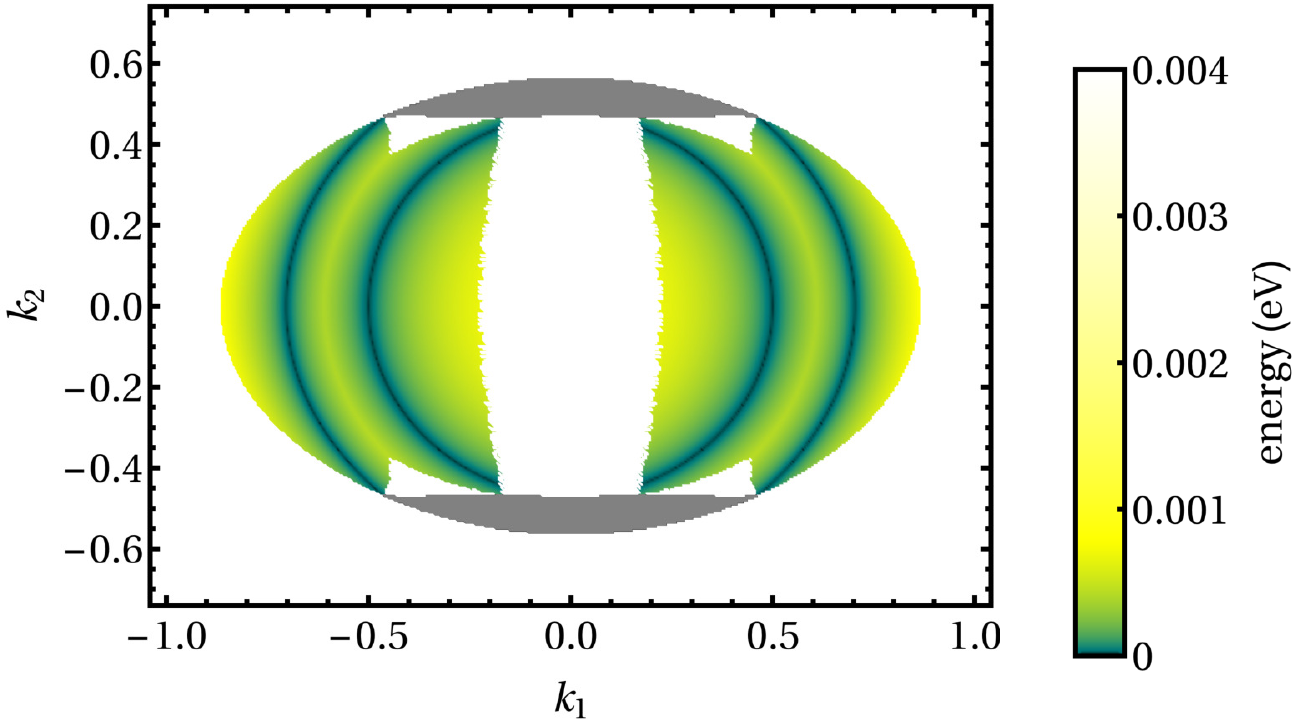}
\end{center}
\caption{Dispersion of the surface bands at the $(010)$ surface for
the TRSB (a) $E_{1g}$ and (b) $E_{2g}$ states. The thickness of the
slab is $W=3000$. Only a close-up of the region of the normal-state
Fermi sea is shown. The spectrum at each momentum is symmetric, the
color refers to the absolute value of the corresponding two energies
$\pm \epsilon(k_1,k_2)$. The projections of the Bogoliubov Fermi
pockets are shown in gray. In the white regions, no surface states
are found. The white region in the center of panel (b) is probably
an artifact of the finite thickness.} 
\label{fig.hex}
\end{figure}

\subsection{Surface states}

We next consider the surface states at the $(010)$ surface, i.e.,
the one normal to the $k_y$-axis. We do not find surface bands in the
normal state. This is expected since the normal-state surface bands
for the cubic model originate from the  topologically nontrivial
band-touching point at $\mathbf{k}=0$, which does not exist
for the $D_{6h}$ point group. Figure \ref{fig.hex} shows the
surface dispersion for the  $E_{1g}$ and $E_{2g}$ pairing
states. Surface bands only appear where the normal-state Fermi sea
has been gapped out, which is consistent with the absence of surface
bands in the normal state. For $E_{1g}$ pairing, two Fermi arcs
emanate from each of the two spheroidal Fermi pockets, consistent
with Chern number $\mathrm{Ch}_1=\pm 2$. Note that the Fermi pockets
are so thin that they are essentially invisible if viewed from the
edge. For the $E_{2g}$ case, we find that the localization length of
surface states becomes very large for $k_1$ approaching zero so that
for small $k_1$ they become indistinguishable from bulk states even
for the large
thickness of $W=3000$ used here. The white region in the center of
Fig.\ \ref{fig.hex} (b) is thus expected to be at least partially an
artifact of the finite thickness. There are four arcs emanating from
each inflated node, consistent with the Chern numbers
$\mathrm{Ch}_1=\pm 4$.

\section{Topological invariants}

The Bogoliubov Fermi pockets  are protected by a $\mathbb{Z}_2$
invariant, which we have identified as the relative sign of the
Pfaffian of the unitarily transformed BdG Hamiltonian on the two
sides of the Fermi surface \cite{ABT17}. For a single band,
pairing states with the same symmetries as the ones discussed above
have point and line nodes that are protected by topological
invariants. Specifically, the point nodes have nonzero Chern
numbers, while the line node for the chiral
$T_{2g}$ state with $\mathbf{l}=(1,i,0)$ is protected by a mirror
symmetry \cite{ChS14,CTS16,BWR16,BzS17}. It is natural to ask
whether these invariants survive in the multiband case, where the
nodes are inflated into Bogoliubov Fermi surfaces.
In the following, we illustrate our general considerations using
the example provided by the $j=3/2$ pairing states of the cubic
superconductor discussed above.

\subsection{Chern invariant}
\label{sub.Chern}

For $CP$ symmetry satisfying $(CP)^2 = +1$, point nodes have a
$2\mathbb{Z}$ invariant, which is given by the first Chern number
for a closed surface $\mathcal{S}$ surrounding the node
\cite{ZSW16}. This Chern number is given
by~\cite{XCN10,She12,CTS16,BzS17}
\begin{equation}
\mathrm{Ch}_1 = \frac{1}{2\pi} \sum_{n\:\mathrm{occ.}}
  \oint_{\mathcal{S}} d^2\mathbf{s}(\mathbf{k})
  \cdot \big[ \nabla_\mathbf{k} \times \mathbf{A}_n(\mathbf{k})
  \big] \,,
\label{1.Chern.2}
\end{equation}
where $d^2\mathbf{s}(\mathbf{k})$ is a vectorial surface element
in momentum space and $\mathbf{A}_n(\mathbf{k})$ is the Berry
connection for the \textit{n}-th band,
\begin{equation}
\mathbf{A}_n(\mathbf{k}) = i\, \langle u_n(\mathbf{k})| 
  \nabla_\mathbf{k} | u_n(\mathbf{k})\rangle \,,
\end{equation}
in terms of the Bloch states $|u_n(\mathbf{k})\rangle$. The sum
in \eq{1.Chern.2} is over the occupied bands. Note that
\eq{1.Chern.2} only holds if the occupied bands are nondegenerate
on $\mathcal{S}$ \cite{CTS16}, which is the case here since TRS
is broken. It is worth emphasizing that this is not a
classification of nodal points but rather of closed surfaces
$\mathcal{S}$ in momentum space  for which the gap does not close
anywhere on $\mathcal{S}$. A nonzero Chern number guarantees that
the gap closes somewhere in the enclosed volume but this need not
happen at a single point.

Following Berry \cite{Ber84}, one can rewrite the Chern number as
a Kubo-type expression,
\begin{align}
\mathrm{Ch}_1 &= \frac{i}{2\pi}
  \sum_{n\:\mathrm{occ.}} \sum_{m\neq n}
  \oint_{\mathcal{S}} d^2\mathbf{s}(\mathbf{k}) \notag \\
&{}\cdot \frac{\langle u_n(\mathbf{k})|
  (\nabla_\mathbf{k}\mathcal{H}) |u_m(\mathbf{k})\rangle
  \times \langle u_m(\mathbf{k})| (\nabla_\mathbf{k}\mathcal{H})
  |u_n(\mathbf{k})\rangle}
  {(E_{\mathbf{k}n} - E_{\mathbf{k}m})^2} \,,
\end{align}
where $E_{\mathbf{k}n}$ is the eigenenergy of the Bloch state
$|u_n(\mathbf{k})\rangle$. This form is useful for the numerical
evaluation since it is independent of the choice of phases of the
Bloch states.
With this, the eight spheroidal Fermi pockets for the
$E_g$ state with $\mathbf{h}=(1,i)$, shown in
Fig.\ \ref{fig.bulk.Eg}(a), have $\mathrm{Ch}_1 = \pm 2$. The
sign of $\mathrm{Ch}_1$ for neighboring pockets is opposite.

\begin{figure}[tbh]
\begin{center}
\includegraphics[scale=0.45]{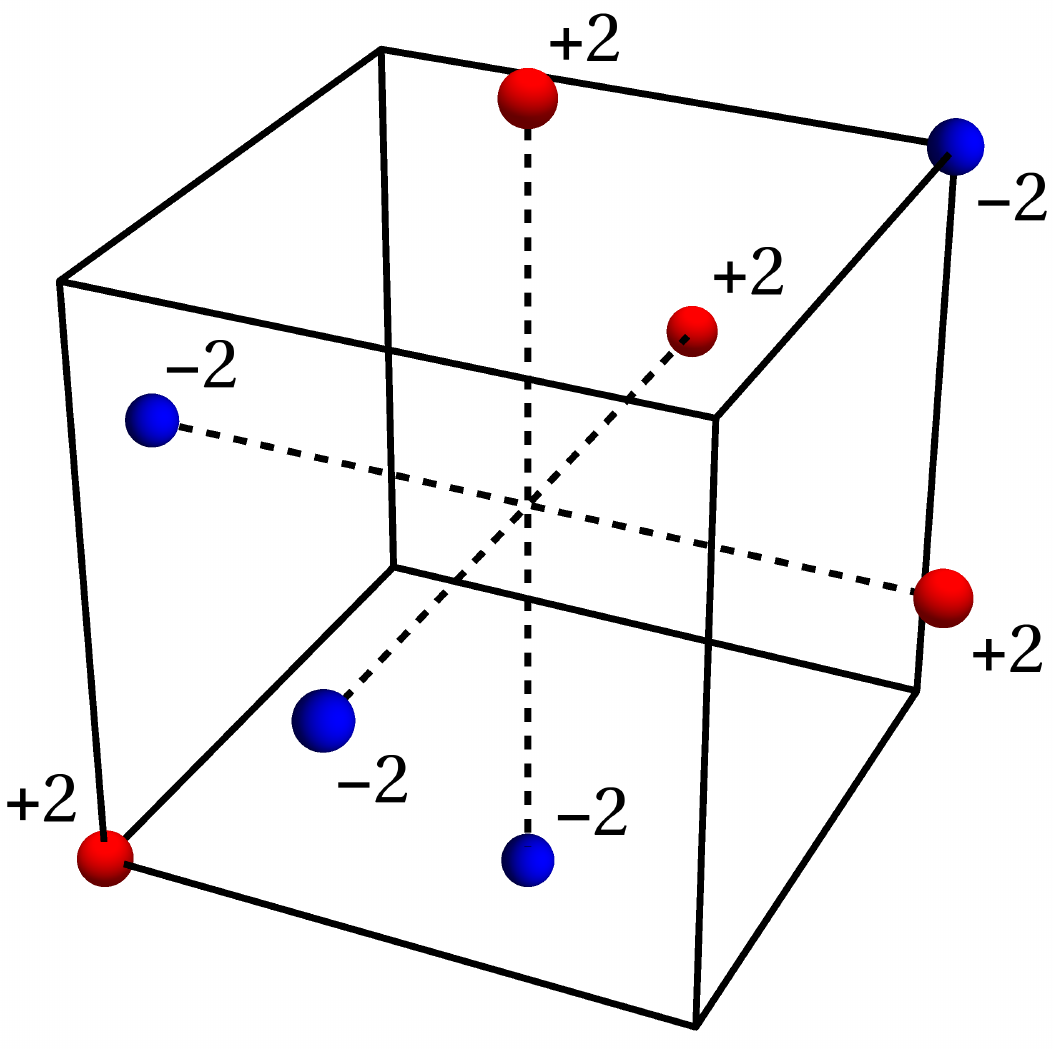}
\end{center}
\caption{Schematic of the locations and Chern numbers of the
inflated point nodes for the cyclic $T_{2g}$ pairing state
with $\mathbf{l}=(1,e^{2\pi i/3},e^{-2\pi i/3})$ and sufficiently
small pairing amplitude. Compare Fig.~\ref{fig.bulk.T2gsmall}.}
\label{fig.T2g.Chern}
\end{figure}

The surfaces enclosing the upper (lower) spheroidal Fermi pocket
for the chiral $T_{2g}$ state, shown in
Fig.\ \ref{fig.bulk.T2g}(a), are found to have Chern number
$\mathrm{Ch}_1 = -2$ ($+2$). Surfaces enclosing the whole
toroidal pocket have $\mathrm{Ch}_1 = 0$.
For the cyclic $T_{2g}$ state, shown in
Fig.\ \ref{fig.bulk.T2gsmall}, there are two
distinct classes of inflated point nodes. Their Chern numbers of
$\mathrm{Ch}_1=\pm 2$ are indicated in Fig.\ \ref{fig.T2g.Chern}.
The Chern numbers of the three pockets on the cubic axes
and next to one of the pockets at a corner are equal to
each other but opposite to the one of the pocket at the corner.
Hence, the Chern numbers of the four Bogoliubov Fermi pockets
that merge for larger pairing amplitudes add up to $\pm 4$, which
are thus the values for the large pockets of complicated shape
shown in Fig.~\ref{fig.bulk.T2g}(d). 

Finally, in the mixed-irrep $\Gamma_{x^2-y^2}+i\Gamma_{xy}$
state, the large upper (lower) pocket seen in
Fig.\ \ref{fig.bulk.mix}(a) has
Chern number $\mathrm{Ch}_1=+4$ ($-4$).  Our results for
the Chern numbers of the inflated point nodes are
consistent with the Berry curvature obtained in~\cite{RGF17}.

In summary, the spheroidal pockets for all considered pairing
states are protected by Chern invariants, in agreement with a
recent analysis by Bzdu\v{s}ek and Sigrist \cite{BzS17}.
Consequently, the pockets can shrink to points but not vanish
unless they annihilate pairwise. There is no corresponding
protection of the toroidal pocket for the chiral $T_{2g}$
state. The Chern numbers are all even, as expected for
$CP$-symmetric superconductors \cite{ZSW16}. They are also
consistent with the observed number of Fermi arcs of surface
states, namely two for pockets with $\mathrm{Ch}_1=\pm 2$ and
four in the case of $\mathrm{Ch}_1=\pm 4$.

We thus find that the inflated point nodes (spheroidal pockets)
are protected by two distinct invariants: an even Chern number
and a Pfaffian \cite{KST14,ZSW16,ABT17}. Bzdu\v{s}ek and Sigrist
\cite{BzS17} have recently formulated a comprehensive theory of
nodal points, lines, and surfaces protected by two invariants,
which they have dubbed ``multiply charged nodes.''

\subsection{Additional Pfaffians}
\label{sub.Pfaffian}

Neither the Altland-Zirnbauer class D nor $CP$ symmetry squaring
to $+1$ leads to invariants protecting line nodes in three
dimensions \cite{ZSW16}. Hence, lattice symmetries are required
for constructing an invariant for the line node in the
single-band version of the chiral $T_{2g}$ state. In the
following, we show that this also holds for the toroidal Fermi
pocket in the multiband case.

The BdG Hamiltonian satisfies mirror symmetry in the $xy$ plane:
\begin{equation}
U_{\sigma z}\, \mathcal{H}(k_x,k_y,-k_z)\, U_{\sigma z}^\dagger
  = \mathcal{H}(k_x,k_y,k_z) \,,
\end{equation}
with
\begin{equation}
U_{\sigma z} = \left(\begin{array}{@{}c@{~}c@{}}
  i\, e^{i\pi J_z} & 0 \\ 0 & -i\, e^{-i\pi J_z}
  \end{array}\right)
  = \tau_0 \otimes \sigma_0 \otimes \sigma_z \,,
\end{equation}
where the first factor in the Kronecker product refers to Nambu
space and the other two to spin-$3/2$ space \cite{endnote.U1}.
In the $k_z=0$ plane, this is a symmetry at fixed momentum.
Hence, it is possible to block diagonalize
$\mathcal{H}(k_x,k_y,0)$:
\begin{equation}
\mathcal{H}(k_x,k_y,0) \to \left(\begin{array}{@{}c@{~}c@{}}
  \mathcal{H}_+(k_x,k_y) & 0 \\ 0 & \mathcal{H}_-(k_x,k_y)
  \end{array}\right) ,
\label{2.block.2}
\end{equation}
where $\mathcal{H}_\pm(k_x,k_y)$ belongs to mirror eigenvalue
$\pm 1$ of $U_{\sigma z}$. Since $U_{\sigma z}$ is already
diagonal in our basis, only a reordering of rows and columns
is required to bring $\mathcal{H}(k_x,k_y,0)$ into
block-diagonal form.

As discussed in \Sec{sub.BFsurfaces}, the BdG Hamiltonian can
be transformed into an antisymmetric matrix
$\tilde{\mathcal{H}}(\mathbf{k})$ with the unitary
matrix $\Omega$ of \eq{eq:defOmega}.
Since $\Omega$ commutes with $U_{\sigma z}$, $\Omega$ also
maps the block-diagonal form of the $k_z=0$ Hamiltonian into
a block-diagonal antisymmetric matrix
\begin{equation}
\left(\begin{array}{@{}c@{~}c@{}}
  \tilde{\mathcal{H}}_+(k_x,k_y) & 0 \\
  0 & \tilde{\mathcal{H}}_-(k_x,k_y)
  \end{array}\right) .
\end{equation}
We can now calculate the Pfaffians of
$\tilde{\mathcal{H}}_\pm(k_x,k_y)$ separately:
\begin{align}
\mathrm{Pf}\, \tilde{\mathcal{H}}_+(k_x,k_y)
  &= \langle\underline{\epsilon}_{\bf k},
  \underline{\epsilon}_{\bf k}\rangle \,, \\
\mathrm{Pf}\, \tilde{\mathcal{H}}_-(k_x,k_y)
  &= \langle\underline{\epsilon}_{\bf k},
  \underline{\epsilon}_{\bf k}\rangle
  + 4 \Delta_0^2 \,,
\end{align}
where $\mathbf{k}=(k_x,k_y,0)$. The full Pfaffian in
\eq{eq:Pfaffian} can be decomposed into a product of
two Pfaffians in the $k_xk_y$ plane,
\begin{align}
P(k_x,k_y,0) &= \mathrm{Pf}\,\tilde{\mathcal{H}}(k_x,k_y,0)
  \notag \\
&= \mathrm{Pf}\,\tilde{\mathcal{H}}_+(k_x,k_y)\;
  \mathrm{Pf}\,\tilde{\mathcal{H}}_-(k_x,k_y) \,.
\end{align}
For each mirror sector, we find zero-energy states wherever
the corresponding Pfaffian changes sign. These sign changes
generically define closed lines in the $k_xk_y$ plane. Since
the two Pfaffians change their sign at different $(k_x,k_y)$,
the lines do not coincide; they correspond to the two
intersections of the toroidal Fermi pocket with the $k_xk_y$
plane. We conclude that these two intersections are
separately protected by $\mathbb{Z}_2$ invariants, namely the
relative signs of the two Pfaffians
$\mathrm{Pf}\,\tilde{\mathcal{H}}_\pm(\mathbf{k})$. This
implies that the toroidal pocket cannot be transformed into
several spheroidal pockets, or a string of sausages, by
symmetry-preserving changes of the Hamiltonian.
However, it is possible to shrink the inner edge to a point
and annihilate it, which transforms the toroidal pocket into
a spheroidal one. Then, the outer edge could also be
contracted to a point and annihilated, which would gap out
the whole pocket. This is possible since it is not protected
by a nonzero Chern number.

An analogous argument can be made based on twofold rotation
symmetry about the $z$-axis.
The symmetry can be expressed as
\begin{equation}
U_{\sigma z}\, \mathcal{H}(-k_x,-k_y,k_z)\,
  U_{\sigma z}^\dagger = \mathcal{H}(k_x,k_y,k_z) \,,
\end{equation}
with the same matrix $U_{\sigma z}$.
This is a symmetry at fixed momentum for $k_x=k_y=0$, i.e.,
on the $k_z$-axis. Also here we find separate Pfaffians
protecting the two intersections of the $k_z$-axis with each
of the spheroidal Fermi pockets. The spheroidal pockets are
thus triply protected---by a Chern number and by two
Pfaffians.

\section{Phenomenological theory}

The physics underlying Bogoliubov Fermi surfaces can be
included in an intuitive extension of the usual Landau
free-energy functional \cite{SiU91} to the TRSB pairing
states considered here. Specifically, we have seen how the
pseudomagnetic field responsible for inflating the point
and line nodes into
Bogoliubov Fermi surfaces is related to the projection of
the nonunitary part of the gap product into the low-energy
states: the time-reversal-odd part of the nonunitary
gap product corresponds to a magnetic order parameter
and the pseudospin polarization of the low-energy states
results in a corresponding physical magnetization. Thus,
the expectation value of the magnetic order parameter is
nonzero in the TRSB nonunitary superconducting states. As
shown in~\Sec{sec:general_model}, this is directly
responsible for the appearance of the Bogoliubov Fermi
surfaces.

The induced magnetic order can be included in the Landau
expansion of the free energy as a subdominant order
parameter. For example, in the case of $E_{g}$
pairing in the cubic superconductor, the expansion
reads
\begin{align}
F_{E_{g}} &= \alpha\, |\bm{h}|^2 + \beta_1\, |\bm{h}|^4
  + \beta_2\, |\bm{h}\times\bm{h}^\ast|^2 \notag \\
&\quad{}+ \alpha_{\text{AIAO}}\, \varphi^2
  + i\kappa_{\text{AIAO}}\, \hat{\bf
  z}\cdot(\bm{h}\times\bm{h}^\ast)\, \varphi \,,
\end{align}
where we refine our notation for the vector order
parameter as
$\bm{h}=\Delta_0{\bf h}
  = h_{3z^2-r^2}\hat{\bf x}+h_{x^2-y^2}\hat{\bf y}$,
so as to be able
to define the cross product. The choice $\beta_2<0$
stabilizes the TRSB pairing state. The first line
corresponds to the free energy of the purely
superconducting state, while the second describes
the coupling between the pairing and the magnetic order
parameter $\varphi$, which, following the correspondence
to the pyrochlore lattice, we designate as the AIAO
order. For the AIAO order to be subdominant, we
require $\alpha_{\text{AIAO}}>0$. It can then
only appear if
$\hat{\bf z}\cdot(\bm{h}\times\bm{h}^\ast)$ is nonzero,
as is realized in the TRSB state. Although the term
coupling the TRSB $E_g$ pairing and the AIAO order is
generally allowed on symmetry grounds, the coupling
a nonzero constant $\kappa_{\text{AIAO}}$ is necessary for
the TRSB state to support Bogoliubov Fermi surfaces.

Similarly, the free energy of the $T_{2g}$ states is
\begin{align}
F_{T_{2g}} &= \alpha\, |\bm{l}|^2
  + \beta_1\, |\bm{l}|^4
  + \beta_2\, |\bm{l}\cdot\bm{l}|^2
  + \beta_3\, \sum_{n>m}|l_n|^2|l_m|^2 \notag \\
&\quad{}+ \alpha_{\text{SI}}\, |{\bf M}|^2
  + i\kappa_{\text{SI}}\,
  (\bm{l}\times\bm{l}^\ast)\cdot{\bf M} \,,
\end{align}
where $\bm{l}=\Delta_0{\bf l}$ is the vector order
parameter. In the first line, the condition
$0<\beta_3<4\beta_2$ stabilizes the chiral
state, whereas the cyclic state requires that
$\beta_3<0<\beta_2$~\cite{SiU91}; all other choices
yield a time-reversal-symmetric pairing state.
The second line describes the coupling to the
subdominant magnetic order parameter 
${\bf M}$, which we call the SI order, again referencing
the magnetic phases of the pyrochlore lattice. A nonzero
coupling constant $\kappa_{\text{SI}}$ indicates that a
TRSB superconducting state will develop Bogoliubov Fermi
surfaces. We note that when all components of the
induced SI order parameter are nonzero (as in the case
of the cyclic state), an additional AIAO order is
generally present due to particle-hole asymmetry in the
normal-state dispersion~\cite{GRDS17}. This AIAO state
will be of order $\Delta_0^6$, however, and is therefore
negligible in the weak-coupling limit.

The case for the hexagonal superconductor is
analogous, but illustrates an interesting interplay
between the magnetic and orbital orders. For example, in
the case of the $E_{2g}$ pairing, we can expand the free
energy as
\begin{align}
F_{E_{2g}} &= \alpha |\bm{r}|^2 + \beta_1|\bm{r}|^4
  + \beta_2 |\bm{r} \times\bm{r}^\ast|^2 \notag \\
&\quad {}+ \alpha_{M}M^2 + \alpha_{O}O^2
  + \alpha_{MO}MO \notag \\
&\quad {}+ i\kappa_{O}\, \hat{\bf z}
  \cdot(\bm{r}\times\bm{r}^\ast)\, O \,,
\end{align}
where
$\bm{r}
  = \Delta_{x^2-y^2}\hat{\bf x}+\Delta_{xy}\hat{\bf y}$
is the vector order parameter of the superconducting
state. As for the $E_{1g}$ irrep of the cubic
superconductor, the TRSB state
$\bm{r}_{\pm}=\Delta_0\, (1,\pm i)$ is stabilized
by $\beta_2<0$. On the second line, $M$ and $O$ represent
the magnetic and orbital orders, respectively. Note that
since $M$ and $O$ belong to the same irrep ($A_{2g}$), a
bilinear coupling between the two is allowed on symmetry
grounds, and is in general nonzero due to the
spin-orbit-coupling term proportional to
$\chi_{2}\otimes\sigma_3$ in the normal-state
Hamiltonian \eq{eq:hexham}. Finally, in the last
line we have the coupling between the superconducting and
orbital orders. When the superconducting state breaks TRS,
the induced orbital order in turn induces a magnetization
via the bilinear term in the second line.

The examples discussed above illustrate an important
concept in the theory of ``intertwined'' orders
\cite{FKT15,FOS18}: a multidimensional ``primary'' order,
here represented by the vector order parameters of the
superconducting states, can combine to form a
``composite'' order, in this case the nonunitary part of
the gap product. A concrete example of this in a
related system was recently given in~\cite{BAA18},
where a nonunitary chiral $d$-wave superconducting state
on the honeycomb lattice was shown to generate a loop
current order. Since the composite and the primary orders
break different symmetries, in principle these orders
can appear at different temperatures. This raises the
intriguing possibility that the induced magnetic
or orbital order preempts the superconductivity.

\section{Summary and conclusions}

In this paper, we have presented a general theory
of Bogoliubov Fermi surfaces, which generically appear
in multiband inversion-symmetric (even-parity) superconducting states 
with spontanously broken TRS. We have focused on the
case of electrons with four-valued internal degrees of
freedom. Our results do not depend on any specific
origin of these degrees of freedom. Moreover, the
generalization to a larger number of internal degrees
of freedom is straightforward.

The four-fold internal degree of freedom allows for exotic
  \emph{internally anisotropic pairing} states. Even for an $s$-wave
  momentum-independent pairing potentials, these states can transform
  nontrivially under 
lattice symmetries due to the dependence on
the internal degrees of freedom. We have shown that this typically
implies that the pairing potential is nonunitary, but 
nonunitarity alone is not sufficient for the existence
of Bogoliubov Fermi surfaces. Rather, a
time-reversal-odd part of the nonunitary gap product is
required, as we have discussed in detail. The Bogoliubov
Fermi surfaces are topologically protected by a
$\mathbb{Z}_2$ invariant, which we have given explicitly
in terms of the Pfaffian of the BdG Hamiltonian
transformed into antisymmetric form \cite{ABT17}. The
physics can be understood based on an effective
low-energy single-band model. In this model, TRSB
superconductivity generates a pseudomagnetic field that
is closely linked to the time-reversal-odd gap product.
The pseudomagnetic
field inflates point and line nodes into Bogoliubov
Fermi surfaces.

The Bogoliubov Fermi surfaces originating from inflated
point nodes retain their topological protection by
nonzero Chern numbers, providing an example for multiply
protected nodes \cite{BzS17}. In addition, at
high-symmetry planes and lines in the Brillouin zone,
additional topological invariants can be constructed in
terms of Pfaffians. These invariants further restrict
the possible deformations of Bogoliubov Fermi surfaces
by changing the Hamiltonian without breaking symmetries,
as we have discussed. Furthermore, we have constructed a
phenomenological Landau theory, which includes the
magnetic order that is induced by the pseudomagnetic field in the
TRSB superconducting state. Intriguingly, in this
formulation, the magnetic order may appear as a composite order
parameter based on fluctuations of the primary
superconducting order parameter, constituting an
example of intertwined orders~\cite{FKT15,FOS18,BAA18}.

Our general findings have been illustrated for two
specific models: a cubic system of electrons with total
angular momentum $j=3/2$, which generically appear close
to the $\Gamma_8$ band-touching point, and a hexagonal
superconductor with internal spin and orbital degrees of
freedom. For each model, we have shown the Bogoliubov
Fermi surfaces in the TRSB superconducting states
expected from symmetry analysis and Landau theory. We
have also obtained the pseudomagnetic field and the
physical magnetization of the low-energy quasiparticles.
Moreover, we have plotted the dispersion of surface
states, which exhibit Fermi arcs consistent with the
Chern numbers $0$, $\pm 2$, $\pm 4$ of the Fermi
pockets. The hexagonal model is particularly
interesting because one of its pure-irrep TRSB pairing
states show double Weyl points in the single-band limit,
which become inflated into comparatively large
Bogoliubov Fermi pockets.

As the next steps, it is necessary to work out detailed
experimental signatures of the Bogoliubov Fermi
surfaces. Experiments probing the finite density of states at the
Fermi energy and the magnetization of low-energy quasiparticles are
most promising. For example, the magnetization could lead to a
magneto-optical Kerr effect.
The magneto-optical Kerr effect
  has been observed in a number of heavy Fermion superconductors
  \cite{sch14,sch15,lev18}, suggesting that this class of materials
  represents an ideal class to search for Bogoliubov Fermi
  surfaces. One promising candidate is URu$_2$Si$_2$, for which the
  finite-field normal state to superconducting transition is first
  order at low temperatures \cite{kas07}. This suggests a pseudospin
  singlet pairing state. In addition URu$_2$Si$_2$
  exhibits a finite polar Kerr signal in the superconducting state
  \cite{sch15} and  there is evidence for a residual density of states
  in zero-field thermal conductivity data \cite{kas07}. Prior to the
  theoretical prediction of Bogoliubov Fermi surfaces, this residual
  density of states has been interpreted as a consequence of impurity
  scattering \cite{kas07}. Our theory suggests that it is worthwhile
  to experimentally revisit
  this interpretation. A second promising candidate material is
  thoriated UBe$_{13}$ which is also observed to break time-reversal
  symmetry \cite{hef90}. In this material, there are specific heat
  measurements revealing a residual density of states that can be
  reversibly changed by more than a factor of two through the
  application of pressure \cite{zie04}. This suggests that this
  residual density states is intrinsic and not a consequence of
  impurity scattering. Bogoliubov Fermi surfaces provides a natural
  explanation for this observation and it would be of interest to
  experimentally revisit this material as well.

\begin{acknowledgments}

The authors thank D. S. L. Abergel, T. Bzdu\v{s}ek, L. Savary, A. P. Schnyder, 
  J. W. F. Venderbos, G. Volovik, and 
  V. M. Yakovenko for stimulating discussions. 
C.\,T. acknowledges financial support by the Deutsche
Forschungsgemeinschaft, in part through Research Training
Group GRK 1621 and Collaborative Research Center
SFB 1143. D.\,F.\,A. acknowledges financial support
through the UWM growth initiative. P.\,M.\,R.\,B acknowledges
  the hospitality of the TU Dresden, where part of this work was completed.

\end{acknowledgments}

\appendix

\section{Pseudospin basis}
\label{app.ps}

The presence of TRS $T$ and IS $P$
in the normal-state Hamiltonian allows us to label the
doubly degenerate eigenstates by a pseudospin
index $s$. The pseudospin basis represents a manifestly
covariant Bloch basis (MCBB) if the pseudospin index can
be chosen so as to transform like a spin $1/2$ under the
symmetries of the lattice. We can define an MCBB as
follows. Let $\phi_{{\bf k},\pm,s}$ be the
orthonormalized four-component eigenvectors of the
normal-state Hamiltonian $H_0({\bf k})$ to
eigenvalues $E_{{\bf k},\pm}$, i.e.,
\begin{equation}
H_{0}({\bf k})\phi_{{\bf k},\pm,s}
  = E_{{\bf k},\pm}\phi_{{\bf k},\pm,s}\,.
\end{equation}
Consider a symmetry operation $g$ of the point group
such that
\begin{equation}
U_gH_0({\bf k})U_g^\dagger = H_0(g{\bf k}) \,,
\end{equation}
where $U_g$ is the unitary matrix for the symmetry
operation in the four-component basis. The eigenvectors
$\phi_{{\bf k},\pm,s}$ define an MCBB if the matrix with
columns composed of these vectors,
\begin{equation}
\Phi_{\bf k} = (\phi_{{\bf k},+,\uparrow},
  \phi_{{\bf k},+,\downarrow},
  \phi_{{\bf k},-,\uparrow},
  \phi_{{\bf k},-,\downarrow}) \,,
\end{equation}
satisfies
\begin{equation}
\Phi_{g{\bf k}}^\dagger U_{g}\Phi_{{\bf k}}
  = {s}_0\otimes u_g \,,
\end{equation}
where $u_{g}$ is the equivalent symmetry operation for
a spin-$1/2$ system.

\subsection{Cubic superconductor}

The MCBB adopted to obtain the plots of the
pseudomagnetic field for the superconducting
states of the cubic model is defined by the
eigenvectors of the normal-state Hamiltonian in
\eq{h.normal.3},
\begin{align}
\psi_{{\bf k},+,\uparrow} &= \frac{1}{\sqrt{2}}
  \left(\begin{array}{c}
  -ie^{-i\phi}\sin\theta \cos\frac{\zeta-\xi}{2} \\
  -i\cos \frac{\zeta+\xi}{2}
  +\cos\theta \sin\frac{\zeta-\xi}{2} \\
  -ie^{-i\phi}\sin\theta \sin\frac{\zeta-\xi}{2} \\
  -i\sin\frac{\zeta+\xi}{2}
  +\cos\theta \cos\frac{\zeta-\xi}{2}
  \end{array}\right), \\
\psi_{{\bf k},+,\downarrow} &= \frac{1}{\sqrt{2}}
  \left(\begin{array}{c}
  i\sin\frac{\zeta+\xi}{2}
  +\cos\theta \cos\frac{\zeta-\xi}{2} \\
  -ie^{i\phi}\sin\theta \sin\frac{\zeta-\xi}{2} \\
  i\cos\frac{\zeta+\xi}{2}
  +\cos\theta \sin\frac{\zeta-\xi}{2} \\
  -ie^{i\phi}\sin\theta \cos\frac{\zeta-\xi}{2}
\end{array}\right),\\
\psi_{{\bf k},-,\uparrow} &= \frac{1}{\sqrt{2}}
  \left(\begin{array}{c}
  -ie^{-i\phi}\sin\theta \cos\frac{\zeta+\xi}{2}\\
  i\cos\frac{\zeta-\xi}{2}
  -\cos\theta \sin\frac{\zeta+\xi}{2} \\
  ie^{-i\phi}\sin\theta \sin\frac{\zeta+\xi}{2} \\
  -i\sin\frac{\zeta-\xi}{2}
  +\cos\theta \cos\frac{\zeta+\xi}{2}
\end{array}\right), \\
\psi_{{\bf k},-,\downarrow} &= \frac{1}{\sqrt{2}}
  \left(\begin{array}{c}
  i\sin\frac{\zeta-\xi}{2}
  +\cos\theta \cos\frac{\zeta+\xi}{2}\\
  ie^{i\phi}\sin\theta \sin\frac{\zeta+\xi}{2} \\
  -i\cos\frac{\zeta-\xi}{2}
  -\cos\theta \sin\frac{\zeta+\xi}{2}\\
  -ie^{i\phi}\sin\theta \cos\frac{\zeta+\xi}{2}
  \end{array}\right) ,
\end{align}
where the angles are defined as
\begin{align}
\phi &= \arctan \frac{\epsilon_{{\bf k},yz}}
  {\epsilon_{{\bf k},xz}} \,, \\
\theta &= \arctan \frac{\sqrt{\epsilon_{{\bf k},yz}^2
  +\epsilon_{{\bf k},xz}^2}}
  {\epsilon_{{\bf k},xy}} \,, \\
\xi &= \arctan \frac{\epsilon_{{\bf k},3z^2-r^2}}
  {\epsilon_{{\bf k},x^2-y^2}} \,, \\
\zeta &= \arctan
  \frac{\sqrt{\epsilon_{{\bf k},x^2-y^2}^2
  +\epsilon_{{\bf k},3z^2-r^2}^2}}
  {\sqrt{\epsilon^2_{{\bf k},yz}
  +\epsilon_{{\bf k},xz}^2+\epsilon_{{\bf k},xy}^2}}
\end{align}
and $\epsilon_{{\bf k},\mu}$ is the coefficient
of the matrix $\gamma_\mu$
defined in Eqs.~(\ref{eq:gammaJ1})--(\ref{eq:gammaJ5})
in the Hamiltonian \eq{h.normal.3}.
It can be verified that one or more of these angles are
ill defined along the $[100]$ and $[111]$ and
symmetry-related directions, where the
pseudospin-$1/2$ description breaks down.  Along
these high-symmetry directions
$\hat{\bf n}_{\text{hs}}$, the Hamiltonian commutes with
$\hat{\bf n}_{\text{hs}}\cdot{\bf J}$ so that
the eigenstates transform under rotations like $j=3/2$
particles. Away from these directions, however,
cubic anisotropy lowers the symmetry of the eigenstates,
permitting a pseudospin-$1/2$ description.

Expressed in the MCBB defined above, the interband
pairing potentials, i.e., $\psi_{{\bf k},I}$ and
${\bf d}_{\bf k}$ in~\eq{eq:Deltaps}, have the compact
forms
\begin{align}
\psi_{\bm{k},I}
  &= [\vec{\eta}_{{\bf k},E_g}
  \times \hat{\epsilon}_{{\bf k},E_g}]\cdot{\bf e}_z\,, \\
{\bf d}_{\bf k}
  &= \vec{\eta}_{{\bf k},T_{2g}}
  \times \hat{\epsilon}_{{\bf k},T_{2g}}
  - \frac{\abs{\vec{\epsilon}_{{\bf k},E_g}}}
  {\abs{\vec{\epsilon}_{\bf k}}}\,
  (\vec{\eta}_{{\bf k},T_{2g}}
  \cdot \hat{\epsilon}_{{\bf k},T_{2g}})
  \hat{\epsilon}_{{\bf k},T_{2g}} \notag \\
&\quad {}+ \frac{1}{\abs{\vec{\epsilon}_{\bf k}}}
  (\vec{\eta}_{{\bf k},E_g}
  \cdot \hat{\epsilon}_{{\bf k},E_g})
  \vec{\epsilon}_{{\bf k},T_{2g}}\,,
\end{align}
where we use the short-hand notation
\begin{align}
\vec{v} &= (v_{3z^2-r^2}, v_{x^2-y^2}, v_{xy}, v_{xz},
  v_{yz})\,, \\
\vec{v}_{E_g} &= v_{3z^2-r^2}{\bf e}_x
  + v_{x^2-y^2}{\bf e}_y\,, \\
\vec{v}_{T_{2g}} &= v_{xy}{\bf e}_x+ v_{xz}{\bf e}_y
  + v_{yz}{\bf e}_z\,, \\
\hat{v} &= \vec{v} / \abs{\vec v}
\end{align}
and $v$ is either $\epsilon$ or $\eta$.

\subsection{Hexagonal superconductor}

The MCBB adopted to obtain the plots of the
pseudomagnetic field for the superconducting
states of the hexagonal model is defined by the
eigenvectors of the normal-state Hamiltonian in
\eq{eq:hexham},
\begin{align}
\psi_{{\bf k},+,\uparrow} &= \frac{1}{\sqrt{2}}
    \left(\begin{array}{c}
    e^{-i \phi}\cos\frac{\zeta}{2} (\cos\tfrac{\xi}{2}
    + e^{i\theta} \sin\tfrac{\xi}{2}) \\
    \sin\tfrac{\zeta}{2}(\sin\tfrac{\xi}{2}
    - e^{i\theta}\cos\tfrac{\xi}{2}) \\
    ie^{-i \phi}\cos\frac{\zeta}{2} (\cos\tfrac{\xi}{2}
    - e^{i\theta} \sin\tfrac{\xi}{2}) \\
    i\sin\tfrac{\zeta}{2}(\sin\tfrac{\xi}{2}
    + e^{i\theta}\cos\tfrac{\xi}{2})
  \end{array}\right), \\
\psi_{{\bf k},+,\downarrow} &= \frac{1}{\sqrt{2}}
    \left(\begin{array}{c}
    \sin\tfrac{\zeta}{2}(e^{-i\theta}\cos\tfrac{\xi}{2}
    - \sin\tfrac{\xi}{2}) \\
    e^{i \phi}\cos\frac{\zeta}{2} (e^{-i\theta}
    \sin\tfrac{\xi}{2} + \cos\tfrac{\xi}{2}) \\
    i\sin\tfrac{\zeta}{2}(e^{-i\theta}\cos\tfrac{\xi}{2}
    + \sin\tfrac{\xi}{2})\\
    ie^{i \phi}\cos\frac{\zeta}{2}
    (e^{-i\theta} \sin\tfrac{\xi}{2}-\cos\tfrac{\xi}{2})
  \end{array}\right), \\
\psi_{{\bf k},-,\uparrow} &= \frac{1}{\sqrt{2}}
    \left(\begin{array}{c}
    e^{-i \phi}\cos\frac{\zeta}{2} (\sin\tfrac{\xi}{2}
    -e^{i\theta}\cos\tfrac{\xi}{2}) \\
    -\sin\tfrac{\zeta}{2}(\cos\tfrac{\xi}{2}
    +e^{i\theta}\sin\tfrac{\xi}{2}) \\
    ie^{-i \phi}\cos\frac{\zeta}{2} (\sin\tfrac{\xi}{2}
    +e^{i\theta}\cos\tfrac{\xi}{2})\\
    i\sin\tfrac{\zeta}{2}(e^{i\theta}\sin\tfrac{\xi}{2}
    -\cos\tfrac{\xi}{2})
  \end{array}\right), \\
\psi_{{\bf k},-,\downarrow} &= \frac{1}{\sqrt{2}}
    \left(\begin{array}{c}
    \sin\tfrac{\zeta}{2}(\cos\tfrac{\xi}{2}
    +e^{-i\theta}\sin\tfrac{\xi}{2}) \\
    e^{i \phi}\cos\frac{\zeta}{2} (\sin\tfrac{\xi}{2}
    -e^{-i\theta}\cos\tfrac{\xi}{2}) \\
    i\sin\tfrac{\zeta}{2}
    (e^{-i\theta}\sin\tfrac{\xi}{2}
    -\cos\tfrac{\xi}{2}) \\
    -ie^{i \phi}\cos\frac{\zeta}{2} (\sin\tfrac{\xi}{2}
    +e^{-i\theta}\cos\tfrac{\xi}{2})
  \end{array}\right),
\end{align}
where the angles are defined as
\begin{align}
\phi &= \arctan \frac{\epsilon_{{\bf k},22}}
  {\epsilon_{{\bf k},21}} \,, \\
\theta &= \arctan \frac{\epsilon_{{\bf k},10}}
  {\epsilon_{{\bf k},30}}  \,, \\
\xi &= \arctan \frac{\sqrt{\epsilon_{{\bf k},21}^2
  + \epsilon_{{\bf k},22}^2}}
  {\sqrt{\epsilon_{{\bf k},10}^2
  + \epsilon_{{\bf k},30}^2}} \,, \\
\zeta &= \arctan \frac{\sqrt{\epsilon_{{\bf k},10}^2
  + \epsilon_{{\bf k},30}^2+\epsilon_{{\bf k},21}^2
  + \epsilon_{{\bf k},22}^2}}{\epsilon_{{\bf k},23}}
\end{align}
and $\epsilon_{{\bf k},\mu\nu}$ is the
coefficient of the matrix
$\chi_{\mu}\otimes\sigma_{\nu}$ in \eq{eq:hexham}.

Similarly to the cubic superconductor, the pseudospin
description breaks down along six- and three-fold
rotation axes. The $E_{1g}$ orbitals transform under
these rotations as if they have angular momentum
$L_z=\pm 1$. Combining this with the spin degrees of
freedom, one can therefore construct states with an
effective total angular momentum $j_z=\pm 3/2$, in
addition to $j_z=\pm 1/2$ states. Away from these
lines, however, the hexagonal crystal anisotropy
quenches the orbital angular momentum, and so a
pseudospin-$1/2$ description is possible.

Expressed in the MCBB defined above, the
interband pairing potentials, i.e.,
$\psi_{{\bf k},I}$ and ${\bf d}_{\bf k}$ in
\eq{eq:Deltaps}, have the compact forms
\begin{align}
\psi_{\bm{k},I} &=
  \frac{\sqrt{|\vec{\epsilon}_{\bf k}|^2
  -\epsilon_{{\bf k},23}^2}}
  {|\vec{\epsilon}_{\bf k}|}\, \eta_{{\bf k},23}
  \notag \\
&\quad {}+ \frac{\epsilon_{{\bf k},23}}
  {|\vec{\epsilon}_{\bf k}|}\,
  \frac{\vec{\eta}_{{\bf k},E_{1g}}\cdot
  \vec{\epsilon}_{{\bf k},E_{1g}}
  +\vec{\eta}_{{\bf k},E_{2g}}\cdot
  \vec{\epsilon}_{{\bf k},E_{2g}}}
  {\sqrt{|\vec{\epsilon}_{\bf k}|^2
  -\epsilon_{{\bf k},23}^2}}\,, \\
d_{{\bf k},x} &= \frac{\left(
  [\vec{\epsilon}_{{\bf k},E_{2g}} \times
  \vec{\epsilon}_{{\bf k},E_{1g}}]
  \cdot{\bf e}_z\right)}
  {\sqrt{|\vec{\epsilon}_{\bf k}|^2
  -\epsilon_{{\bf k},23}^2}}\,
  \bigg(\frac{\vec{\eta}_{{\bf k},
  E_{2g}}\cdot\hat{\epsilon}_{{\bf k},
  E_{2g}}}{|\vec{\epsilon}_{{\bf k},E_{2g}}|} \notag \\
& \quad {}- \frac{\vec{\eta}_{{\bf k},E_{1g}}
  \cdot\hat{\epsilon}_{{\bf k},
  E_{1g}}}{|\vec{\epsilon}_{{\bf k},E_{1g}}|}\bigg)
  \notag \\
& \quad {}- \left(\vec{\epsilon}_{{\bf k},E_{2g}}
  \cdot\vec{\epsilon}_{{\bf k},E_{1g}}\right)
  \left([\vec{\eta}_{{\bf k},E_{1g}}
  \times \hat{\epsilon}_{{\bf k},E_{1g}}]
  \cdot{\bf e}_z\right) ,\\
d_{{\bf k},y} &=
  \frac{\vec{\epsilon}_{{\bf k},E_{2g}}\cdot
  \vec{\epsilon}_{{\bf k},E_{1g}}}
  {\sqrt{|\vec{\epsilon}_{\bf k}|^2
  -\epsilon_{{\bf k},23}^2}}\,
  \bigg(\frac{\vec{\eta}_{{\bf k},E_{2g}}
  \cdot\hat{\epsilon}_{{\bf k},E_{2g}}}
  {|\vec{\epsilon}_{{\bf k},E_{2g}}|} \notag \\
& \quad {}- \frac{\vec{\eta}_{{\bf k}, E_{1g}}\cdot
  \hat{\epsilon}_{{\bf k},E_{1g}}}
  {|\vec{\epsilon}_{{\bf k},E_{1g}}|}\bigg) \notag \\
& \quad {}+ \big([\hat{\epsilon}_{{\bf k},E_{2g}}
  \times \hat{\epsilon}_{{\bf k},E_{1g}}]
  \cdot{\bf e}_z\big)\big([\vec{\eta}_{{\bf k},E_{1g}}
  \times \hat{\epsilon}_{{\bf k},E_{1g}}]
  \cdot{\bf e}_z\big)\,, \\
d_{{\bf k},z} &= \vec{\eta}_{{\bf k},E_{2g}}
  \cdot\hat{\epsilon}_{{\bf k},E_{2g}}\,,
\end{align}
where we use the short-hand notation
\begin{align}
\vec{v} &= (v_{23}, v_{21}, v_{22}, v_{30},
  v_{10})\,, \\
\vec{v}_{E_{1g}} &= v_{21}{\bf e}_x
  + v_{22}{\bf e}_y\,, \\
\vec{v}_{E_{2g}} &= v_{30}{\bf e}_x
  + v_{10}{\bf e}_y\,, \\
\hat{v} &= \vec{v} / \abs{\vec v}
\end{align}
and $v$ is either $\epsilon$ or $\eta$.

\section{Physical magnetization}
\label{app.magnet}

In this Appendix, we derive \eq{mphys.2} for the
contribution to the physical magnetization due to states
close to the normal-state Fermi surface at momentum
$\mathbf{k}$. We start from the expectation value
$\langle\mathbf{J}\rangle$ of angular momentum in the
$-$ band state at $\mathbf{k}$. The $-$ band is
split by the pseudomagnetic field
$\delta\mathbf{h}_{\mathbf{k},-}$.
$\langle\mathbf{J}\rangle$ is the expectation value in
the lower-energy state resulting from this splitting,
which according to \eq{eq:deltaHminus} is the state with
pseudospin {antiparallel} to
$\delta\mathbf{h}_{\mathbf{k},-}$. This state reads
\begin{equation}
|\psi_{\mathbf{k},-}\rangle = \sin\frac{\theta}{2}\,
  |\mathbf{k},-,\uparrow\rangle
  - e^{i\phi} \cos\frac{\theta}{2}\,
  |\mathbf{k},-,\downarrow\rangle \,,
\end{equation}
where $\theta$ and $\phi$ are the spherical coordinates
describing the direction of
$\delta\mathbf{h}_{\mathbf{k},-}$. The expectation value
is then, in components,
\begin{widetext}
\begin{align}
\langle\psi_{\mathbf{k},-}| J_\mu
  |\psi_{\mathbf{k},-}\rangle
  &= \left( \langle\mathbf{k},-,{\uparrow}|\,
    \sin\frac{\theta}{2}
  + \langle\mathbf{k},-,{\downarrow}|\, e^{-i\phi}
    \cos\frac{\theta}{2} \right) J_\mu
  \left( \sin\frac{\theta}{2}\,
    |\mathbf{k},-,\uparrow\rangle
  - e^{i\phi} \cos\frac{\theta}{2}\,
  |\mathbf{k},-,\downarrow\rangle \right) \notag \\
&= \frac{1}{2} \sum_{ss'} \left(\begin{array}{cc}
  1 - \cos\theta & - e^{i\phi} \sin\theta \\
  - e^{-i\phi} \sin\theta & 1 + \cos\theta
  \end{array}\right)_{\!ss'} \,
  \langle\mathbf{k},-,s| J_\mu |\mathbf{k},-,s'\rangle \,.
\end{align}
\end{widetext}
The contribution from the unit matrix $s_0$ in
pseudospin space vanishes due to $PT$ symmetry of the
normal state. The expression then reads
\begin{align}
\langle&\psi_{\mathbf{k},-}| J_\mu
  |\psi_{\mathbf{k},-}\rangle \notag \\
&= -\frac{1}{2} \sum_{ss'} \left(
  \widehat{\delta\mathbf{h}}_{\mathbf{k},-}
  \cdot \mathbf{s}^T\right)_{ss'}
  \langle\mathbf{k},-,s| J_\mu
  |\mathbf{k},-,s'\rangle \,,
\end{align}
where $\widehat{\delta\mathbf{h}}_{\mathbf{k},-}$ is the
unit vector in the direction of the pseudomagnetic field.
With the help of the $4\times 4$ pseudospin operator
$\check{\mathbf{s}}$, see \eq{eq:checks}, we can rewrite
the matrix element as
\begin{align}
\langle&\psi_{\mathbf{k},-}| J_\mu
  |\psi_{\mathbf{k},-}\rangle \notag \\
  &= -\frac{1}{2} \sum_{ss'}
  \widehat{\delta\mathbf{h}}_{\mathbf{k},-} \cdot
  \langle\mathbf{k},-,s| \check{\mathbf{s}}^T
  |\mathbf{k},-,s'\rangle \langle\mathbf{k},-,s|
  J_\mu |\mathbf{k},-,s'\rangle \notag \\
&= -\frac{1}{2} \sum_{ss'}
  \widehat{\delta\mathbf{h}}_{\mathbf{k},-} \cdot
  \langle\mathbf{k},-,s'| \check{\mathbf{s}}
  |\mathbf{k},-,s\rangle
  \langle\mathbf{k},-,s| J_\mu
  |\mathbf{k},-,s'\rangle \notag \\
&= -\frac{1}{2}\,
  \widehat{\delta\mathbf{h}}_{\mathbf{k},-} \cdot
  \Tr \mathcal{P}_{\mathbf{k},-} \check{\mathbf{s}}
  \mathcal{P}_{\mathbf{k},-} J_\mu \,.
\end{align}
One can show that the result satisfies
$|\langle\psi_{\mathbf{k},-}| \mathbf{J}
  |\psi_{\mathbf{k},-}\rangle|\le 3/2$, as expected.

To obtain the contribution to the magnetization from
states in the vicinity of $\mathbf{k}$ at the
Fermi surface, we sum
$\langle\psi_{\mathbf{k}+\delta\mathbf{q},-}| J_\mu
  |\psi_{\mathbf{k}+\delta\mathbf{q},-}\rangle$
over $\delta\mathbf{q}$, where $\delta\mathbf{q}$ is
orthogonal to the Fermi surface at $\mathbf{k}$. The
sum is only over those momenta for which the
lower-energy state of the pseudospin-split $-$ band
is occupied and the upper state is empty. The energy
shifts due to the pseudomagnetic field are
$\pm |\delta\mathbf{h}_{\mathbf{k}
  +\delta\mathbf{q},-}|$,
see \eq{eq:deltaHminus}. For weak pairing, we can
neglect the dependence of
$\delta\mathbf{h}_{\mathbf{k}+\delta\mathbf{q},-}$
and $\langle\psi_{\mathbf{k}+\delta\mathbf{q},-}|
  J_\mu |\psi_{\mathbf{k}+\delta\mathbf{q},-}\rangle$
on $\delta\mathbf{q}$. Then we simply have to multiply
$\langle\psi_{\mathbf{k},-}| J_\mu
  |\psi_{\mathbf{k},-}\rangle$ by the width
$2q_\mathrm{max}$ of the momentum shell within which
only one band is occupied, where $q_\mathrm{max}$
satisfies
\begin{equation}
q_\mathrm{max}\, |\mathbf{v}_{\mathbf{k},-}|
  = |\delta\mathbf{h}_{\mathbf{k},-}| \,.
\end{equation}
Here,
$\mathbf{v}_{\mathbf{k},-}
  = \partial E_{\mathbf{k},-}/\partial \mathbf{k}$
is the Fermi velocity. The contribution to the physical
magnetization then reads
\begin{equation}
m_{\mathbf{k},\mu} = - \frac{1}
  {|\mathbf{v}_{\mathbf{k},-}|}\,
  \delta\mathbf{h}_{\mathbf{k},-} \cdot
  \Tr \mathcal{P}_{\mathbf{k},-} \check{\mathbf{s}}
  \mathcal{P}_{\mathbf{k},-} J_\mu \,,
\end{equation}
which is \eq{mphys.2}.


\begin{thebibliography}{99}
\bibitem{endnote.Eg}Rotations about the threefold
axis $[111]$ result in the degenerate solutions
$\mathbf{h} = (\alpha,\pm i \alpha)$ and
$\mathbf{h} = (\alpha^2,\pm i \alpha^2)$ with
$\alpha = e^{2\pi i/3}$. However, these are related to
$\mathbf{h} = (1,\pm i)$ by global phase changes.
\end{thebibliography}
\end{document}